\documentclass[twocolumn]{aastex631}
\usepackage{natbib}
\usepackage{graphicx}
\usepackage{graphics}
\usepackage{tabularx}
\usepackage{lipsum}
\usepackage{subfigure}
\usepackage{amsmath,amsfonts,amsthm,bm} 
\usepackage{float}
\usepackage{bm}
\usepackage{courier}
\usepackage{color}
\usepackage{colortbl}
\usepackage{threeparttable}
\usepackage{hyperref}
\hypersetup{colorlinks=true,
linkcolor=red,
citecolor=blue}
\usepackage{blindtext}
\usepackage{placeins}
\usepackage[super]{nth}
\usepackage{multirow}
\usepackage{enumitem}
\singlespace

\newcommand{\EV}[1] {\langle#1\rangle}
\newcommand{\mbf}[1] {$\mathbf #1$}

\begin{document}

\title{The Atacama Cosmology Telescope: Map-Based Noise Simulations for DR6}

\correspondingauthor{Zachary~Atkins}
\email{zatkins@princeton.edu}

\author[0000-0002-2287-1603]{Zachary~Atkins}
\affiliation{Joseph Henry Laboratories of Physics, Jadwin Hall, Princeton University, Princeton, NJ, USA 08544}

\author[0000-0003-2856-2382]{Adriaan~J.~Duivenvoorden}
\affiliation{Joseph Henry Laboratories of Physics, Jadwin Hall, Princeton University, Princeton, NJ, USA 08544}
\affiliation{Center for Computational Astrophysics, Flatiron Institute, New York, NY 10010, USA}

\author[0000-0002-1297-3673]{William~R.~Coulton}
\affiliation{Center for Computational Astrophysics, Flatiron Institute, New York, NY 10010, USA}

\author[0000-0001-7805-1068]{Frank~J.~Qu}
\affiliation{DAMTP, Centre for Mathematical Sciences, University of Cambridge, Wilberforce Road, Cambridge CB3 OWA, UK}

\author[0000-0002-1035-1854]{Simone~Aiola}
\affiliation{Center for Computational Astrophysics, Flatiron Institute, New York, NY 10010, USA}

\author[0000-0003-0837-0068]{Erminia~Calabrese}
\affiliation{School of Physics and Astronomy, Cardiff University, The Parade, Cardiff, Wales CF24 3AA, UK}

\author[0000-0001-6702-0450]{Grace~E.~Chesmore}
\affiliation{Department of Physics, University of Chicago, 5720 South Ellis Avenue, Chicago, IL 60637, USA}

\author[0000-0002-9113-7058]{Steve~K.~Choi} 
\affiliation{Department of Physics, Cornell University, Ithaca, NY 14853, USA}
\affiliation{Department of Astronomy, Cornell University, Ithaca, NY 14853, USA}

\author[0000-0002-3169-9761]{Mark~J.~Devlin}
\affiliation{Department of Physics and Astronomy, University of Pennsylvania, 209 South 33rd Street, Philadelphia, PA, USA 19104}

\author[0000-0002-7450-2586]{Jo~Dunkley}
\affiliation{Joseph Henry Laboratories of Physics, Jadwin Hall, Princeton University, Princeton, NJ, USA 08544}
\affiliation{Department of Astrophysical Sciences, Peyton Hall, Princeton University, Princeton, NJ, USA 08544}

\author[0000-0002-4765-3426]{Carlos~Herv\'ias-Caimapo}
\affiliation{Instituto de Astrof\'isica and Centro de Astro-Ingenier\'ia, Facultad de F\'isica, Pontificia Universidad Cat\'olica de Chile, Av. Vicu\~na Mackenna 4860, 7820436 Macul, Santiago, Chile}

\author[0000-0002-1697-3080]{Yilun~Guan}
\affiliation{Dunlap Institute for Astronomy and Astrophysics, University of Toronto, 50 St. George St., Toronto, ON M5S 3H4, Canada}

\author[0000-0002-2613-2445]{Adrien~La~Posta}
\affiliation{Université Paris-Saclay, CNRS/IN2P3, IJCLab, 91405 Orsay, France}

\author[0000-0002-0309-9750]{Zack~Li}
\affiliation{Canadian Institute for Theoretical Astrophysics, University of Toronto, Toronto, ON, Canada M5S 3H8}

\author[0000-0002-6849-4217]{Thibaut~Louis}
\affiliation{Université Paris-Saclay, CNRS/IN2P3, IJCLab, 91405 Orsay, France}

\author[0000-0001-6740-5350]{Mathew~S.~Madhavacheril}
\affiliation{Department of Physics and Astronomy, University of Pennsylvania, 209 South 33rd Street, Philadelphia, PA, USA 19104}

\author{Kavilan~Moodley}
\affiliation{Astrophysics Research Centre, University of KwaZulu-Natal, Westville Campus, Durban 4041, South Africa}
\affiliation{School of Mathematics, Statistics and Computer Science, University of KwaZulu-Natal, Westville Campus, Durban 4041, South Africa}

\author[0000-0002-4478-7111]{Sigurd~Naess}
\affiliation{Institute of Theoretical Astrophysics, University of Oslo, Norway}

\author[0000-0002-8307-5088]{Federico~Nati}
\affiliation{Department of Physics, University of Milano-Bicocca, Piazza della Scienza 3, 20126 Milano (MI), Italy}

\author[0000-0001-7125-3580]{Michael~D.~Niemack}
\affiliation{Department of Physics, Cornell University, Ithaca, NY 14853, USA}
\affiliation{Department of Astronomy,Cornell University, Ithaca, NY 14853, USA}

\author[0000-0002-9828-3525]{Lyman~Page}
\affiliation{Joseph Henry Laboratories of Physics, Jadwin Hall, Princeton University, Princeton, NJ, USA 08544}

\author{Roberto~Puddu}
\affiliation{Instituto de Astrof\'isica and Centro de Astro-Ingenier\'ia, Facultad de F\'isica, Pontificia Universidad Cat\'olica de Chile, Av. Vicu\~na Mackenna 4860, 7820436 Macul, Santiago, Chile}

\author[0000-0003-4006-1134]{Maria~Salatino}
\affiliation{Department of Physics, Stanford University, Stanford, California 94305}
\affiliation{Kavli Institute for Astroparticle Physics and Cosmology, Stanford, California 94305}

\author[0000-0002-8149-1352]{Crist\'obal~Sif\'on}
\affiliation{Instituto de F\'isica, Pontificia Universidad Cat\'olica de Valpara\'iso, Casilla 4059, Valpara\'iso, Chile}

\author[0000-0002-7020-7301]{Suzanne~T.~Staggs}
\affiliation{Joseph Henry Laboratories of Physics, Jadwin Hall, Princeton University, Princeton, NJ, USA 08544}

\author{Cristian~Vargas}
\affiliation{Instituto de Astrof\'isica and Centro de Astro-Ingenier\'ia, Facultad de F\'isica, Pontificia Universidad Cat\'olica de Chile, Av. Vicu\~na Mackenna 4860, 7820436 Macul, Santiago, Chile}

\author[0000-0002-2105-7589]{Eve~M.~Vavagiakis}
\affiliation{Department of Physics, Cornell University, Ithaca, NY 14853, USA}

\author[0000-0002-7567-4451]{Edward~J.~Wollack}
\affiliation{NASA Goddard Space Flight Center, 8800 Greenbelt Rd, Greenbelt, MD 20771 USA}

\begin{abstract}
The increasing statistical power of cosmic microwave background (CMB) datasets requires a commensurate effort in understanding their noise properties. The noise in maps from ground-based instruments is dominated by large-scale correlations, which poses a modeling challenge. This paper develops novel models of the complex noise covariance structure in the Atacama Cosmology Telescope Data Release 6 (ACT DR6) maps. We first enumerate the noise properties that arise from the combination of the atmosphere and the ACT scan strategy. We then prescribe a class of Gaussian, map-based noise models, including a new wavelet-based approach that uses directional wavelet kernels for modeling correlated instrumental noise. The models are empirical, whose only inputs are a small number of independent realizations of the same region of sky. We evaluate the performance of these models against the ACT DR6 data by drawing ensembles of noise realizations. Applying these simulations to the ACT DR6 power spectrum pipeline reveals a $\sim 20\%$ excess in the covariance matrix diagonal when compared to an analytic expression that assumes noise properties are uniquely described by their power spectrum. Along with our public code, \texttt{mnms}, this work establishes a necessary element in the science pipelines of both ACT DR6 and future ground-based CMB experiments such as the Simons Observatory (SO).
\end{abstract}

\section{Introduction} \label{sec: intro}
Increasingly precise cosmic microwave background (CMB) instruments and data products have the potential to place state-of-the-art constraints on astrophysical, cosmological, and theoretical models. Currently, ground-based instruments collect data at both small angular scales, including the Atacama Cosmology Telescope (ACT) \citep{ACT_telescope, Thornton2016, AdvACT} and the South Pole Telescope (SPT)  \citep{SPT_3G_1, SPT_3G_2}; and large angular scales, including POLARBEAR \citep{Polarbear_2}, CLASS \citep{CLASS_2}, and BICEP/Keck \citep{BICEP3_2}. While these instruments observe the sky with $\mathcal{O}(10^3 - 10^4)$ detectors, next-generation projects, such as the Simons Observatory (SO) \citep{SO_instrument}, the BICEP Array \citep{BICEP_Array}, and AliCPT \citep{AliCPT} will increase this count by an order of magnitude once fully operational. Thereafter, the CMB-S4 experiment \citep{S4, Barron2022} plans to deploy yet another order of magnitude increase in detector count, to several hundred thousand. With growing statistical power, the need for accurate modeling of instrument systematics and noise properties will become more urgent.

Accurate, computationally tractable noise models are a critical component for statistical inference with these new data. For instance, accurate noise models are necessary for an unbiased reconstruction of the CMB power spectrum covariance matrix \citep[e.g.,][]{MASTER, Efstathiou2021, Li2021}; for the subtraction of dominant bias terms in CMB lensing reconstruction \citep[e.g.,][]{Kesden2003_N0, Hanson2009, cross-estimator}; to check for biases and quantify the variance of internal linear combination (ILC) foreground mitigation pipelines \citep[e.g.,][]{WMAP_NILC, Mv20}; for unbiased Bayesian/parametric component separation pipelines \citep[e.g.,][]{Eriksen2008, Dunkley2009, Svalheim2020}; for optimal Wiener filtering or inverse-covariance filtering used in, for example, bispectrum \citep{Smith2011} or lensing \citep{Smith2007} estimation. A unique difficulty in the noise modeling of ground-based CMB instruments, however, is the dominance of complicated, spatially-red atmospheric noise from broad microwave water lines \citep[e.g.,][]{D13, Takakura2017, Morris2022} at large scales. Importantly, increasing detector counts do not circumvent this noise component. Therefore, our experience with current CMB experiments offers a realistic preview of the noise modeling challenge facing future surveys. 

Existing noise modeling frameworks either do not scale well to ground-based CMB data volumes, or simply ignore prominent features in the data. A reasonably successful, though resource-intensive, approach has been to model instrumental noise in the time-domain, and then propagate noise timestream realizations into map space \citep[e.g.,][]{FFP10, NPIPE, Chown2018, Bleem2022} through a mapmaking pipeline. For \textit{Planck}, 25 million CPU hours were spent processing the FFP8 (full focal plane) simulations \citep{FFP8}. As the size of current and future ground-based experiments' time-ordered data (TOD) exceeds that of \textit{Planck},\footnote{The total TOD volume included in ACT DR6 is $\sim$190\,TB. For \textit{Planck}, the total TOD volume is $\sim$13\,TB. SO is expected to generate $\sim$2\,TB of TOD per \textit{day} of observations.} this route will grow more costly. Many analyses also make use of analytic noise models from mapmaker products, such as low-resolution, dense pixel-pixel noise covariance matrices from \textit{WMAP} \citep{Jarosik2007} or per-pixel noise variance maps from \textit{Planck} \citep{Planck-overview:2018}. For high-resolution ground-based experiments, the former approach is numerically intractable and the latter approach neglects the strongly correlated atmospheric noise. Other analyses have taken a hybrid approach to model both spatially-varying, uncorrelated noise variance and large-scale noise correlations simultaneously \citep[e.g.,][]{DeBelsunce2021}. More recently, ACT \citep{L17, C20} and SPT \citep{Millea2021} have developed noise models that are more appropriate for ground-based instruments. In particular, both collaborations used a model that captures spatially-varying, uncorrelated noise variance, as well as a stationary, two-dimensional (2D) ``stripy" component (defined in either Fourier space or spherical harmonic space) resulting from the scan strategy.\footnote{Henceforth, we will refer to ``spherical harmonic space" by a shorthand, ``harmonic space." Also, while ``stationary" typically describes 1D signals over time, here we take it to mean ``invariant under translations in the 2D space discussed" --- map space, Fourier space, or harmonic space.} Nevertheless, by limiting the non-white noise to a stationary representation, these models do not generalize to surveys with wide sky coverage, such as the complete ACT survey, or the SO and CMB-S4 planned surveys. 

ACT's sixth data release (ACT DR6) will deliver microwave sky maps in three frequency bands (f090, f150, and f220), covering $\sim$40\% of the sky to a cumulative depth of $\lesssim15\,\mu$K-arcmin ($\lesssim20\,\mu$K-arcmin) in temperature (polarization). In this paper, we develop novel noise models for ACT DR6, which aim to capture the observed, complicated map space noise structure inherent to ground-based CMB instruments. These models are empirical: they are built directly from ACT maps and make few \textit{a priori} assumptions about particular map properties. We begin with an adapted version of a \textit{tiled noise model} from \citet{N20}, which builds a stationary, 2D noise model within interleaved tiles on the sky. We generalize the formalism of this model into a class of Gaussian noise models in arbitrary bases; this allows us to prescribe an \textit{isotropic wavelet noise model} and a \textit{directional wavelet noise model}. The latter two wavelet approaches draw on existing wavelet formalism in Fourier space \citep[e.g.,][]{Freeman1991} and harmonic space \citep[e.g.,][]{Wiaux2008, sd-wavelets, directional-sd-wavelets, curvelet-sd-wavelets}, which have been applied to CMB data \citep[e.g.,][]{WMAP_NILC, Aghanim2016, Spin-SILC, Coulton23}, large-scale structure \citep[e.g.,][]{Allys2020, Cheng2020} and Galactic foregrounds \citep[e.g.,][]{Regaldo-SaintBlancard2021}. While the isotropic model uses the wavelets of \citet{Wiaux2008, sd-wavelets}, we also prescribe a new, tunable, directional wavelet set in Fourier space. In addition, our directional wavelets allow the lossless transform to and from wavelet space. To our knowledge, these models represent the first application of 2D wavelet analysis in the context of astrophysical instrument noise modeling. The models' primary user interface is the drawing of simulated noise maps, and we make our code --- \texttt{mnms}\footnote{\texttt{mnms} (Map-based Noise ModelS) is available at \url{https://github.com/simonsobs/mnms}. The repository provides information on a publicly-available release of noise models and simulations. It also provides code tutorials and documentation.} --- publicly available. These simulations are also consumed by downstream ACT DR6 analyses, to be described in upcoming publications \citep[e.g.,][]{Qu23, Coulton23}.

This paper is structured as follows. In \S\ref{sec: act_instrument}, we summarize the ACT instrument and DR6 data products that enter the noise modeling. In \S\ref{sec: data_noise}, we define the map-based noise and establish its most salient properties in the ACT data. In \S\ref{sec: noise_models}, we detail the construction of the three noise models, including their mathematical formalism. In \S\ref{sec: eval}, we test the noise models' ability to reproduce the complex noise covariance structure in the data, and apply them to the ACT DR6 CMB power spectrum pipeline. We discuss implications and future directions for noise modeling, and conclude in \S\ref{sec: disc_conc}.

Unless otherwise stated, in this paper, we refer to scalar quantities by italicized variables of any case, vector quantities as boldface variables of lower case, and matrix quantities as boldface variables of upper case. We also define multiplication ($*$) and division (/) of vectors to signify element-wise operations.

\begin{table}
    \centering
    \movetableright=-42pt
    \begin{tabular}{c|c|c|c}
        Array & Band & Frequencies & Beam \\
        & & (GHz) & (arcmin) \\
        \hline
        \hline
        \multirow{2}{*}{PA4} & f150 & 124 -- 172 & 1.4 \\
        & f220 & 182 -- 277 & 1.0 \\
        \multirow{2}{*}{PA5} & f090 & 77 -- 112 & 2.0 \\
        & f150 & 124 -- 172 & 1.4 \\
        \multirow{2}{*}{PA6} & f090 & 77 -- 112 & 2.0 \\
        & f150 & 124 -- 172 & 1.4 \\
    \end{tabular}
    \caption{Properties of the Advanced ACT detector arrays used in DR6. We will refer to particular polarized maps by their array and band (e.g., PA4 f220) throughout the text. The frequency ranges contain central 99\% of the band power. Beam sizes refer to the average beam full-width at half-maximum (FWHM).}
    \label{tab: arrays}

    \centering
    \begin{tabular}{c|c|c|c}
        Dec., R.A. & Shape & Pix. Size & $\ell_{max}$\\
        (deg) & (Rows, Cols.) & (arcmin) & (Nyq.) \\
        \hline
        \hline
        $-63$ to 23, & \multirow{2}{*}{10,320, 43,200} & \multirow{2}{*}{0.5} & \multirow{2}{*}{21,600} \\
        180 to $-180$ & & &
    \end{tabular}
    \caption{Properties of the ACT DR6 maps for all arrays, bands, and Stokes polarization components. In addition to the data maps, these also apply to the inverse-variance and cross-linking maps. The $\ell_{max}$ gives the Nyquist spherical-harmonic bandlimit of the pixelization.}
    \label{tab: maps}
\end{table}

\section{The ACT Instrument and Products} \label{sec: act_instrument}

\begin{figure*}
    \centering
    \includegraphics[width=\textwidth]{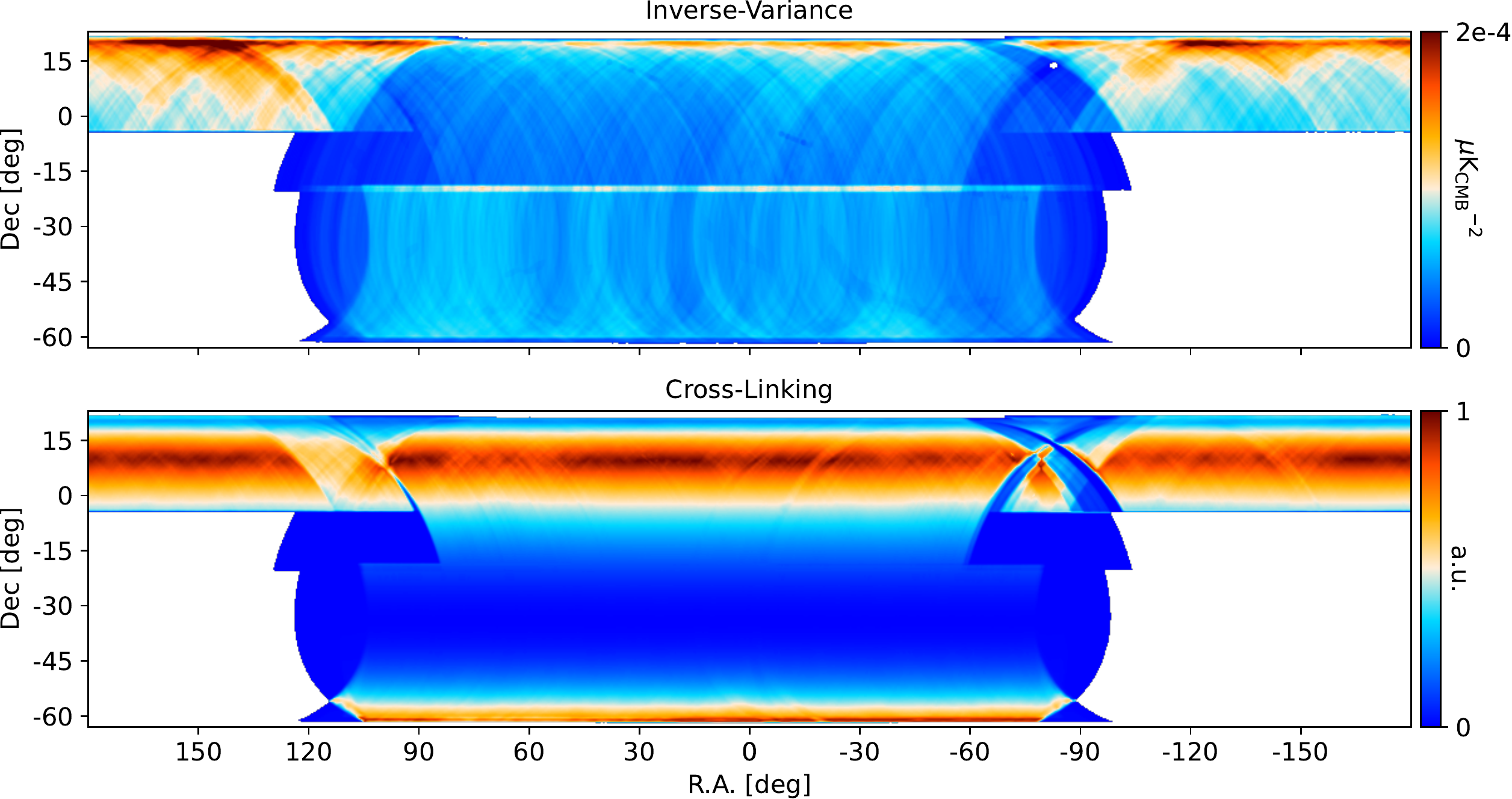}
    \caption{\textit{Top:} The inverse-variance map, \mbf{h}, for PA5 f090, first split. Larger values correspond to deeper regions of the map. The ACT scanning strategy is also discernible. Border ticks give the declination (Dec) and right-ascension (R.A.) in degrees. \textit{Bottom:} The cross-linking map, $1-\mathbf f_p$, for the PA5 f090 coadd (see Equation \ref{eq: coadd_def}). Values close to 0 indicate poor cross-linking, and vice versa for values close to 1, in arbitrary units (henceforth, a.u.).}  
    \label{fig: ivar_xlink}
\end{figure*}

The Atacama Cosmology Telescope (ACT) observed the polarized microwave sky from an altitude of $\sim$5,200\,m in the Atacama Desert, Chile. Its receiver has gone through several upgrades since its initial deployment in 2007; this paper --- and DR6 products generally --- focus on data collected since its most recent upgrade, Advanced ACT, in 2016 \citep{AdvACT}.\footnote{ACT ended observations in late 2022.} In particular, we focus on data collected by three dichroic detector modules, known as ``polarization arrays" (PAs, for short): PA4 (deployed 2016 -- 2022) observing in the f150 and f220 frequency channels \citep{PA4}, PA5 (deployed 2017 -- 2022) observing in the f090 and f150 frequency channels, and PA6 (deployed 2017 -- 2019), observing the same frequency channels as PA5 \citep{PA56}. Additional properties of these detector arrays are given in Table \ref{tab: arrays}. In 2019, PA6 was exchanged with a new detector array, PA7, observing in the f030 and f040 frequency channels. These ``low frequency" measurements are not discussed in this paper. For arrays PA4 -- 6, we use the DR6 ``night only" maps, which includes only data taken between 23:00 and 11:00 UTC; however, the methods we develop are equally applicable to ACT daytime data and null test maps.\footnote{Distinguishing features of datasets such as their beam, overall depth, sky footprint, etc. make no difference to our empirical modeling. Indeed, the methods are not even restricted to maps of the CMB.} We make use of three ACT DR6 map-based products in this paper as follows.

\textit{Sky Maps:} The mapmaking for DR6 will be described in a forthcoming paper; the following is a summary. The detector timestreams of each array are reduced into Stokes I, Q, and U sky maps by a  maximum-likelihood mapmaking algorithm \citep{D13, N14, L17, A20}. The ACT mapmaker assumes the following data model:
\begin{align} \label{eq: mapmaking_model}
    \mathbf d_t = \mathbf P \mathbf m + \mathbf n_t
\end{align}
where \mbf{d_t} are TODs, \mbf{P} is a projection matrix mapping the pixelized sky \mbf{m} from map space to TOD space (using telescope pointing), and \mbf{n_t} is additive, Gaussian noise in each detector sample. Each data vector in Equation \ref{eq: mapmaking_model} contains intensity and polarization components. The maximum-likelihood map under this model is:
\begin{align} \label{eq: mapmaking_soln}
    \mathbf m = (\mathbf P^T \mathbf N_t^{-1} \mathbf P)^{-1}\mathbf P^T \mathbf N_t^{-1} \mathbf d_t
\end{align}
where \mbf{N_t} is the covariance of \mbf{n_t}. TODs are grouped from the same array and frequency band into the same vector \mbf{d_t}, thus producing per-array, per-frequency, polarized (Stokes I, Q, U) maps. In ACT DR6, Equation \ref{eq: mapmaking_soln} is solved in eight disjoint ``splits" of the timestreams: one-day-long TOD blocks are distributed exclusively into eight sets, such that the noise in each split map is independent. As in ACT DR4 \citep{A20, MK21}, the maps are produced in the \textit{Plate-Carr\'{e}e} cylindrical projection. Further properties of the DR6 maps are given in Table \ref{tab: maps}.\footnote{This paper uses the first science-grade version of the ACT DR6 maps, labeled dr6.01. These maps are used in the upcoming ACT DR6 lensing analysis \citep{Qu23} and component separation analysis \citep{Coulton23}. Since these maps were generated, the ACT collaboration has made some refinements to the mapmaking that improve the large-scale transfer function and polarization noise levels. We expect to use a second version of the maps for further science analyses and for the DR6 public data release, and will update these noise simulations as those maps are produced and released.}

\textit{Inverse-Variance Maps:} In addition to maps of the microwave sky signal, the mapmaking algorithm produces auxiliary products used later in this paper. Of particular importance are ``inverse-variance" maps, which give the \textit{inverse} of the white-noise covariance between Stokes components in each pixel. Following the ACT convention \citep{C20, Mv20}, we label them \mbf{h}. The inverse-variance maps are analogous to detector hit-count maps, but also include the effect of intrinsic detector properties per detector hit \citep{MK21}. Because the TOD noise power spectrum is red, with a ``white-noise floor" at high frequencies \citep{D13, Morris2022}, the inverse-variance maps are only accurate at small scales. Although the Stokes covariance in each pixel is formally a $3\times 3$ matrix, for the same reasons as in DR4 (see \citet{A20}), DR6 includes only the Stokes intensity-intensity (II) element \citep{A20}. This has little effect on our noise models, however. As we will show in \S\ref{sec: noise_models}, our models still account for correlated Stokes components in the DR6 maps. An example inverse-variance map is shown in the top panel of Figure \ref{fig: ivar_xlink}

\textit{Cross-Linking Maps:} The mapmaker also produces ``cross-linking maps," which we use as a reference in \S\ref{sec: noise_properties} and to guide our modeling decisions in \S\ref{sec: noise_models}. These are maps which contain information about the ACT receiver's local scanning direction \citep[see e.g.,][for a more complete description]{A20, C20, MK21}. In particular, the ``polarization fraction," \mbf{f_p}, of the cross-linking map encodes the degree to which a given pixel is observed by many different scanning directions. The cross-linking maps are produced in such a way that $\mathbf f_p=0$ corresponds to maximal cross-linking (many crossing scans), and $\mathbf f_p=1$ corresponds to no cross-linking (all scans in the same direction). Therefore, we use the quantity $1-\mathbf f_p$ in later sections, such that high values of $1-\mathbf f_p$ correspond to high cross-linking and vice versa. An example cross-linking map is shown in the bottom panel of Figure \ref{fig: ivar_xlink}. Both the inverse-variance and cross-linking maps are produced with the same projection and parameters (see Table \ref{tab: maps}) as the sky maps.

\section{Map-Based Noise in ACT Data} \label{sec: data_noise} 
We begin our analysis by exploring the map-based noise in the ACT dataset. To do so, we first define ``map-based noise" and prescribe how we measure it.

\subsection{Definition and Measurement} \label{sec: splits_math}
Each pixelized, polarized (Stokes I, Q and U) split map has an independent noise realization: 
\begin{align} \label{eq: split_def}
    \mathbf m_i \equiv \mathbf s + \mathbf n_i 
\end{align}
where \mbf{m_i} is the map solution in TOD split $i$, \mbf{s} is the underlying sky signal (assumed constant over splits), and \mbf{n_i} is additive noise in each pixel (we suppress indices labeling the array, frequency, and polarization). The form of Equation \ref{eq: split_def} is not arbitrary --- it follows from the additive time-domain noise in the definition of \mbf{d_t} (Equation \ref{eq: mapmaking_model}) and the linearity of the mapmaking solution (Equation \ref{eq: mapmaking_soln}). That is, \mbf{n_i} is the mapped time-domain noise.

This paper aims to model the noise, \mbf{n_i}, in each ACT split. Of course, we can never measure \mbf{n_i} directly, or else we could produce noiseless maps. Rather, we must use a proxy for the noise. In particular, we use differences of map splits, relying on the assumption that each \mbf{m_i} in Equation \ref{eq: split_def} measures an identical signal. We give the prescription by which we construct these difference maps as follows.

The first step is to construct an unbiased estimate of the signal, \mbf{s}, in the split maps. We do this by taking the inverse-variance-weighted average of \mbf{m_i} over splits $i$, which we call the map-based ``coadd," \mbf{c}:
\begin{align} \label{eq: coadd_def}
    \mathbf c = \frac{\sum_{i}{\mathbf h_i*\mathbf m_i}}{\mathbf\Sigma}
\end{align}
where \mbf{h_i} is the inverse-variance map for split $i$ and $\mathbf\Sigma \equiv \sum_{i}{\mathbf h_i}$. Next, we subtract \mbf{c} from each of the \mbf{m_i} to construct a ``raw" difference map, \mbf{d_i}:
\begin{align} \label{eq: diff_def}
    \mathbf d_i &\equiv \mathbf m_i - \mathbf c.
\end{align}
If \mbf{c} were exactly \mbf{s}, then \mbf{d_i} would equal \mbf{n_i} in that split. Instead, because \mbf{c} contains an average of the noise from each split, so too does \mbf{d_i}. We can see this by combining Equations \ref{eq: split_def}, \ref{eq: coadd_def}, and \ref{eq: diff_def}:
\begin{align} \label{eq: diff_def_2}
    \mathbf d_i = \mathbf n_i - \frac{\sum_{j}{\mathbf h_j*\mathbf n_j}}{\mathbf\Sigma}
\end{align}
The upshot, though, is that \mbf{d_i} is a decent proxy for \mbf{n_i}: it contains \textit{only} noise, and ``mostly" (which we will quantify in Equation \ref{eq: diff_noise_corr}) noise from split $i$. Indeed, this is the proxy used in previous ACT analyses \citep{C20, Mv20, N20}.

In this paper, we perform one more step by applying a correction factor to the raw difference maps. Specifically, we correct for the fact that the variance in \mbf{d_i} is less than the variance in \mbf{n_i}. To evaluate $Var(\mathbf d_i)$, first recall from \S\ref{sec: act_instrument} that $Var(\mathbf n_i) = 1/\mathbf h_i$ only in the white noise, uncorrelated-pixel limit. We work in that limit for now, and also define the difference map inverse-variance along the same lines: $Var(\mathbf d_i) \equiv 1/\mathbf w_i$. In Appendix \ref{apx: diff_maths}, we show that \mbf{w_i} satisfies:
\begin{align} \label{eq: diff_noise_var}
    \frac{1}{\mathbf w_i} = \frac{1}{\mathbf h_i} - \frac{1}{\mathbf\Sigma}.
\end{align}
That is, the left-hand side --- the variance of \mbf{d_i} --- is less than that of \mbf{n_i} by exactly $1/\mathbf \Sigma$. This motivates the following correction to the raw difference maps:
\begin{align} \label{eq: diff_def_3}
    \bm \nu_i \equiv \mathbf d_i*\sqrt{\frac{\mathbf w_i}{\mathbf h_i}}
\end{align}
such that the variance of $\bm\nu_i$ is the same as that of \mbf{n_i}:
\begin{equation} \label{eq: var_nu_var_n}
    \begin{aligned}
        Var(\bm \nu_i) = Var(\mathbf d_i)&*\frac{\mathbf w_i}{\mathbf h_i} = \\
        = \frac{1}{\mathbf w_i}&*\frac{\mathbf w_i}{\mathbf h_i} = \frac{1}{\mathbf h_i} = Var(\mathbf n_i).
    \end{aligned}   
\end{equation}
Both when we examine map-based noise properties of a given split (\S\ref{sec: noise_properties}), and build noise models (\S\ref{sec: noise_models}), we will always be referring to the \textit{corrected} difference map noise measurements, $\bm\nu_i$, in Equation \ref{eq: diff_def_3}.

\begin{figure}
    \centering
    \includegraphics[width=\columnwidth]{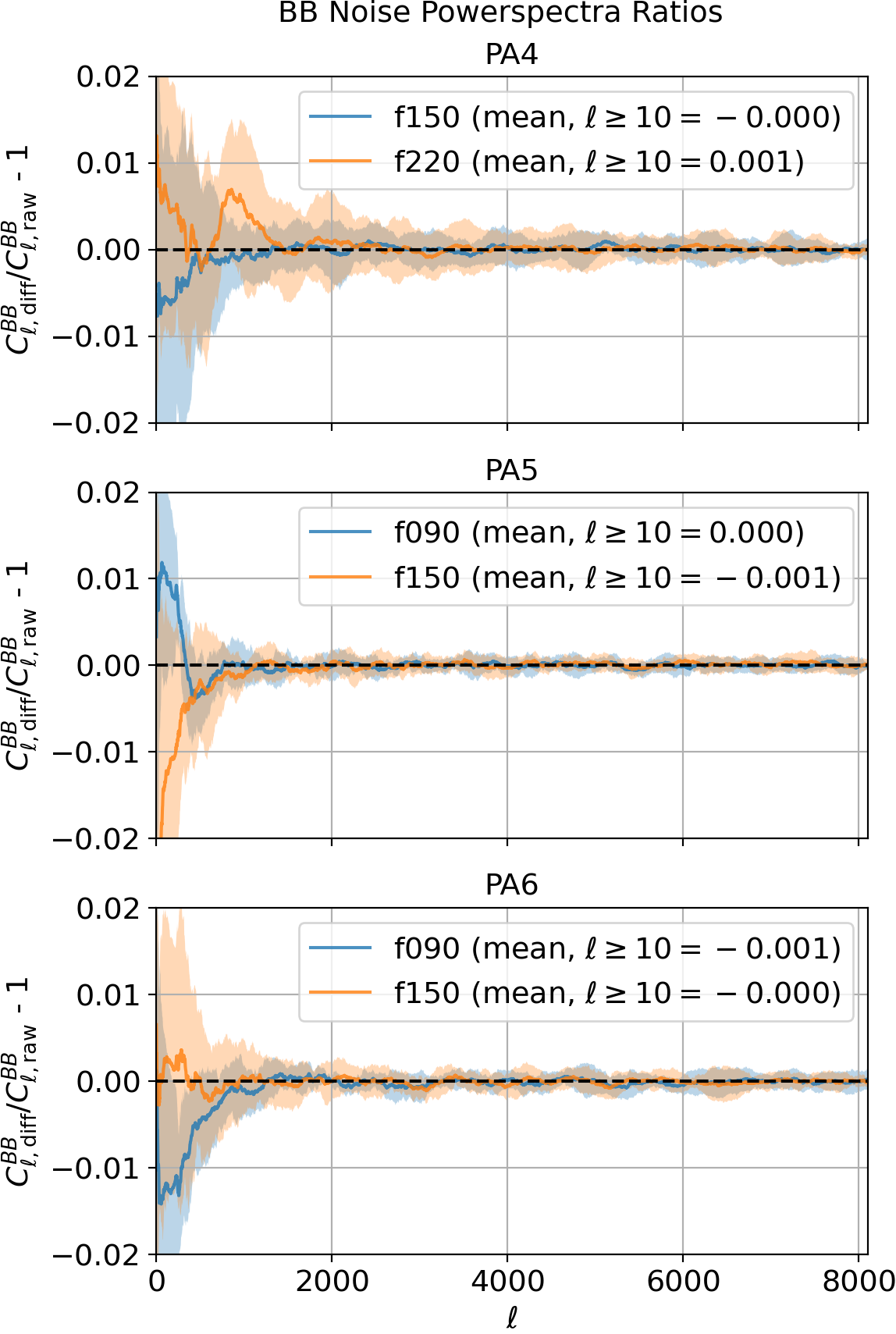}
    \caption{Ratios of BB pseudospectra between the corrected difference maps, $\bm\nu_i$, and the individual split maps, \mbf{m_i}, for each array and frequency channel in ACT DR6. All pseudospectra are measured within this paper's ``pseudospectrum mask" described in \S\ref{sec: implementation_masks}. The dashed black line is the nominal result (ratio of unity), the solid lines (shaded bands) are the means ($1\sigma$ range) over the eight splits of each dataset. To mitigate any expected BB signal, in addition to masking the Galaxy, all maps are point-source subtracted and filtered using the ACT DR4 ground pickup filter \citep{C20}, where we remove Fourier modes with $|k_x|\leq 90$ and $|k_y|\leq 50$. Spectra are smoothed with a tophat kernel of size $\Delta\ell=250$ to better visualize their overall trend.}
    \label{fig: bb_ratios}
\end{figure}

We emphasize that $\bm\nu_i$ and \mbf{n_i} are distinct but related quantities. Their correlation is given by (Appendix \ref{apx: diff_maths}):
\begin{align} \label{eq: diff_noise_corr}
    \bm\rho_i = \sqrt{1 - \frac{\mathbf h_i}{\mathbf\Sigma}}
\end{align}
which approaches 1 as the number of splits increases, assuming splits contain reasonably uniform TOD volumes. For example, given eight approximately uniform splits, Equation \ref{eq: diff_noise_corr} yields a correlation of $\sim$93.5\%. Because we can never directly measure \mbf{n_i} in a map, we resort to using $\bm\nu_i$ as a highly correlated tracer. 

Fortunately, we can test the two assumptions that the uncorrected difference maps (Equation \ref{eq: diff_def}) contain pure noise, and that the corrected difference maps (Equation \ref{eq: diff_def_3}) have accurate variance. From the best-fit model of the signal from ACT DR4 \citep{C20}, the expected BB power from the lensed CMB and Galactic dust (with the Galaxy masked, as in this paper's pseudospectrum mask, see \S\ref{sec: implementation_masks}) should be at the sub-percent level compared to the ACT DR6 per-split noise. In other words, the BB power spectra from raw, \textit{non-differenced} split maps, \mbf{m_i}, should be a reasonably clean measure of the map-based noise; in particular, they should match BB power spectra from the corrected-power difference maps, $\bm\nu_i$. Such a test is not trivial: \textit{Planck} \citep{FFP8} found a several percent discrepancy in their High Frequency Instrument (HFI) channels when performing a similar comparison, which was attributed to systematics in their pre-processing. 

As shown in Figure \ref{fig: bb_ratios}, ACT DR6 exhibits good agreement. All arrays and frequencies achieve an average difference map ($\bm\nu_i$) to raw map (\mbf{m_i}) BB spectra ratio of within 0.1\% of nominal. Recall, in constructing the difference maps (Equation \ref{eq: diff_def_3}), we had to work in the white noise, uncorrelated-pixel limit where the inverse-variance maps, \mbf{h_i}, are expected to be accurate. As discussed more in \S\ref{sec: noise_properties}, this limit is reached only on small angular scales. However, the performance of the $\sqrt{\mathbf w_i/\mathbf h_i}$ correction factor is more robust: the BB ratios are consistent with nominal at the $\sim1\%$ level for scales as large as $\ell\sim 300$. While the observed percent-level deviations for $\ell\lesssim 300$ do matter when evaluating our noise models in \S\ref{sec: eval}, overall this result strongly supports the scheme by which we measure the noise in a given map split.

\begin{figure*}
    \centering
    \includegraphics[width=\textwidth]{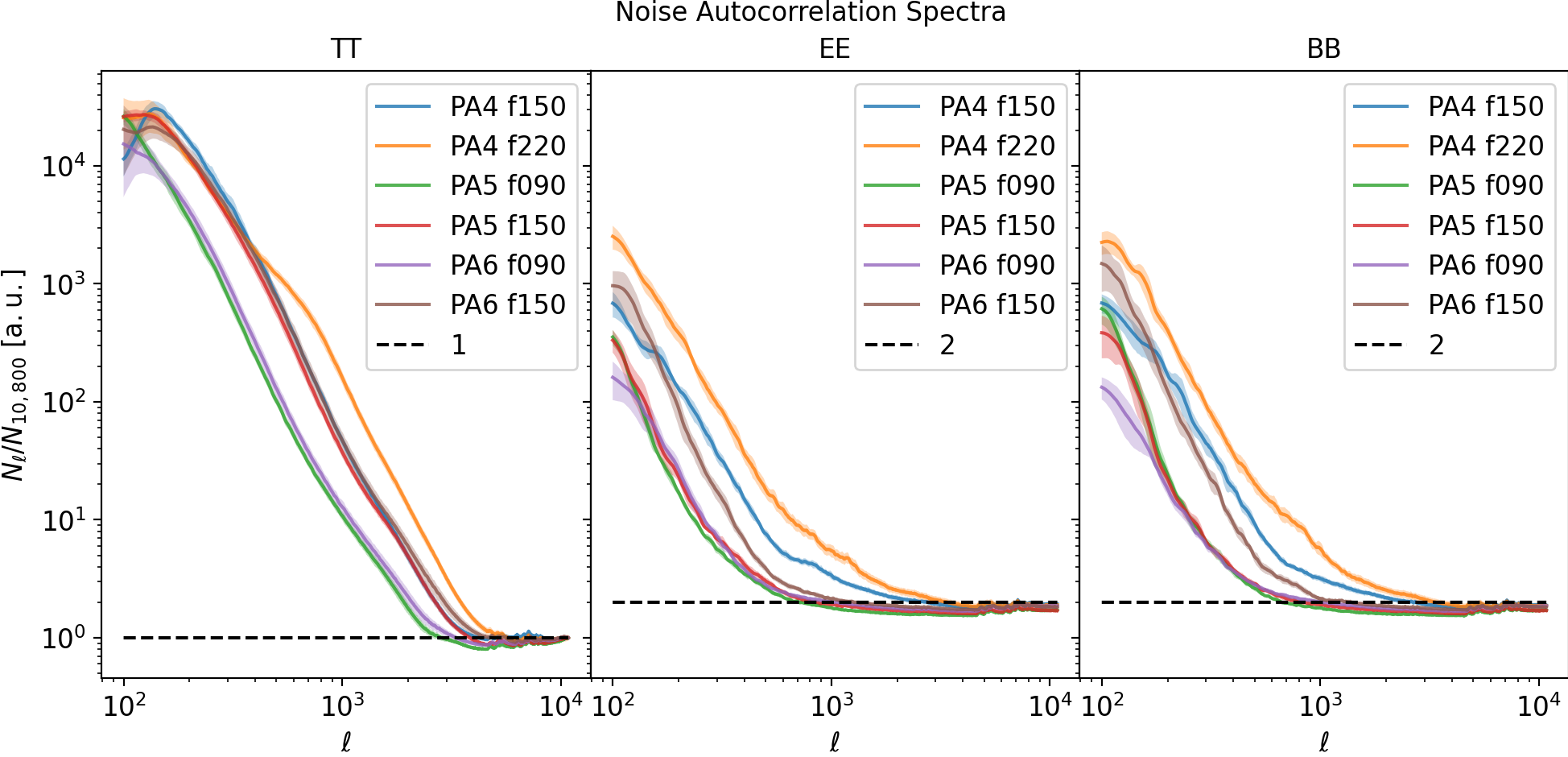} \\
    \vspace{10pt}
    \includegraphics[width=\textwidth]{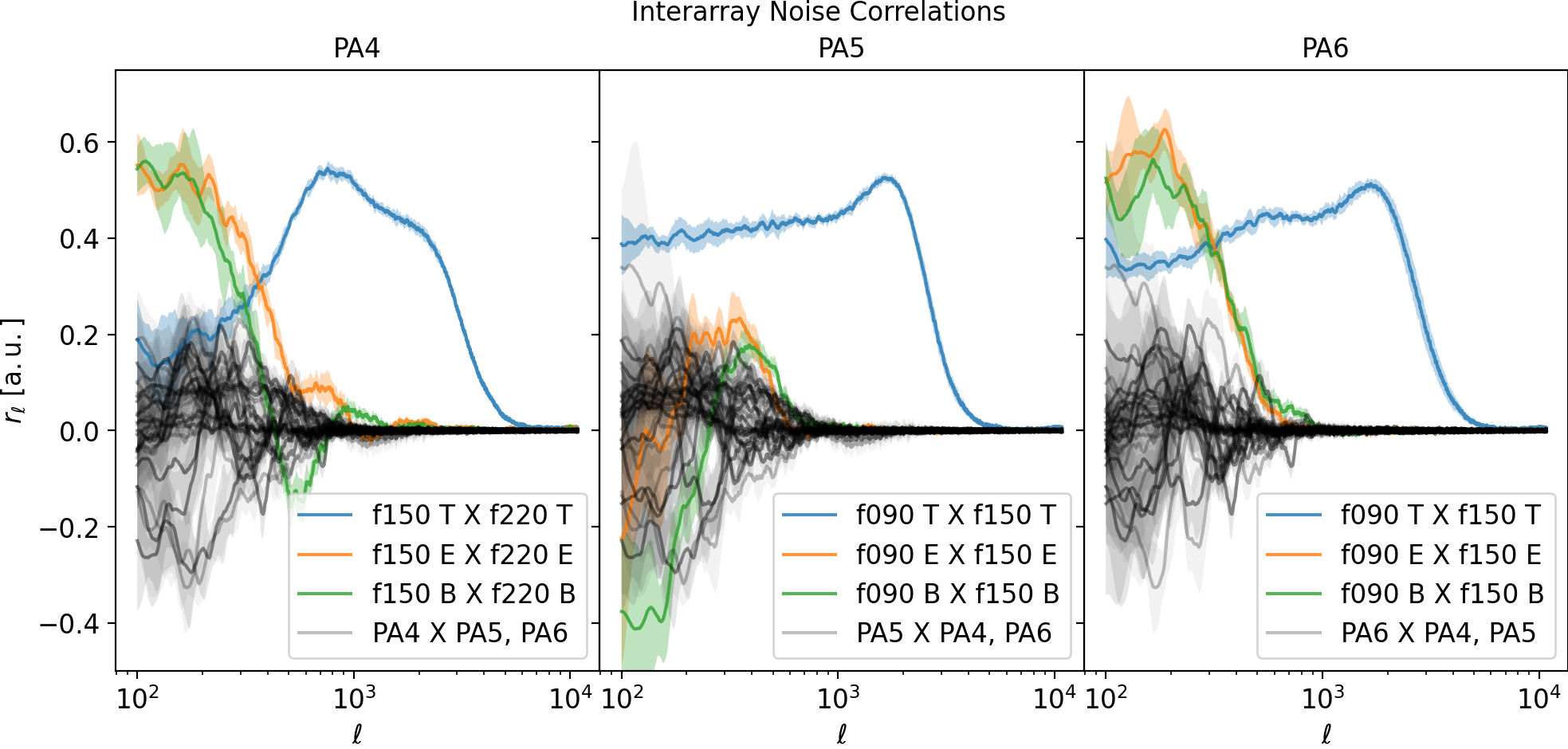}
    \caption{\textit{Top, large-scale noise correlations:} Noise power spectra by polarization autocorrelation (TT, EE, BB), for each detector array and frequency band. The power spectra are evaluated on the corrected difference maps (Equation \ref{eq: diff_def_3}), after applying the apodized pseudospectrum mask described in \S\ref{sec: implementation_masks}. Shaded bands give the $1\sigma$ range over the eight split maps for each array/frequency. All spectra are smoothed with a tophat kernel of size $\Delta\ell=25$. Both the TT and polarization (EE, BB) spectra are normalized by the value of the TT spectrum at $\ell=10,800$ to suppress the effect of variable data volume between arrays. Polarization spectra plateau at a value of $\sim$2, owing to each polarized component receiving approximately half the TOD volume as intensity. Both the intensity and polarization spectra are steep and red at large scales. \textit{Bottom, correlated frequencies:} Correlation spectra ($r_{\ell}$, Equation \ref{eq: correlation_spectra}) evaluated from the mean cross-power spectrum of the eight difference maps, for each detector array. The same mask is applied as in the top panel, and shaded regions again correspond to the $1\sigma$ range over the eight split maps. For legibility, we only plot the cross-spectra between different frequency bands --- all crosses are between the same polarization components (i.e., TT, EE, or BB). The colored series indicate cross-spectra between frequency bands on the same physical detector array, while the grey series indicate the cross-spectra between bands on different physical detector arrays. The only significant cross-spectra come from the same physical detector array.}
    \label{fig: N_ell_r_ell}
\end{figure*}

\subsection{ACT Map-Based Noise Properties} \label{sec: noise_properties}
We motivate our noise model design choices by enumerating five salient features of the ACT map-based noise. We take these five features to define the scope of our modeling in \S\ref{sec: noise_models}; however, we do not expect this list to be exhaustive in reality. Nevertheless, we do expect the noise properties in this section to be (a) the most significant for downstream analyses, and (b) common to ground-based, scanning microwave telescopes.

\subsubsection{Large-Scale Noise Correlations}
Large-scale noise correlations are a dominant noise feature in ground-based microwave observatories \citep{D13, Henning2018}. Their origin is typically attributed to fluctuating, nearly unpolarized emission by atmospheric water vapor \citep{Errard2015, Morris2022}, which can leak into polarized timestreams via optical and detector systematics \citep{Takakura2017}. We see their effect in the power spectra of the DR6 difference maps, as shown in the top panel of Figure \ref{fig: N_ell_r_ell}. Both the intensity and polarized map-based noise power spectra are steep and red, spanning several orders of magnitude in power from their white-noise level at small scales to their peaks at large scales. The correlated noise in intensity maps is higher than in polarization, mainly due to a smaller transition scale --- or, higher $\ell_{\mathrm{knee}}$ --- from red, correlated noise to white noise. The power-law index of the correlated noise --- the slope in log-log space, as plotted in the top panel of Figure \ref{fig: N_ell_r_ell} --- is similar between intensity and polarization. As is expected for thermal water emission, the intensity spectra increase monotonically with the frequency band of the map. This scaling is not as clear in the polarization spectra, highlighting the contribution from non-atmospheric effects to the large-scale polarized noise. Likewise, the convergence of the PA4 f220 TT spectrum to the f150 spectra at low-$\ell$ requires more work to understand its origin.

\subsubsection{Correlated Frequencies}
Because the ACT detector arrays are dichroic, maps of each frequency band on a given detector array observe correlated fluctuations. We probe this by forming ``correlation spectra" of the two frequency maps:
\begin{align} \label{eq: correlation_spectra}
    r_{\ell}^{ab} \equiv \frac{C_{\ell}^{ab}}{\sqrt{C_{\ell}^{aa}C_{\ell}^{bb}}}
\end{align}
where $C^{ab}$ denotes the cross-spectrum of two fields $a$ and $b$. The bottom panel of Figure \ref{fig: N_ell_r_ell} shows how the correlation spectra vary by detector array, polarization, and angular scale. The TT correlation spectra between frequency maps on the \textit{same} array stand out in particular: the two frequencies are $\sim$25\% -- 50\% correlated at scales where the atmospheric noise is dominant ($\ell \lesssim 3,000$, see the top panel Figure \ref{fig: N_ell_r_ell}). Likewise, the polarization (EE, BB) correlation spectra between frequency maps, again on the \textit{same} array, appear significant to similar levels on larger scales (though more for PA4 and PA6 than PA5). Owing to the increased distance between detectors, correlations of any component between frequency maps on \textit{different} physical detector arrays are considerably reduced. Other details apparent in Figure \ref{fig: N_ell_r_ell} are less intuitive. It is not clear why the TT correlation peaks at $\ell \approx 1,000$ but declines for larger scales especially in PA4, nor why PA5 polarization correlations are considerably different from those of PA4 and PA6. Further work is needed to understand these features. Nevertheless, our models treat these empirical noise correlations in the ACT maps as a given.

\subsubsection{Map-Depth Inhomogeneity}
Map-depth inhomogeneity\footnote{Herafter, we take ``inhomogeneities" to mean quantities that vary over 2D position on the sky, and ``anisotropies" to mean quantities that vary over ``azimuthal direction" in the maps, as centered on any given 2D position. These differ from their conventional definitions in cosmology.} is a commonly modeled property of map-based noise in the literature. For ACT, this information is encoded in the inverse-variance maps \mbf{h} discussed in \S\ref{sec: act_instrument}. The inverse-variance maps contain considerable morphology, as is seen in the top panel of Figure \ref{fig: ivar_xlink}. In particular, we see map depth varies spatially on both large and small scales, generally being deeper closer to the celestial equator and shallowest near the Galactic plane. As discussed, we expect this spatial modulation to only be accurate for the high-$\ell$ ACT noise. Furthermore, we can discern the telescope scan strategy inducing coherent arcs across the observed footprint. Between declinations of $\sim0^{\circ}$ to $15^{\circ}$, the ACT scan strategy tends to have an ``x-like" directional pattern, while declinations of $\sim-45^{\circ}$ to $-15^{\circ}$ have a uniformly ``vertical" scanning pattern. The scanning pattern is reflected in the bottom panel of Figure \ref{fig: ivar_xlink}, which contains a cross-linking map, $1-\mathbf f_p$ (see \S\ref{sec: act_instrument}). Regions where the cross-linking is low are those where the ACT scans are uniformly vertical; regions where the cross-linking is high are those where the ACT scans are ``x-like."

\begin{figure*}
    \centering
    \includegraphics[width=\textwidth]{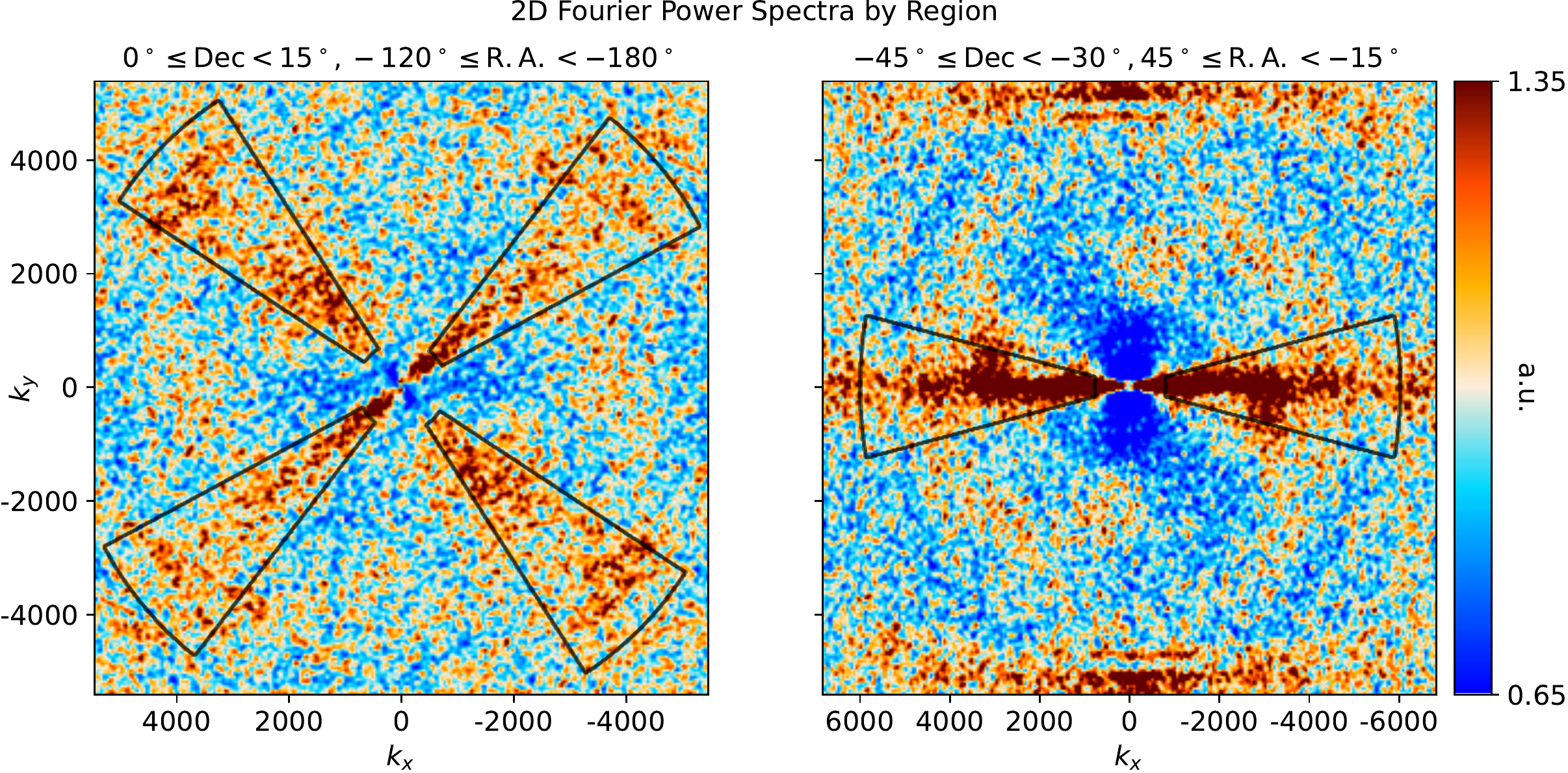}
    \caption{\textit{Spatially-varying noise anisotropy:} TT power spectra in 2D Fourier space for PA5 f090, first split. The 2D spectra are normalized radially by average power in annuli of $\Delta k=50$ and smoothed by a uniform 2D tophat kernel of sidelength $\Delta k=100$. The left spectrum is measured in the region from $0^\circ$ to $15^\circ$ Dec and $-120^\circ$ to $-180^\circ$ R.A.. The right spectrum is measured in the region from $-45^\circ$ to $-30^\circ$ Dec and $45^\circ$ to $-15^\circ$ R.A.. The edge of both regions is apodized with a $2^\circ$-wide cosine taper before taking Fourier transforms. Coordinates at the image perimeter give the vertical and horizontal Fourier modes. Because the 2D spectra are normalized by their overall radial profile, the colorscale is unitless. As is explained in the text, the noise exhibits stripy patterns that align with the local scan strategy. In Fourier space, these patterns manifest as bars perpendicular to the local scan strategy. We outline the primary features from the scanning strategy in black. The left (right) panel outlined region contains an average power excess of 8\% (25\%). Secondary features --- such as detector-detector correlations on the top and bottom of the right panel --- are also visible. The regions of low noise power near the center result from the removal of the radial, or ``isotropic," part of the power spectrum.} 
    \label{fig: 2d_ps_by_region}
\end{figure*}

\subsubsection{Spatially-Varying Noise Anisotropy}
Unfiltered ACT data have historically shown stripy noise features \citep{L17, C20}, and the DR6 maps are no exception. Like the large-scale correlations (Figure \ref{fig: N_ell_r_ell}), stripy noise patterns are a manifestation of the atmospheric noise. Given a particular low-frequency atmospheric signal in a TOD, as the telescope scans across the sky, the signal will manifest as a one-dimensional stripe in the map. Thus, not only are the largest spatial scales dominated by atmospheric signals, but those signals also tend to ``trace" the ACT scan pattern. We confirm this intuition by examining the 2D Fourier noise power in the previously mentioned declination bands, as shown in Figure \ref{fig: 2d_ps_by_region}. In order to emphasize the ``anisotropic" noise power, we first divide-out the ``isotropic" radial profile (i.e., the 1D spectra from Figure \ref{fig: N_ell_r_ell}) that would otherwise dominate the full 2D power. Near the celestial equator (the left panel), we recover a strong ``x-like" pattern in the relative 2D noise power, as expected from our by-eye assessment of the ACT scan pattern. Likewise, in the southern region with vertical, poorly cross-linked scans (the right panel), we find the 2D noise power to be concentrated in a horizontal ``bar" in Fourier space. In other words, the 2D noise power exhibits long-wavelength (small wavenumber) correlations in the y-direction, but is approximately featureless (white) in the x-direction, as expected. 

Furthermore, the poorly cross-linked region exhibits stronger overall anisotropies than the well cross-linked region. To demonstrate this, in Figure \ref{fig: 2d_ps_by_region}, we measure the power excess within the dominant ``x-like" or ``bar" features in the 2D spectra. The outlined ``x-like" portion of the well cross-linked 2D spectrum contains an average noise power excess of 8\%, whereas the same as measured in the outlined ``bar" of the poorly cross-linked 2D spectrum is 25\%. Thus, the ACT map-based noise contains anisotropies whose structure and prominence varies spatially. 

\begin{figure*}
    \includegraphics[width=\textwidth]{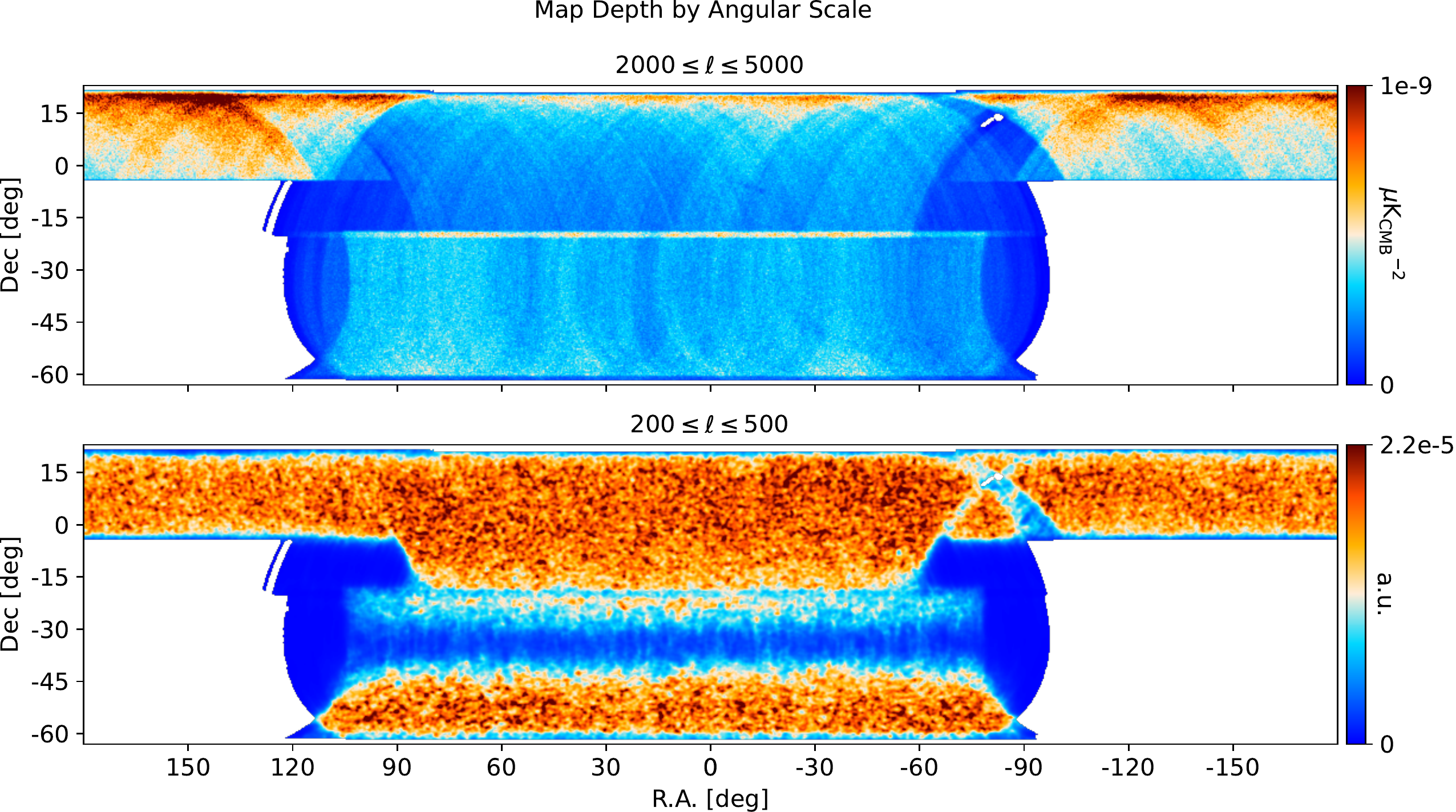}
    \caption{\textit{Scale-dependent map depth:} ``Inverse-variance maps" as measured directly from the PA5 f090 difference maps. \textit{Top:} After filtering $\bm\nu_i$ for small scales, $2000\leq\ell\leq5000$, and averaging over Stokes Q and U in only the first split. \textit{Bottom:} After filtering $\bm\nu_i\sqrt{\mathbf h_i}$ for large scales, $200\leq\ell\leq500$, and averaging over Stokes Q and U in all eight splits. We square the filtered maps to assess local noise variance, smooth them for legibility (Gaussian $\mathrm{FWHM}=4\pi/5000\, \mathrm{rad}\approx0.14^o$ and $\mathrm{FWHM}=2\pi/500\, \mathrm{rad}\approx0.72^o$, respectively), and take their inverse. We also multiply by pixel area to account for declination-dependent loss of power from the filtering. Because we invert the variance, small values indicate \textit{larger} noise power. As is explained in the text, the small-scale noise power is modulated by the mapmaker inverse-variance, and the large-scale noise power traces the mapmaker cross-linking.}
    \label{fig: ivar_by_scale}
\end{figure*}

\subsubsection{Scale-Dependent Map Depth}
As discussed above, the ACT map-based noise is spatially modulated by the inverse-variance maps at small scales. At large scales where the atmospheric correlations dominate, however, the map depth spatial morphology resembles the amount of cross-linking.\footnote{An equivalent description of scale-dependent map depth is a spatially-dependent noise power spectrum.} This scale-dependent behavior can be understood as a consequence of the ACT mapmaking solution itself. At small scales where the noise is uncorrelated between detectors, variance in the maximum likelihood map in Equation \ref{eq: mapmaking_soln} decreases only with the number of detector hits. This does not apply to large-scale correlated modes observed simultaneously by many detectors. Instead, variance in the estimation of large scales decreases with the number of independent scans that observe the mode (heuristically, for atmospheric correlations on the order of the array size, the array acts as if it has only one detector, not hundreds). All scans are not equally effective at reducing variance in the estimate, however. In a single scan, large-scale, correlated signals on the sky are degenerate in Equation \ref{eq: mapmaking_soln} with low-frequency, correlated atmosphere in the TOD. Observations of a fixed sky position with differing scan directions act to lift this degeneracy, while parallel scans do not. Therefore, in addition to averaging down with the number of independent scans, the large-scale map-based noise averages down with increased map \textit{cross-linking} as well.

To isolate this cross-linking-like morphology, we should first normalize the difference maps by the number of array scans at each point. Maps of these statistics are not readily available, so we use the inverse-variance maps as a proxy.\footnote{This essentially assumes that the per-pixel number of array scans is proportional to the per-pixel number of detector hits, which is not exact but more than good enough for our purposes here.} To be precise, we first multiply the difference maps, $\bm\nu_i$, by $\sqrt{\mathbf h_i}$, such that the variance of $\bm\nu_i$ on small-scales is uniform. Borrowing from Equation \ref{eq: var_nu_var_n}, we have \textit{on small-scales}:
\begin{equation} \label{eq: var_nu_var_n_2}
    \begin{aligned}
        Var(\bm \nu_i \sqrt{\mathbf h_i}) = Var(\bm \nu_i)&*\mathbf h_i = \\
        =\frac{1}{\mathbf h_i}&*\mathbf h_i = 1.
    \end{aligned}
\end{equation}
Then we filter $\bm\nu_i\sqrt{\mathbf h_i}$ for \textit{large-scales} and look for residual structure in the noise power across the map. Figure \ref{fig: ivar_by_scale} presents the small- and large-scale behavior of the data: the top panel shows the \textit{inverse} of the variance of $\bm\nu_i$ after filtering for small scales, while the bottom panel shows the \textit{inverse} of the variance of $\bm\nu_i\sqrt{\mathbf h_i}$ after filtering for large scales as just discussed. Comparing to Figure \ref{fig: ivar_xlink}, the empirical small scale power clearly traces the mapmaker-produced inverse-variance map, while the empirical large-scale power resembles the mapmaker-produced cross-linking map --- with greater noise power (less inverse-variance) corresponding to less cross-linking, and vice versa. In sum, different angular scales have their own distinct ``inverse-variance maps," resulting directly from the ACT scanning strategy.

Note that all of the above noise properties are each manifestations of the same underlying process: the propagation of correlated time-domain noise and the ACT scanning strategy through the mapmaking. Further study of atmospheric emission \citep{Morris2022} and instrumental noise sources \citep{D13} may reveal new, subtle map-based noise properties not explored in this paper.

\section{Noise Models} \label{sec: noise_models}
In this section we describe three different noise models that capture the ACT map-based noise properties from \S\ref{sec: noise_properties}. Implicit in Equation \ref{eq: split_def} is the assumption that the particular map based noise realizations in our data, \mbf{n_i}, are zero-mean random vectors. We also make the assumption that the \mbf{n_i} are Gaussian distributed.\footnote{This is well motivated because the time-domain noise is already well described by Gaussian noise. The noise properties stay Gaussian when the linear mapmaking equation (Equation \ref{eq: mapmaking_soln}) is applied. Furthermore, each each pixel is is built up from many independent time samples ($\sim97\%/99\%/98\%$ of pixels have greater than 100 time sample hits in each split map of PA4/5/6) which is expected to further reduce any non-Gaussianity.} Therefore, a statistically complete ``noise model" for each \mbf{n_i} is given by its covariance matrix, \mbf{N_i}. We build estimates of this matrix using the difference maps, $\bm\nu_i$, and use it to draw noise simulations. 

The long-distance noise correlations from the previous section suggest that \mbf{N_i} is dense in the map basis. Combined with the large pixel count in the ACT DR6 maps ($N_{\mathrm{pix}} = 445,824,000$, see Table \ref{tab: maps}), the construction of an explicit $N_{\mathrm{pix}}\times N_{\mathrm{pix}}$ \mbf{N_i} is intractable \citep{N20}. Instead, we leverage the statistical properties in \S\ref{sec: noise_properties} to build a sparse, but still approximately complete, covariance matrix. The key to a sparse representation is the isolation of stationary regions in some projection of the noise. For example, given map space fields $a$ and $b$ at locations \mbf{x} and \mbf{x'}, and their Fourier transforms $\tilde{a}$ and $\tilde{b}$ at modes \mbf{k} and \mbf{k'}, the cross-correlation theorem states:
\begin{equation}  \label{eq: fourier_trick}
    \begin{aligned}
        &\mathrm{If:} \\
        &\quad \EV{a(\mathbf x)b^*(\mathbf x')} \equiv C^{ab}(\mathbf x, \mathbf x') = C^{ab}(\mathbf x - \mathbf x') \\
        &\mathrm{then: } \\
        &\quad \EV{\tilde{a}(\mathbf k)\tilde{b}^*(\mathbf k')} \equiv \tilde{C}^{ab}(\mathbf k, \mathbf k') = \tilde{C}^{ab}(\mathbf k)\delta(\mathbf k - \mathbf k')
    \end{aligned}
\end{equation}
where $C^{ab}$ is the map space covariance of $a$ and $b$, and $\tilde{C}^{ab}$ is its Fourier transform.\footnote{Due to the unitary property of the Fourier transform, Equation \ref{eq: fourier_trick} is equally valid if we exchange map space and Fourier space.} In other words, if the covariance of two fields is \textit{stationary} in map space --- that is, exhibits translational invariance (even while being correlated) --- then in Fourier space it is \textit{uncorrelated}. In particular, only the diagonal of the Fourier covariance --- the (cross) power spectrum $\tilde{C}^{ab}$ --- needs to be measured. Therefore, for each noise model discussed in this section, we will first filter the ACT DR6 noise maps to promote their stationarity, and then transform them to permit a sparse representation of the covariance matrix. 

Following a brief discussion of some additional notation in \S\ref{sec: noise_models_notation}, we begin in \S\ref{sec: noise_models_tile} with an updated version of the ``tiled" noise model from \citet{N20}, including the algorithms by which we build the model and draw simulations. We describe the same for two novel noise models: an ``isotropic wavelet" (or ``wavelet") noise model in \S\ref{sec: noise_models_wav} and a ``directional wavelet" (or ``directional") noise model in \S\ref{sec: noise_models_fdw}. Throughout \S\ref{sec: noise_models_tile}, \S\ref{sec: noise_models_wav}, and \S\ref{sec: noise_models_fdw}, we will introduce proper matrix notation for all operations performed on our map vectors. This will aid users incorporating the noise covariance matrices into their analysis program, as well as clarify how we draw noise simulations. Lastly, for all three models, we describe map preprocessing steps and additional implementation details in section \S\ref{sec: implementation}.

\subsection{Linear Algebra Notation} \label{sec: noise_models_notation}
Before diving in, we establish some notation accompanying our map vectors and matrices. So far, we have understood our proxy noise maps, $\bm\nu_i$, to have $N_{\mathrm{pix}}$ elements. Henceforth, we will add an additional index to the maps which iterates over frequency bands (2) and polarization indices (3) on a given detector array. That is, we should now take the vector $\bm\nu_i$ to refer to a concatenation of the (6) polarized frequency maps from each array, with $6N_{\mathrm{pix}}$ elements. In this way, our noise models will be able to capture the observed frequency-frequency cross-correlations from Figure \ref{fig: N_ell_r_ell}.\footnote{While we do not observe significant polarization-polarization noise cross-correlations in DR6, in keeping with previous ACT convention \citep{C20, Mv20}, we allow for them in the models.} When needed, we will refer to a particular frequency and polarization subset of the data by its superscript index, $a$, ranging from 0 to 5 (e.g., for PA4, the $a=0-2$ blocks of $\bm\nu_i^a$ would refer to the PA4 f150 band, I, Q, and U, respectively). This notation extends to other map vectors like \mbf{h_i}, and any matrix objects we construct from the outer-product of two vectors --- when needed, we will refer to matrix block elements using superscript indices $a$ and $b$. Each map vector and noise model will continue to be distinct for each detector array and split, $i$.

We also define matrices for the map transforms --- discrete Fourier transforms (DFTs) and spherical harmonic transforms (SHTs) --- used in our modeling and simulation algorithms. We denote the forward DFT by the matrix \mbf{F}, which is unitary ($\mathbf F^{-1} = \mathbf F^{\dag}$). Following \citet{libsharp, Seljebotn2019}, we denote spherical harmonic synthesis (from harmonic to map space) by the $N_{\mathrm{pix}}\times N_{a_{lm}}$ matrix \mbf{Y}, and spherical harmonic analysis (from map to harmonic space) by $\mathbf Y^{\dag}\mathbf W$, where \mbf{W} is a square diagonal matrix containing per-pixel ``quadrature weights" used in the numerical integration. In general, it is the case that $N_{\mathrm{pix}} \neq N_{a_{lm}}$: SHTs are, unlike DFTs, rectangular operations. Unless otherwise stated, we always perform SHTs to the Nyquist bandlimit of the input pixelization. Following the default SHT implementation in \texttt{pixell}\footnote{\url{https://github.com/simonsobs/pixell}} and \texttt{healpy}\footnote{\url{https://github.com/healpy/healpy}} \citep{HEALPix, healpy}, when applied to a spin-2 pair of Stokes Q and U maps, we will take $\mathbf Y^{\dag}\mathbf W$ to be the linear combination of spin-2 SHTs that outputs E- and B-mode  harmonic coefficients \citep{zaldarriaga_1997}.\footnote{Unlike cosmological modeling, the noise modeling done in this work does not obviously benefit from a description in terms of the parity even and odd E and B fields. It is likely that a reformulation in terms of spin-0 SHTs of Q and U would have worked equally well.} Thus, if \mbf{s} is a polarized map with Stokes I, Q, and U components, the vector $\mathbf a = \mathbf Y^{\dag}\mathbf W \mathbf s$ consists of I (also referred to as T), E, and B harmonic coefficients.

Finally, we make use of a compact notation to describe products of matrices and vectors with incompatible shapes. The notation is similar to the ``broadcasting" of vector and matrix objects used in \texttt{numpy}, the main scientific computing package for the Python programming language \citep{numpy}. The use of this notation helps save ink, prevents redundant definitions of the same linear operation, and directly corresponds to the implementation of these operations in code. For example, under this notation it makes equal sense to write $\mathbf F\mathbf m$ for the DFT of an intensity-only map \mbf{m} as it does $\mathbf F\bm\nu_i$, where, as discussed, $\bm\nu_i$ is a six-component map. In the latter case, the DFT matrix \mbf{F} is understood to be promoted to a $6\times6$ square block-diagonal matrix with the original \mbf{F} along the block-diagonal. Likewise, $\mathbf a = \mathbf Y^{\dag}\mathbf W \mathbf m$ is a spin-0 harmonic transform, while $\mathbf a_i = \mathbf Y^{\dag}\mathbf W \bm\nu_i$ is a six-component transform (e.g., for PA4, the $a=0-2$ blocks of $\mathbf a_i^a$ would refer to the PA4 f150 band, T, E, and B respectively). 

\subsection{Tiled Noise Model} \label{sec: noise_models_tile}
Previous ACT analyses \citep[e.g.,][]{L17} noted that the stripy noise pattern appeared to be spatially constant across the released maps. In ACT DR4, the relatively small (compared to DR6) sky footprint of each mapped region was referred to as a sky ``patch" \citep{A20, MK21}, and thus downstream analyses, notably \citet{C20} and \citet{Mv20}, built a sparse, 2D Fourier space model of the noise covariance in each patch. In addition, these analyses weighed the Fourier space models by the mapmaker inverse-variance to capture the spatial dependence of the map depth. Thus, within a given patch, the ACT DR4 noise models captured isotropic noise correlations (the ``radial" direction in 2D Fourier space), correlated frequencies (by taking cross-spectra in Equation \ref{eq: fourier_trick}), map depth inhomogeneity (by inverse-variance weighting difference maps), and spatially-\textit{constant} anisotropy (constant across the patch).

The main goal of the tiled noise model is to reproduce the spatially-\textit{varying} anisotropy in the ACT noise, as shown in Figure \ref{fig: 2d_ps_by_region}. Our model is based on the ``constant correlation" model of \citet{N20}, which built an inverse-variance weighted, 2D Fourier power spectrum in interleaved $4^\circ\times4^\circ$ tiles that are cut out of the larger ACT maps. In this sense, the noise model in each tile resembles that for each DR4 patch. We highlight important changes from \citet{N20}, but for further background, we refer the reader to \S4 of that paper, especially Figures 7 and 10 -- 12.

\subsubsection{Model Estimation} \label{sec: noise_models_tile_algorithm}
We outline how to measure the tiled noise model, starting from a set of difference maps, $\bm\nu_i$. We follow a similar algorithm as \citet{N20}, but make a key change in Step \ref{enum: tile_algorithm, item: decoupling_filter} to improve performance. This also establishes a general prescription for a class of Gaussian map-based noise models, since the wavelet and directional noise models will proceed analogously.

\begin{figure}
    \centering
    \includegraphics[width=\columnwidth]{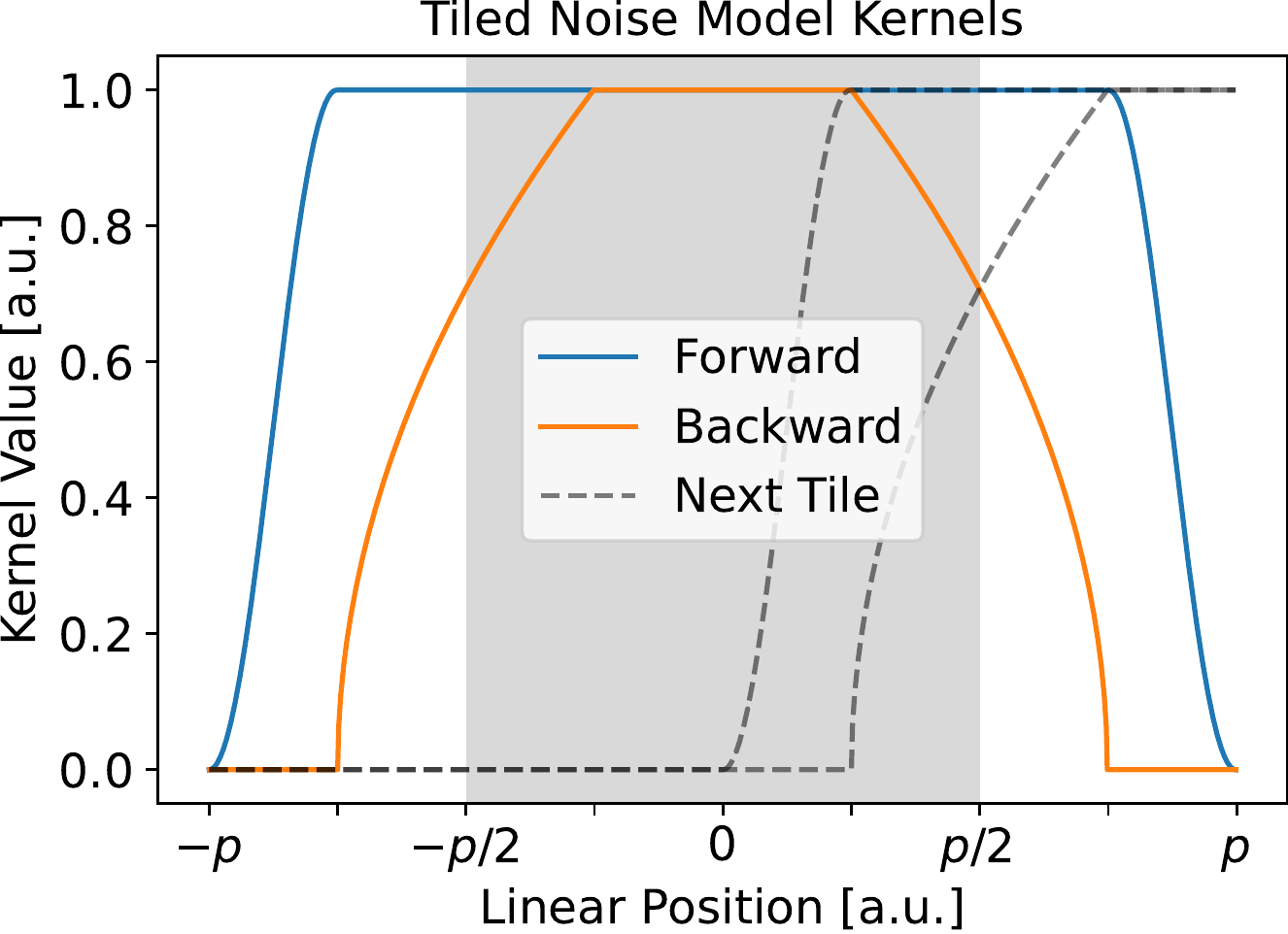}
    \caption{A 1D cross-sectional profile of the ``forward" and ``backward" tile kernel for a tile of pitch $p$. The grey band has width $p$ and is centered on the tile. In this paper, we use $p=4^\circ$. The forward and backward profiles smoothly extend into adjacent tiles; the dashed lines indicate the overlapping kernels for the tile immediately to the right. The full 2D tile kernel is the formed by the outer product of perpendicular 1D profiles.}
    \label{fig: tile_kernel}
\end{figure}

\begin{enumerate}
    \item \label{enum: tile_algorithm, item: stationary_filter} As discussed above, we first want to filter the noise maps to maximize their stationarity. But we know, \textit{a priori}, that their covariance is not translationally invariant in map space, because the map depth has spatial structure. Thus, before anything else, we remove the covariance's dependence on absolute sky position --- due to the map depth, at least --- by multiplying $\bm\nu_i$ with $\sqrt{\mathbf h_i}$. We already saw the effect of this in Equation \ref{eq: var_nu_var_n_2}: the per-pixel variance of $\bm\nu_i*\sqrt{\mathbf h_i}$ (on small scales) is uniform.\footnote{Equations \ref{eq: diff_def_3} and \ref{eq: var_nu_var_n_2} are together equivalent to the first step of the noise prescription of \citet{C20}: $\mathbf d_i*\sqrt{\mathbf w_i}$.} We introduce the matrix form of the operation $\bm\nu_i*\sqrt{\mathbf h_i}$:
    \begin{align} \label{eq: matrix_ivar}
        \mathbf H_i^{\frac{1}{2}} \bm\nu_i
    \end{align}
    where $\mathbf H_i^{1/2}$ is a square diagonal matrix whose diagonal entries are populated by $\sqrt{\mathbf h_i}$. Thus, $\mathbf H_i^{1/2}$ is a ``stationarity filter" acting on $\bm\nu_i$.
\end{enumerate}
    
If we were strictly following the \citet{N20} prescription, we would next break up $\mathbf H_i^{1/2} \bm\nu_i$ into small tiles and evaluate their 2D Fourier power spectra. Doing so in the context of this paper, however, would introduce some negative side-effects. As we will see in Step \ref{enum: tile_algorithm, item: tiling}, the tiling of the maps essentially consists of multiplying the maps with small masks. Multiplication by these local masks in map space effects a convolution by a broad kernel in harmonic space. This would result in a mode-coupling bias --- power being ``smeared" across scales --- since the noise power spectra are very steep. Therefore, we first perform a ``demodulation" step, in which we divide-out the isotropic part of the noise power spectrum (the curves in the top panel of Figure \ref{fig: N_ell_r_ell}) prior to tiling. This results in much flatter power spectra that are less sensitive to mode-coupling. In Appendix \ref{apx: Nl_benefits}, we demonstrate the degree to which mode-coupling induced bias can dominate the tiled noise model, especially for large-scale polarization, without the inclusion of the demodulation step. We proceed as follows.

\begin{enumerate}[resume]
    \item \label{enum: tile_algorithm, item: decoupling_filter} First, we measure the noise power pseudospectra within a large, smooth ``pseudospectrum estimate mask", $\bm\mu_{\mathrm{est}}$.\footnote{We need to use $\bm\mu_{\mathrm{est}}$ because the edges of the ACT footprint are not smooth, so the unmasked pseudospectrum would have undesired ringing artifacts. We discuss our choice of mask in \ref{sec: implementation_masks}.} That is, we measure all the auto- and cross-pseudospectra of the vector $\bm\mu_{\mathrm{est}}*\sqrt{\mathbf h_i}*\bm\nu_i$, which we store in the object $\mathbf C_{\mathrm{est}}$. For example, for PA5, $\mathbf C_{\mathrm{est}}^{0,3}$ (the $a=0$, $b=3$ part of $\mathbf C_{\mathrm{est}}$) is the f090 T$\times$ f150 T cross-pseudospectrum: the only off-diagonal parts of $\mathbf C_{\mathrm{est}}$ are the frequency and polarization crosses indexed by $a$ and $b$. Thus, one can also think of $\mathbf C_{\mathrm{est}}$ as containing $N_{a_{lm}}$ $6\times6$ matrices at each harmonic mode. We note that $\mathbf C_{\mathrm{est}}$ is a symmetric matrix. Then we construct $\mathbf C_{\mathrm{est}}^{-1/2}$, where the matrix exponent acts on each $6\times6$ matrix of $\mathbf C_{\mathrm{est}}$ separately. Finally, we filter the weighted noise maps, $\mathbf H_i^{1/2}\bm\nu_i$, by $\mathbf C_{\mathrm{est}}^{-1/2}$ in harmonic space:
    \begin{align}
        \mathbf Y \mathbf C_{\mathrm{est}}^{-\frac{1}{2}} \mathbf Y^{\dag}\mathbf W \mathbf H_i^{\frac{1}{2}} \bm\nu_i.
    \end{align}
    This new vector is still in map space, and it still has size $6N_{\mathrm{pix}}$. On small scales, it has uniform per-pixel variance. But now, at least as measured within the apodized mask $\bm\mu_{\mathrm{est}}$, it also has flat autospectra,\footnote{The power spectrum of the filtered map as measured in small patches of the sky may not be flat, especially in regions outside $\bm\mu_{\mathrm{est}}$, but should have reduced dynamic range compared to Figure \ref{fig: N_ell_r_ell}, see Appendix \ref{apx: Nl_benefits}.} and has been \textit{decorrelated} --- its cross-spectra have been nulled. As discussed, the goal of flattening the spectra is to mitigate mode-coupling when we tile these maps in Step \ref{enum: tile_algorithm, item: tiling}. The benefit of decorrelating them here will become apparent when we outline the algorithm of drawing a noise simulation: in that case, we will start with a map of easy-to-produce white noise, and by applying the inverse of $\mathbf C_{\mathrm{est}}^{-1/2}$ --- that is, $\mathbf C_{\mathrm{est}}^{1/2}$ --- we will emerge with a map with properly-correlated spectral components.
    
    \item \label{enum: tile_algorithm, item: tiling} After the demodulation step, our goal is still to filter the noise maps to maximize their stationarity. Thus, we break the noise maps into a set of small tiles as motivated by the central assumption of the tiled noise model: that the stripy pattern within small sky patches is constant, but varies over larger distances. We follow the same tiling scheme as \citet{N20}, wherein the ACT footprint is divided into $M\sim\mathcal{O}(1,500)$ equal-sized square tiles with a $4^\circ$ pitch.  In practice, each tile is actually larger than $4^\circ\times4^\circ$ to allow for a $1^\circ$ cosine apodization around its border.\footnote{As with $\bm\mu_{\mathrm{est}}$, the apodization prevents strong ringing when taking the per-tile DFT in the next step. Some tiles extend past the edge of the ACT footprint. For such a tile, we simply redefine its new border to be the intersection of the old border and the ACT footprint, and apodize the new border instead.} As a result, adjacent tiles partially overlap. We fold the apodization into the definition of the tile ``profile," or tile ``kernel:" cutting a tile out of a map consists of multiplying the map by the tile kernel. A cross-section of an arbitrary tile kernel is shown in Figure \ref{fig: tile_kernel}, displaying the apodized edge and overlap with neighboring kernels. We name this operation a ``forward tiling transform;" the ``backward" transform, which combines tiles back into one map, will be defined when we draw simulations.
    
    We write the matrix form of the forward tiling transform. Each tile has significantly fewer pixels than the full map: $N_{\mathrm{pix,tile}} \ll N_{\mathrm{pix}}$. Thus, the first step in applying a given forward tile --- say, tile $j$ --- to a map vector is to ``crop" the map vector, retaining only those pixels contained in tile $j$. The next step is to multiply the cropped map vector by the apodized tile kernel itself. Together, these operations look like:
    \begin{align} \label{eq: single_tile}
        \bm\tau_{j,f}\mathbf P_{j}
    \end{align}
    where $\mathbf P_{j}$ is a $N_{\mathrm{pix,tile}}\times N_{\mathrm{pix}}$ cropping matrix consisting of $N_{\mathrm{pix,tile}}$ rows of unique unit vectors that select the $j^{\mathrm{th}}$ tile pixels, and $\bm\tau_{j,f}$ is a $N_{\mathrm{pix,tile}}\times N_{\mathrm{pix,tile}}$ forward tile matrix, a square diagonal matrix with the $j^{\mathrm{th}}$ tile kernel values along its diagonal. The forward tile transform then consists of all of these tile operations acting on the same map vector:
    \begin{align} \label{eq: forward_tile_transform}
        \mathcal{T}_f = \begin{pmatrix}
        \bm\tau_{0,f}\mathbf P_{0} \\
        \bm\tau_{1,f}\mathbf P_{1} \\
        \vdots \\
        \bm\tau_{M-1,f}\mathbf P_{M-1} \\
        \end{pmatrix}
    \end{align}
    where $\mathcal{T}_f$ is a block-column-vector of $M$ forward tiles. Putting it all together, our input noise maps have been transformed into weighted, demodulated, tiled noise maps:
    \begin{align}
        \mathcal{T}_f \mathbf Y \mathbf C_{\mathrm{est}}^{-\frac{1}{2}} \mathbf Y^{\dag}\mathbf W \mathbf H_i^{\frac{1}{2}} \bm\nu_i.
    \end{align}
    
    \begin{figure}
        \centering
        \includegraphics[width=\columnwidth]{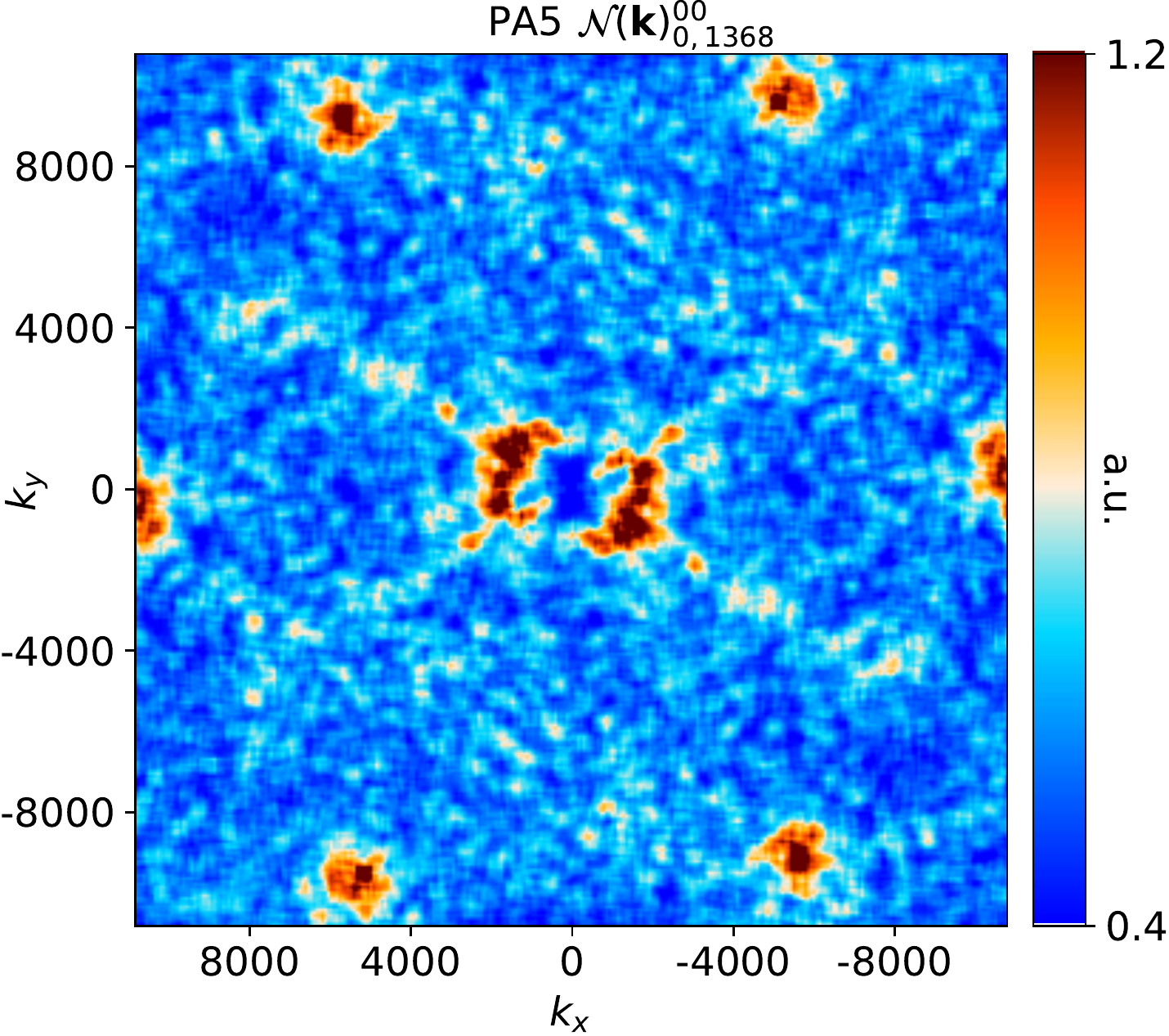}
        \caption{The tiled 2D Fourier power spectrum for PA5 f090, first split ($i=0$). This particular power spectrum corresponds to tile $j=1,368$ in a model bandlimited to $\ell_{\mathrm{max}}=10,800$, and we see all Fourier modes \mbf{k} for the TT ($a=b=0$) component. As in Figure \ref{fig: 2d_ps_by_region}, we can visualize the local anisotropic structure in this tile, including an ``x-like" scan strategy and detector-detector correlations at the image perimeter (small scales). More examples are visible in \citet{N20}, Figure 11.}
        \label{fig: tile_cov}
    \end{figure}
    
    \item Assuming that within each tile the noise covariance is stationary, we take our cue from the cross-correlation theorem (Equation \ref{eq: fourier_trick}). That is, by transforming to Fourier space, we will assume that each Fourier mode in the resulting vector is \textit{uncorrelated} with all other Fourier modes:
    \begin{align} \label{eq: tile_uncorrelated_vector}
        \mathbf u_i \equiv \mathbf F \mathcal{T}_f \mathbf Y \mathbf C_{\mathrm{est}}^{-\frac{1}{2}} \mathbf Y^{\dag} \mathbf W \mathbf H_i^{\frac{1}{2}} \bm\nu_i
    \end{align}
    where \mbf{u_i} is the weighted, demodulated, tiled, 2D Fourier projection of the map-based noise, and \mbf{F} is understood to be a square block-diagonal matrix of DFTs (one for each tile). For each Fourier mode, however, we still retain the six correlated map fields: two frequency bands and three Stokes components.

    \item With the construction of \mbf{u_i}, most of the hard work is complete. The last steps use \mbf{u_i} to build a sparse (diagonal) covariance matrix, just as in Equation \ref{eq: fourier_trick}. For each tile, we evaluate its 2D Fourier noise power spectra:
    \begin{align} \label{eq: tile_ps}
        \mathcal{N}(\mathbf k)_{i,j}^{ab}\delta(\mathbf k - \mathbf k') = \frac{1}{f_{j,2}}\Re\left(\mathbf u_{i,j}^a(\mathbf k) \mathbf u_{i,j}^{b*}(\mathbf k)\right)
    \end{align}
    where \mbf{k} is a given 2D Fourier mode, $\Re$ denotes the real part (relevant when $a\neq b$), $\mathbf u_{i,j}^a(\mathbf k)$ is the $j^{\mathrm{th}}$ tile and $a^{\mathrm{th}}$ component of our transformed noise map for split $i$, and $f_{j,2}$ corrects for the loss of power from the apodization in the tile \citep{C20, Mv20, N20}:
    \begin{align} \label{eq: mask_correction}
        f_{j,2} = \frac{1}{N_{\mathrm{pix, tile}}}\mathrm{Tr}[\bm\tau_{j,f}^{\dag}\bm\tau_{j,f}]
    \end{align}
    which is just the average value of the squared tile kernel. Thus, in each tile, the covariance matrix $\mathcal{N}_{i,j}$ is block-diagonal in each Fourier mode \mbf{k}, with a single block being a $6\times6$ matrix $\mathcal{N}(\mathbf k)_{i,j}$. The total covariance matrix $\mathcal{N}_i$ is block-diagonal in tiles, with a single block being $\mathcal{N}_{i,j}$:
    \begin{align} \label{eq: block_covmat}
        \mathcal{N}_i &= \begin{pmatrix}
        \mathcal{N}_{i,0} & 0 & \hdots & 0 \\
        0 & \mathcal{N}_{i,1} & \hdots & 0 \\
        \vdots & \vdots & \ddots & \vdots \\
        0 & 0 & \hdots & \mathcal{N}_{i,M-1}
        \end{pmatrix}.
    \end{align}
    This highly diagonal form is the sparse representation we originally sought: it has $\mathcal{O}(N_{\mathrm{pix}})$ nonzero elements near its diagonal, rather than being dense with $\mathcal{O}(N_{\mathrm{pix}}^2)$ nonzero elements.
    
    \item \label{enum: tile_algorithm, item: smoothing} As in \citet{C20, Mv20, N20}, we smooth each tile's power spectrum estimate $N(\mathbf k)_{i,j}^{ab}$ over nearby modes \mbf{k} to reduce sample variance. Unlike the smoothing procedure for the two wavelet models (discussed in \S\ref{sec: implementation_smoothing} and Appendix \ref{apx: smoothing}), for the tiled model we cannot reliably distinguish between random fluctuations and true features in the tiled 2D power spectra. We thus select the same smoothing width in Fourier space as was used in \citet{Mv20} and \citet{N20}, $\Delta k=400$ in both the x- and y-directions. We perform the smoothing via convolution with a tophat kernel directly applied to the 2D power spectra.\footnote{Recall we have already divided-out the steep isotropic part of the power spectrum, so we are smoothing over the \textit{anisotropic} features in $N(\mathbf k)_{i,j}^{ab}$ \textit{relative} to the isotropic power spectrum.} The smoothing width is fixed both within a given tile, and for each tile. An example smoothed, tiled 2D Fourier power spectrum is shown in Figure \ref{fig: tile_cov}.
    
    \item Lastly, in preparation for drawing noise simulations, we note that the actual objects we save to disk are not the tiled 2D Fourier covariance matrices, but rather their matrix square-root. This is the matrix analog of obtaining the standard deviation from the variance in preparation for sampling a scalar Gaussian random variable. Thus, the object we save to disk is actually $\mathcal{N}_i^{1/2}$. As in Step \ref{enum: tile_algorithm, item: decoupling_filter}, the only off-diagonal parts of $\mathcal{N}_i$ are the frequency and polarization crosses indexed by $a$ and $b$. Therefore, the matrix exponent acts on each $6\times6$ matrix $\mathcal{N}(\mathbf k)_{i,j}$ separately. Some examples of $\mathcal{N}(\mathbf k)_{i,j}^{1/2}$ are visible in Figure \ref{fig: tile_schematic}. Because it is also used in drawing simulations, we save $\mathbf C_{\mathrm{est}}^{1/2}$ to disk as well.
\end{enumerate}

These steps generate the tiled model square-root covariance $\mathcal{N}_i^{1/2}$ for the purposes of this paper. A schematic depiction of all these steps is given in the top row of Figure \ref{fig: tile_schematic}.

\begin{figure*}
    \centering
    \includegraphics[width=\textwidth]{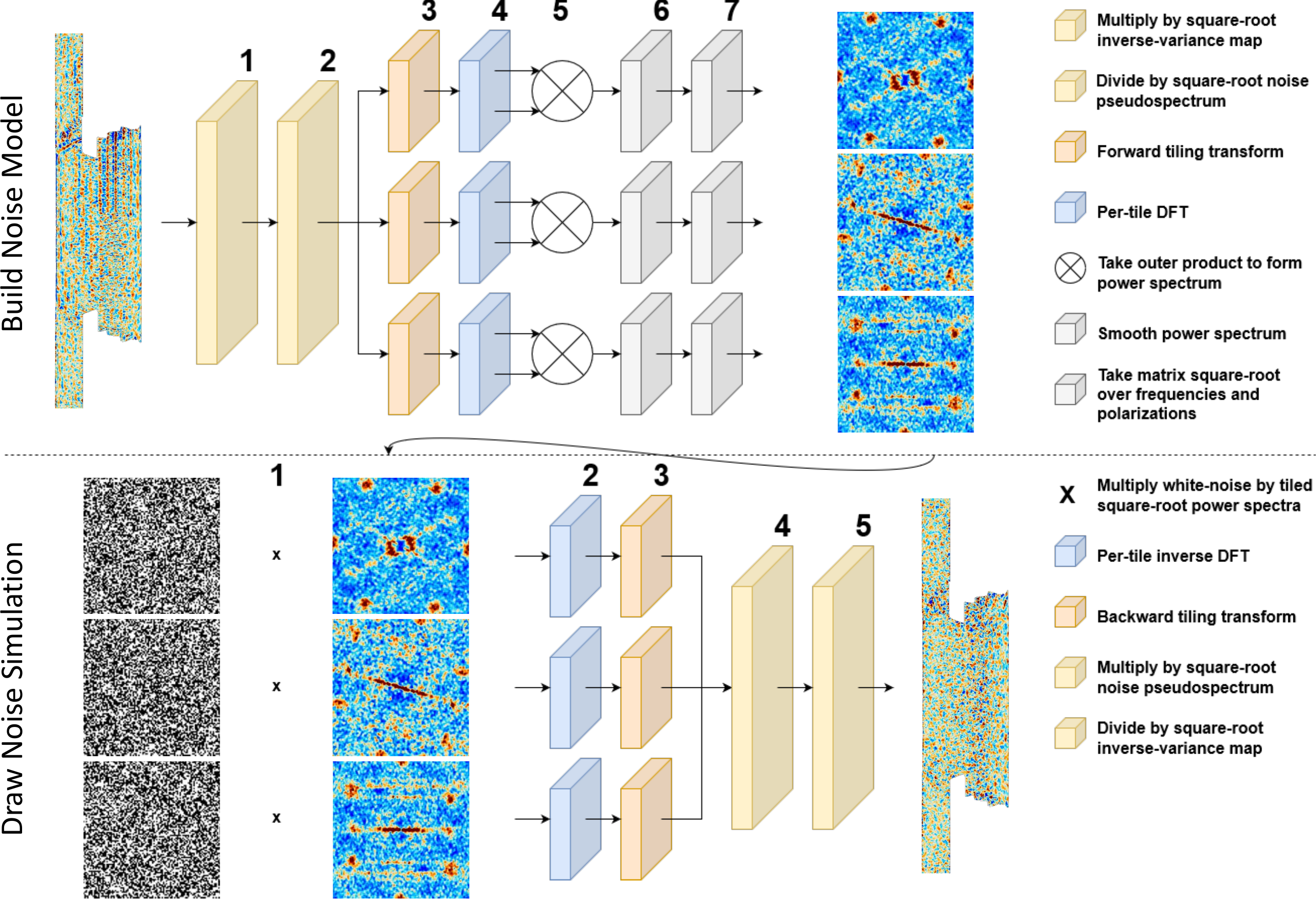}
    \caption{\textit{Top:} The sequence of filters and other operations involved in measuring the sparse covariance in the tiled noise model. Large boxes denote filters applied to the entire map, while small boxes denote filters or operations in the tiled basis. The orange filters represent the ``forward" tiling kernels in Figure \ref{fig: tile_kernel}. Each operation is labelled by the corresponding step number in the model estimation algorithm. The input difference map is PA5 f090, T ($a=0$), first split ($i=0$), and the outputs are 2D square-root Fourier power spectra, TT ($a=b=0$). We see the noise anisotropy varies by tile. \textit{Bottom:} The sequence of filters and other operations involved in drawing a simulation of the map-based noise in the tiled model. The first column denotes independent, Gaussian white noise. Importantly, the orange filters now represent the ``backward" tiling kernels in Figure \ref{fig: tile_kernel}, while all other operations revert those of the top row. Each operation is labelled by the corresponding step number in the simulation algorithm. The output is a tiled noise simulation of PA5 f090, T, first split.}
    \label{fig: tile_schematic}
\end{figure*}

\subsubsection{Simulations and Covariance Matrix} \label{sec: noise_models_tile_formalism}
The ultimate aim of this paper is to create realistic simulations of the map-based noise. Here, we explain how we draw such a such a simulation from the model we have just described. Along the way, we also write down the full map-based noise covariance matrix. The following steps are essentially applicable to the wavelet models as well; in \S\ref{sec: noise_models_wav_formalism} and \S\ref{sec: noise_models_fdw_formalism} we only discuss minor differences from what follows here. Batches of these map-based noise simulations are the main product of this paper. 

There are two key steps in extracting a realization of the map-based noise from the tiled model. We start by drawing a realization in tiled 2D Fourier space. Doing so is tractable thanks to the highly diagonal structure of the tiled 2D Fourier covariance, $\mathcal{N}_i$. Then we need to transform the realization back to map space, which essentially means reverting the various operations that built \mbf{u_i} from $\bm\nu_i$ in Equation \ref{eq: tile_uncorrelated_vector}.

\begin{enumerate} \label{enum: tile_sim}
    \item For each tile, draw Gaussian white noise in tiled 2D Fourier space and take the inner product with the square-root covariance:
    \begin{align}
        \mathcal{N}_i^{\frac{1}{2}}\bm\eta_i
    \end{align}
    where $\bm\eta_i$ is complex-valued white noise with the same shape as \mbf{u_i}. Again, the only nontrivial (off-diagonal) part of this product is in the frequency/polarization index. By definition, $\EV{\bm\eta_i\bm\eta_i^\dag} = \mathbf I$, where \mbf{I} is the identity. Therefore, $\mathcal{N}_i^{1/2}\bm\eta_i$ is a sample from the tiled covariance:
    \begin{align} \label{eq: white_noise_trick}
        \EV{\mathcal{N}_i^{\frac{1}{2}}\bm\eta_i (\mathcal{N}_i^{\frac{1}{2}}\bm\eta_i)^\dag} = \mathcal{N}_i^{\frac{1}{2}}\EV{\bm\eta_i\bm\eta_i^\dag}\mathcal{N}_i^{\frac{1}{2},\dag} = \mathcal{N}_i^{\frac{1}{2}}\mathcal{N}_i^{\frac{1}{2},\dag} = \mathcal{N}_i
    \end{align}
    by construction. This demonstrates why we save to disk the square-root covariance.
    \item With our tiled 2D Fourier space noise realization in hand, we proceed to revert Equation \ref{eq: tile_uncorrelated_vector}. So, first we transform out of tiled Fourier space and back to tiled map space with a per-tile inverse DFT:
    \begin{align}
        \mathbf F^{\dag} \mathcal{N}_i^{\frac{1}{2}}\bm\eta_i.
    \end{align}
    \item Next, we need to stitch tiles back together into a single map. That is, we need to apply a ``backward" tiling transform. At this point in the algorithm, each tile contains an independent realization of its own stripy noise pattern. Therefore, as noted in \citet{N20}, naively pasting each tile next to one another with a step-function edge would produce sharp discontinuities in the synthesized map. Just like the forward transform, the solution is to apodize and overlap their borders such that the transition across tiles is continuous. A good illustration of this is shown in \citet{N20}, Figure 13. Unfortunately, we cannot just reuse the tile kernels from the forward transform: we should specifically design the apodization to preserve the per-pixel noise variance in the tile overlap. We show the exact constraint that must be satisfied in Equation \ref{eq: admissibility} and derive it in Appendix \ref{apx: unbiased_covmat}, but, in short, the forward tile profile does not satisfy it. Instead, we use the backward profile shown in Figure \ref{fig: tile_kernel}, which is the square-root of the bilinear interpolation from \citet{N20}. 
    
    The matrix form of the backward tiling transform is identical to that of the forward tiling transform (Equation \ref{eq: forward_tile_transform}):
    \begin{align} \label{eq: backward_tile_transform}
        \mathcal{T}_b = \begin{pmatrix}
        \bm\tau_{0,b}\mathbf P_{0} \\
        \bm\tau_{1,b}\mathbf P_{1} \\
        \vdots \\
        \bm\tau_{M-1,b}\mathbf P_{M-1} \\
        \end{pmatrix}.
    \end{align}
    The only difference is that the diagonal of each $N_{\mathrm{pix,tile}}\times N_{\mathrm{pix,tile}}$ backward tile matrix, $\bm\tau_{j,b}$, contains the values of the backward tile kernels. The projection matrices $\mathbf P_j$ are unchanged: by applying $\mathcal{T}_b^{\dag}$, they perform a ``paste" operation that is the exact opposite of the ``crop" operation from the forward transform. Thus we have:
    \begin{align}
        \mathcal{T}_b^{\dag} \mathbf F^{\dag} \mathcal{N}_i^{\frac{1}{2}}\bm\eta_i
    \end{align}
    where the vector now has the same shape as an input noise map, $\bm\nu_i$, and likewise lives in the original, untiled map space.
\end{enumerate}

As discussed, when stitched together in the synthesized map, the backward tile profiles should preserve the total per-pixel noise variance. For example, for a given pixel belonging only to one tile, it should be the case that the value of that tile's backward kernel is 1 --- or else we would artificially increase or decrease that pixel's variance. In general --- including for pixels in an overlapping region between multiple tiles --- the \textit{quadrature sum} of all contributing tile kernels should likewise be 1:
\begin{align} \label{eq: admissibility}
    \mathbf I = \sum_{j=0}^{M-1} \mathbf P_j^{\dag} (\bm\tau_{j,b}^{\dag} \bm\tau_{j,b}) \mathbf P_j
\end{align}
where the term in parentheses is just the squared value of the backward tile kernels, and the sum over projection matrices performs the tile ``stitching." We derive Equation \ref{eq: admissibility} more rigorously in Appendix \ref{apx: unbiased_covmat}. Continuing with the simulation algorithm:

\begin{enumerate}[resume]
    \item We need to apply the inverse of the demodulation filter from the model estimation. That is, we multiply the simulation in harmonic space by $\mathbf C_{\mathrm{est}}^{1/2}$:
    \begin{align}
        \mathbf Y \mathbf C_{\mathrm{est}}^{\frac{1}{2}} \mathbf Y^{\dag}\mathbf W \mathcal{T}_b^{\dag} \mathbf F^{\dag} \mathcal{N}_i^{\frac{1}{2}}\bm\eta_i.
    \end{align}
    This reapplies the steep noise auto- and cross-spectra to the simulation.
    
    \item The last step in drawing a noise simulation is to divide by the square-root of the inverse-variance map, or multiply element-wise by $1/\sqrt{\mathbf h_i}$. This is analogous to multiplying by the per-pixel standard deviation. In matrix form, a simulation from the tiled noise model is therefore defined as:
    \begin{align} \label{eq: tile_sim}
        \bm\xi_i \equiv \mathbf H_i^{-\frac{1}{2}} \mathbf Y \mathbf C_{\mathrm{est}}^{\frac{1}{2}} \mathbf Y^{\dag}\mathbf W \mathcal{T}_b^{\dag} \mathbf F^{\dag} \mathcal{N}_i^{\frac{1}{2}}\bm\eta_i.
    \end{align}
\end{enumerate}

These steps generate a tiled model noise simulation, $\bm\xi_i$, for the purposes of this paper. A schematic depiction of all these steps is given in the bottom row of Figure \ref{fig: tile_schematic}. We also have the necessary ingredients to write the full map-based noise covariance matrix in the tiled model. By definition, the map-based noise covariance matrix is:
\begin{align} \label{eq: covariance_matrix}
    \mathbf N_i = \EV{\bm\xi_i \bm\xi_i^\dag}.
\end{align}
Combining Equations \ref{eq: white_noise_trick}, \ref{eq: tile_sim}, and \ref{eq: covariance_matrix} we have:
\begin{align} \label{eq: tile_covmat}
    \mathbf N_i = \mathbf H_i^{-\frac{1}{2}} \mathbf Y \mathbf C_{\mathrm{est}}^{\frac{1}{2}} \mathbf Y^{\dag}\mathbf W \mathcal{T}_b^{\dag} \mathbf F^{\dag} \mathcal{N}_i \mathbf F \mathcal{T}_b \mathbf W\mathbf Y \mathbf C_{\mathrm{est}}^{\frac{1}{2}} \mathbf Y^{\dag} \mathbf H_i^{-\frac{1}{2}}
\end{align}
where we have used that $\mathbf H_i^{-1/2}$, $\mathbf C_{\mathrm{est}}^{1/2}$, and $\mathbf W$ are Hermitian.

This section has given a detailed accounting of how we both measure the tiled noise model and draw a tiled noise simulation. While thorough, it is an investment that will pay off in our discussion of the two wavelet models: in spite of the wavelets' orthogonal approach to modeling the map-based noise, their actual mechanics are highly similar to the tiled model.

\subsection{Isotropic Wavelet Noise Model} \label{sec: noise_models_wav}
Our isotropic wavelet, or ``wavelet," noise model especially targets the scale-dependent map depth of the ACT DR6 noise, as shown in Figure \ref{fig: ivar_by_scale}. The model works by taking Figure \ref{fig: ivar_by_scale} at face value: maps of the per-pixel noise variance appear different depending on the filtered range of angular scales. That is, each range of angular scales could be assigned its own ``inverse-variance" map, where each map still operates in the white noise, uncorrelated-pixel limit. This idea is especially well motivated by the observation that the transition from the small-scale to the large-scale map depth in Figure \ref{fig: ivar_by_scale} is reasonably smooth (recall that the bottom panel shows the morphology after normalizing by the mapmaker inverse-variance). In other words, we do not need especially high-resolution ``filters" in harmonic space to capture the simultaneous scale- and spatial-dependence of the map depth. We therefore assume that the set of scale-dependent inverse-variance maps forms a complete and sparse description of the map-based noise.

Applying the core principle of the cross-correlation theorem (Equation \ref{eq: fourier_trick}), a pixel-wise uncorrelated covariance in map space inspires us to seek stationary regions in harmonic space. We isolate these regions by tiling the harmonic plane, where now we call the harmonic space kernels ``wavelets:" a set of smooth, interleaved profiles defined over $\ell$, each of which filters for a compact range of angular scales.\footnote{Technically speaking, there is not an exact analog of Equation \ref{eq: fourier_trick} in harmonic space; instead, we use it as an approximately true ``guiding principle."} Thus, the swapping of map space and harmonic space is the main \textit{algorithmic} difference between the tiled model and the wavelet model, and we can proceed nearly identically as in \S\ref{sec: noise_models_tile}. We first prescribe how we construct a sparse representation of the map-based noise covariance under the wavelet model in \S\ref{sec: noise_models_wav_algorithm}. For reference, this procedure is similar to the construction of the total data covariance in needlet ILC (NILC) analyses \citep[e.g.,][]{WMAP_NILC, Aghanim2016, Coulton23}. Then, in \S\ref{sec: noise_models_wav_formalism}, we outline how we draw a simulation from the wavelet model. 

\subsubsection{Model Estimation} \label{sec: noise_models_wav_algorithm}
As before, the goal of the model estimation algorithm is to first filter the ACT DR6 noise maps to promote their stationarity, and then transform them to permit a sparse representation of the covariance matrix. In the wavelet model, we work in harmonic space first and then transform into map space.

\begin{figure}
    \centering
    \includegraphics[width=\columnwidth]{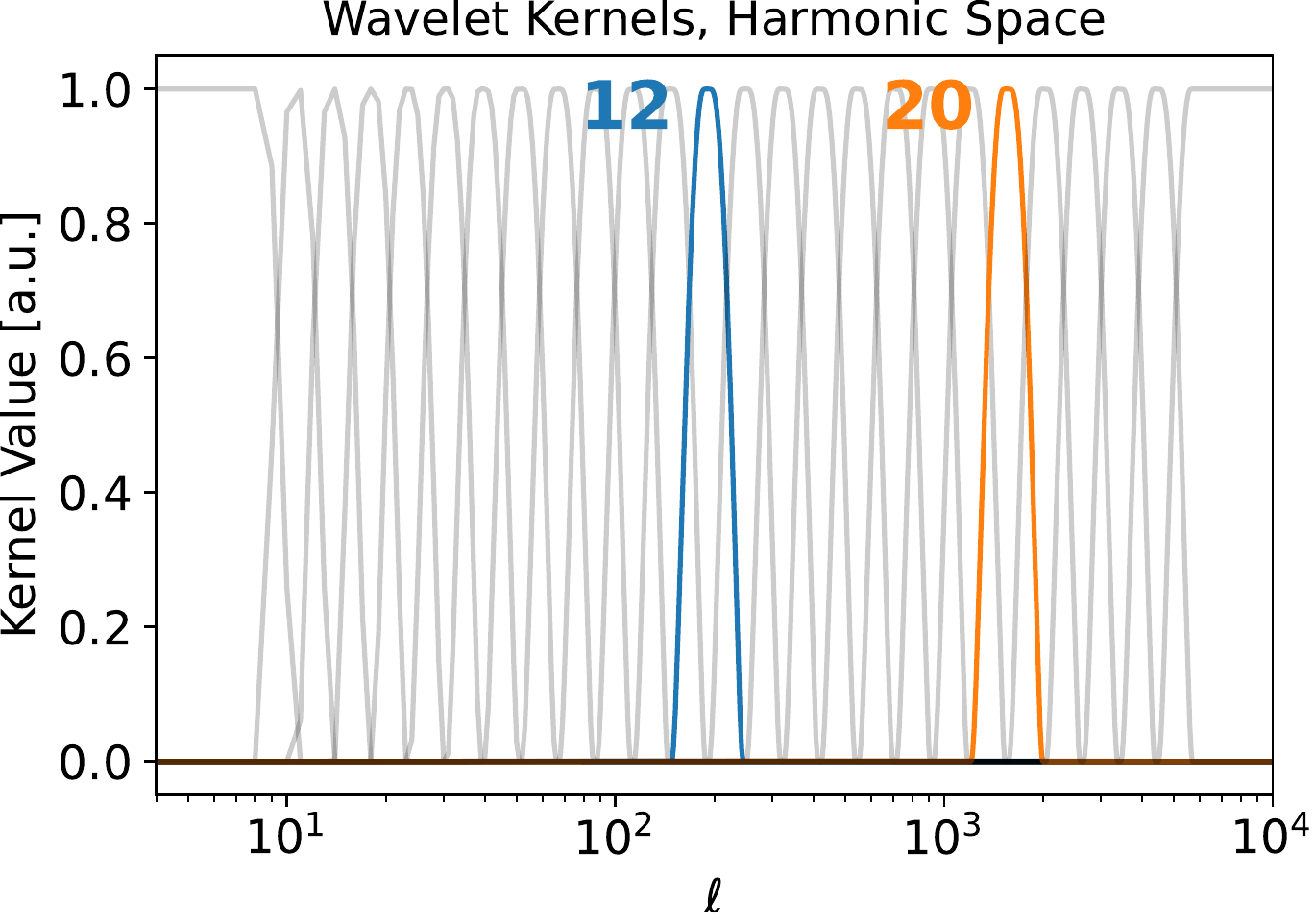}
    \includegraphics[width=0.9\columnwidth]{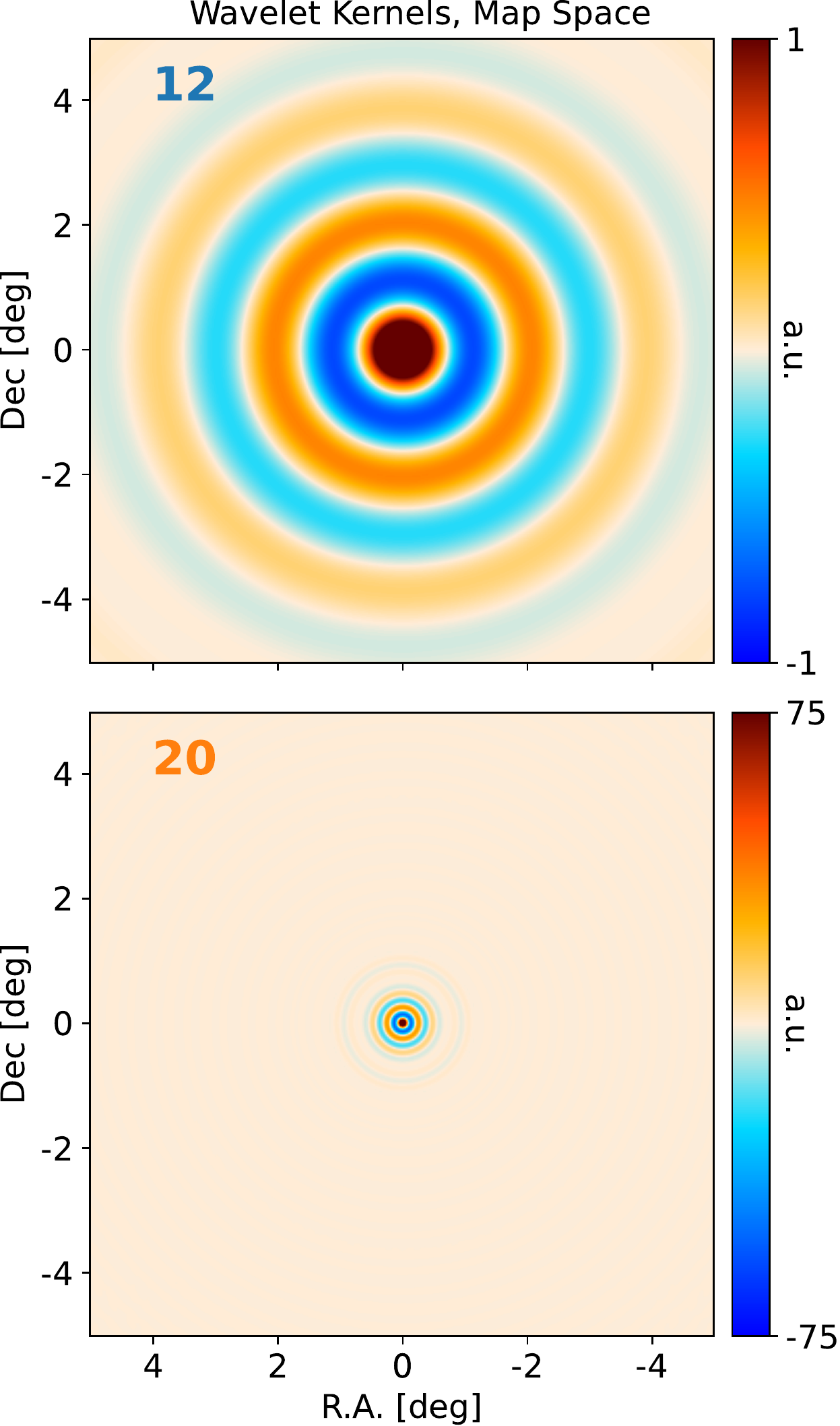}
    \caption{\textit{Top:} Scale-discrete wavelet kernels \citep{Wiaux2008, sd-wavelets} defined in harmonic space. Kernels with indices $j=12$ and $j=20$ are highlighted. \textit{Middle:} Kernel $j=12$ in map space, with support in $\ell$ of $\sim100 - 200$. As expected, this kernel captures scales on the order of $\sim1^\circ - 2^\circ$. \textit{Bottom:} Kernel $j=20$ in map space, with support in $\ell$ of $\sim1,000 - 2,000$. As expected, this kernel captures scales on the order of $\sim0.1^\circ - 0.2^\circ$. Its wider width in harmonic space translates to a narrower width in map space.}
    \label{fig: wav_kernel}
\end{figure}

\begin{enumerate}
    \item We start by considering the harmonic transform of the noise maps. Thus, our input noise vectors are given by:
    \begin{align}
        \mathbf Y^{\dag}\mathbf W \bm\nu_i
    \end{align}
    which is now a $6N_{a_{lm}}$ element vector instead of a $6N_\mathrm{pix}$ element vector.
    
    \item \label{enum: wav_algorithm, item: flattening} Just as in the tiled model, at this stage, we have \textit{a priori} knowledge that the noise covariance is not translationally invariant in harmonic space. Instead, the noise variance depends on the absolute position in harmonic space --- that is, depends on $\ell$ --- due to the steep noise power spectra (Figure \ref{fig: N_ell_r_ell}). Therefore, we still need to apply a ``stationarity filter," only this time in harmonic space instead of map space. We perform the same filtering operation as in \S\ref{sec: noise_models_tile_algorithm}, Step \ref{enum: tile_algorithm, item: decoupling_filter}, wherein we divide-out the square-root noise spectra\footnote{The $\mathbf C_{\mathrm{est}}$ measured in the wavelet model will not be the same as the tiled model, since in the tiled model we had already multiplied the noise maps by the square-root inverse-variance before measuring the spectra.} as measured in the same pseudospectrum estimate mask, $\bm\mu_{\mathrm{est}}$:
    \begin{align}
        \mathbf C_{\mathrm{est}}^{-\frac{1}{2}} \mathbf Y^{\dag}\mathbf W \bm\nu_i.
    \end{align}
    Whereas in the tiled model, this operation functioned as a demodulation filter, now the flattened power spectra help promote the noise stationarity in harmonic space. Again, we further explore the necessity of this filter in Appendix \ref{apx: Nl_benefits}.
\end{enumerate}
    
In principle, we could now perform the harmonic space equivalent of a ``demodulation" step to avoid mode-coupling. Recall, in the (map space) tiled model, this involved a filter operating in harmonic space --- dividing-out the square-root noise pseudospectra. The effect of this was to suppress mixing power across angular scales when tiling the maps. Therefore, for the (harmonic space) wavelet model, demodulation would involve a filter operating in \textit{map space} --- namely, multiplying by the square-root inverse-variance maps (as in Equation \ref{eq: var_nu_var_n_2}). Analogous to the tiled model, this would suppress mixing power across regions of the map when breaking the harmonic plane into wavelets in the next step: multiplication by a local wavelet profile in harmonic space effects a convolution by a broad kernel in map space. Nevertheless, in practice we found this demodulation step was not necessary for performance of the wavelet model, and we chose to forego it here. Although we do not include it, we discuss the drawbacks and benefits of this demodulation step further in \S\ref{sec: disc_conc}.

\begin{figure*}
    \centering
    \includegraphics[width=\textwidth]{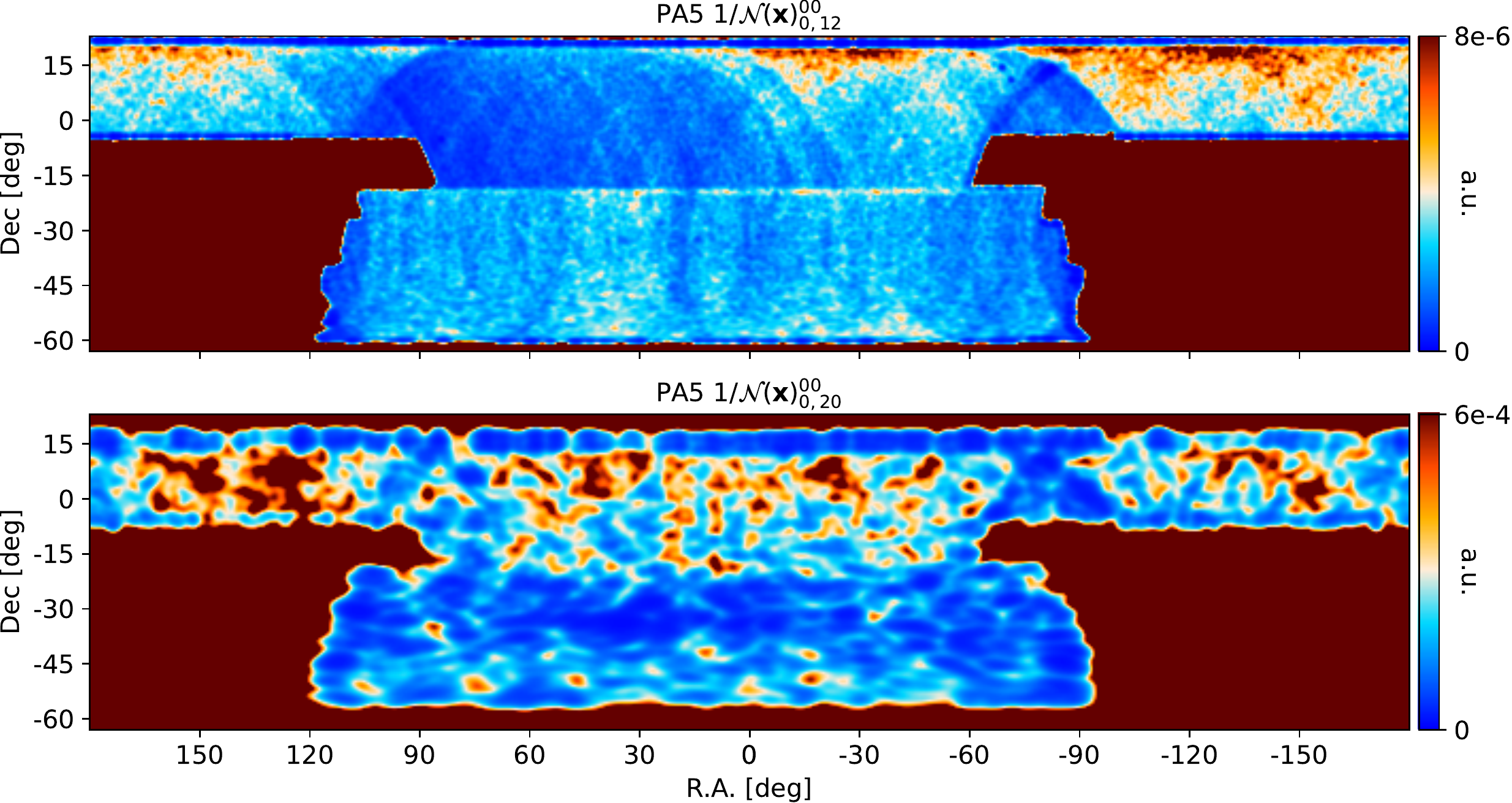}
    \caption{\textit{Top (Bottom):} The inverse of a wavelet noise power map corresponding to the high-$\ell$, $j=20$ (low-$\ell$, $j=12$) wavelet kernels highlighted in Figure \ref{fig: wav_kernel}. The particular component is PA5 f090, TT ($a=b=0$), first split ($i=0$). As expected, when compared to Figure \ref{fig: ivar_by_scale}, we see the small-scale (large-scale) wavelet inverse-variance is suggestive of the morphology of the mapmaker inverse-variance (cross-linking) map.}
    \label{fig: wav_cov}
\end{figure*}
    
\begin{enumerate}[resume]
    \item \label{enum: wav_algorithm, item: tiling} After flattening the spectra, our goal is still to isolate stationary regions in the transformed noise. As discussed, we achieve this by breaking the noise into a set of ``wavelets," which are analogous to tiles defined in harmonic space. For this purpose we use the ``scale-discrete" wavelets of \citet{Wiaux2008, sd-wavelets} with a scaling parameter $\lambda=1.3$ ($\lambda$ sets the logarithmic wavelet width), lowest wavelet scale $J_0=9$ (which sets the $\ell_{\mathrm{max}}$ of the lowest-$\ell$ wavelet), and highest wavelet scale $J=33$ (which determines the overall number of wavelets). These parameters generate the $M=26$ wavelet kernels shown in the top panel of Figure \ref{fig: wav_kernel}.\footnote{We augment the highest-$\ell$ wavelet kernel to extend to arbitrarily high-$\ell$ at a value of 1.} The wavelet log-spacing is well suited to the roughly power law scaling of the noise spectra. In addition, the smooth overlap of the wavelet profiles plays a similar role as the apodized tile edges in \S\ref{sec: noise_models_tile_algorithm}, Step \ref{enum: tile_algorithm, item: tiling}: it prevents strong ringing of the \textit{map-space} wavelet kernels when taking the per-wavelet SHT in the next step. Said another way, it sacrifices some of each wavelet's resolution in harmonic space to increase its resolution in map space. The bottom panels of Figure \ref{fig: wav_kernel} demonstrate just that: the map-space wavelet kernels are compact, and can therefore resolve local noise features in map space. We also observe that because the higher-$\ell$ wavelets are wider in harmonic space, their map-space kernels are commensurately tighter. Thus, the wavelet decomposition is capable of simultaneously localizing noise properties over angular scale and sky position.
    
    We again denote this operation a ``forward tiling transform." As before, the first step in applying a given wavelet profile to a map vector is to ``crop" the noise vector. Only now, we retain all harmonic modes up to the wavelet $\ell_{\mathrm{max}}$, which is to say we \textit{bandlimit} the noise vector for each wavelet. Then, we multiply the set of bandlimited vectors by the corresponding wavelet profile. For a given wavelet $j$, the matrix form is identical to Equation \ref{eq: single_tile}, where \mbf{P_j} is a $N_{a_{lm},\mathrm{wav}}\times N_{a_{lm}}$ bandlimiting matrix, and $\bm\tau_{j,f}$ is a $N_{a_{lm},\mathrm{wav}}\times N_{a_{lm},\mathrm{wav}}$ square diagonal matrix with the $j^{\mathrm{th}}$ wavelet kernel values along its diagonal. Then the forward tiling transform, $\mathcal{T}_f$ has the exact same form as Equation \ref{eq: forward_tile_transform}. Our input noise maps thus become transformed, weighted, tiled noise maps:
    \begin{align}
        \mathcal{T}_f \mathbf C_{\mathrm{est}}^{-\frac{1}{2}} \mathbf Y^{\dag}\mathbf W \bm\nu_i.
    \end{align}
    
    \item We again are guided by the cross-correlation theorem (Equation \ref{eq: fourier_trick}): we now transform into map space, where we assume that each pixel in the resulting vector is \textit{uncorrelated} with all other pixels:
    \begin{align} \label{eq: wav_uncorrelated_vector}
        \mathbf u_i \equiv \mathbf Y \mathcal{T}_f \mathbf C_{\mathrm{est}}^{-\frac{1}{2}} \mathbf Y^{\dag}\mathbf W \bm\nu_i.
    \end{align}
    Thus, \mbf{u_i} contains a set of $M$ bandlimited ``wavelet maps," where each map has six correlated frequency/polarization components, but whose pixels are uncorrelated.
    
    \item Using \mbf{u_i}, we evaluate the pixel-diagonal wavelet power maps:
    \begin{align} \label{eq: wav_ps}
        \mathcal{N}(\mathbf x)_{i,j}^{ab}\delta(\mathbf x - \mathbf x') = \frac{1}{f_{j,2}(\mathbf x)}\left(\mathbf u_{i,j}^a(\mathbf x) \mathbf u_{i,j}^{b*}(\mathbf x)\right)
    \end{align}
    where \mbf{x} is a given wavelet map pixel, $\mathbf u_{i,j}^a(\mathbf x)$ is the $j^{\mathrm{th}}$ wavelet map and $a^{\mathrm{th}}$ component of our transformed noise vector evaluated at \mbf{x}, and $f_{j,2}(\mathbf x)$ corrects for the power suppression due to the modes lost when multiplying by each wavelet profile \citep{Dahlen2008}:
    \begin{align} \label{eq: mask_correction_2}
        f_{j,2}(\mathbf x) = \frac{A(\mathbf x)}{4\pi}\mathrm{Tr}[\bm\tau_{j,f}^{\dag}\bm\tau_{j,f}]
    \end{align}
    where $A$, the pixel area in steradians, accounts for the spherical geometry of the wavelets. The total wavelet covariance matrix, $\mathcal{N}_i$, is thus the same form as Equation \ref{eq: block_covmat}. Each block along the diagonal, $\mathcal{N}_{i,j}$, is itself block-diagonal in map pixel \mbf{x}, with a single block being a $6\times6$ matrix $\mathcal{N}(\mathbf x)_{i,j}$; the elements of $\mathcal{N}(\mathbf x)_{i,j}$ are given by Equation \ref{eq: wav_ps}. Again, this highly diagonal form is our desired sparse representation, containing $\mathcal{O}(N_{a_{lm}})$ nonzero elements near its diagonal.
    
    \item \label{enum: wav_algorithm, item: smoothing} We smooth each wavelet power map $\mathcal{N}(\mathbf x)_{i,j}^{ab}$ over nearby pixels \mbf{x} to reduce sample variance. Some examples of smoothed wavelet power maps are shown in Figure \ref{fig: wav_cov}. We see the wavelet model is suggestive of cross-linking-like noise morphology at large scales, and inverse-variance-like noise morphology at small scales.
    
    Unlike the tiled model, we do not use the same smoothing width for all wavelets. Instead, we select a Gaussian smoothing kernel with a FWHM of $f(\ell_{\mathrm{max}})\pi/\ell_{\mathrm{max}}$ radians, where $f$ is some function of $\ell$ and $\ell_{\mathrm{max}}$ is the maximum support of the given wavelet. Because the $\ell_{\mathrm{max}}$ of our SHTs are set to the Nyquist bandlimit of the corresponding map pixelization, $\pi/\ell_{\mathrm{max}}$ for a given wavelet is the pixel size (in radians) of the corresponding wavelet map. Thus, $f$ counts the number of pixels in the wavelet power maps across the smoothing FWHM. As shown in \S\ref{sec: implementation_smoothing}, Figure \ref{fig: optimal_smoothing}, $f$ takes on values of $\mathcal{O}(10)$. The choice of $f(\ell)$ is discussed further in \S\ref{sec: implementation_smoothing}, as well as in Appendix \ref{apx: smoothing}. 
    
    \item Lastly, as before, we save to disk the matrix square-root of $\mathcal{N}_i$ --- that is, $\mathcal{N}_i^{1/2}$ --- and $\mathbf C_{\mathrm{est}}^{1/2}$. Some examples of $\mathcal{N}(\mathbf x)_{i,j}^{1/2}$ are visible in Figure \ref{fig: wav_schematic}. 
    \begin{figure*}
        \centering
        \includegraphics[width=\textwidth]{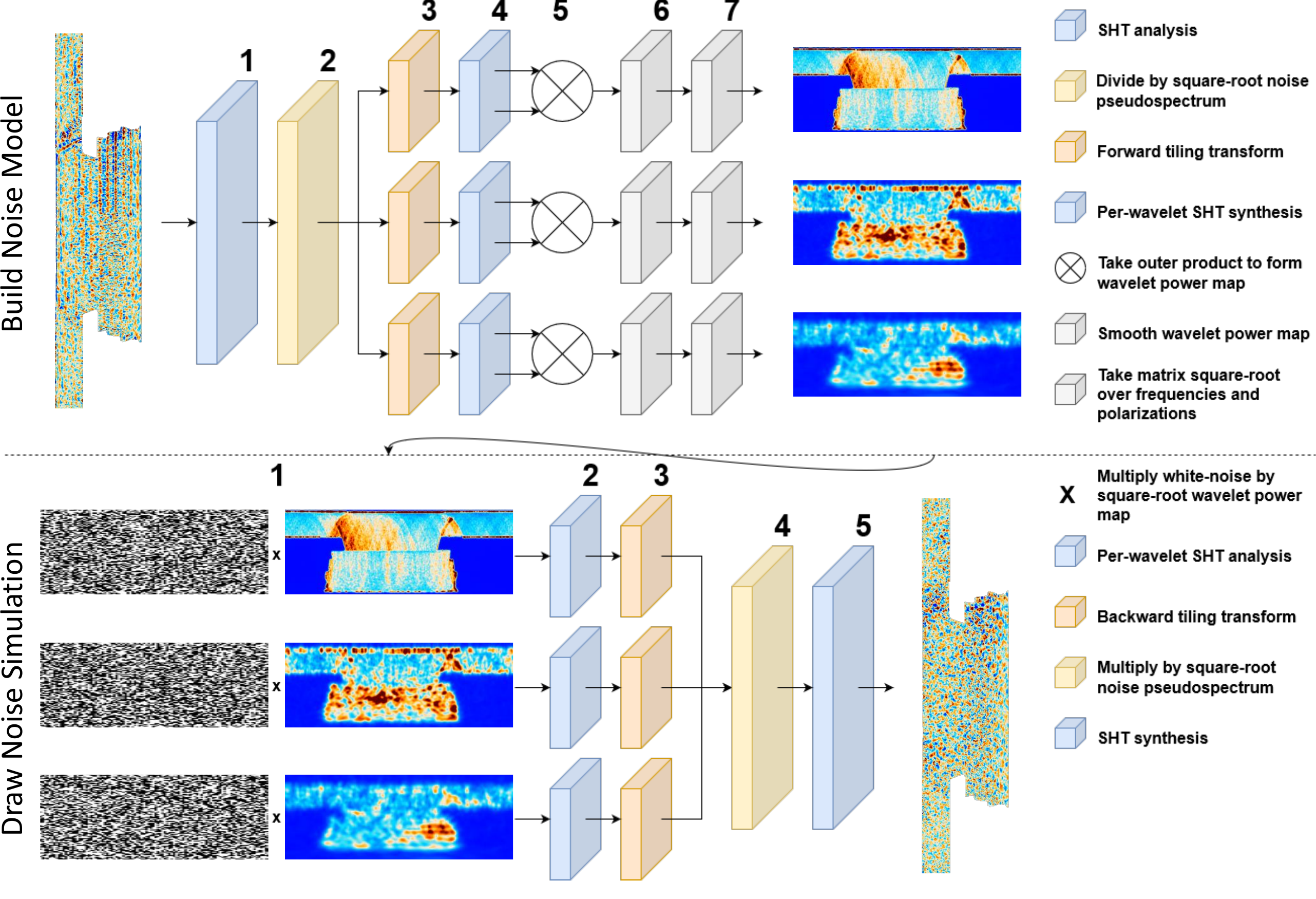}
        \caption{\textit{Top:} The sequence of filters and other operations involved in measuring the sparse covariance in the wavelet noise model. Large boxes denote filters or operations applied to the entire map, while small boxes denote filters or operations after tiling the harmonic plane. The orange filters represent the ``forward" wavelet kernels in Figure \ref{fig: wav_kernel} (``forward" and ``"backward" kernels are the same in this model). Each operation is labelled by the corresponding step number in the model estimation algorithm. The input difference map is PA5 f090, T ($a=0$), first split ($i=0$), and the outputs are square-root wavelet power maps, TT ($a=b=0$). We see the spatial variation in map depth also varies by wavelet scale, with high-$\ell$ to low-$\ell$ running top to bottom. \textit{Bottom:} The sequence of filters and other operations involved in drawing a simulation of the map-based noise in the wavelet model. The first column denotes independent, Gaussian white noise. The orange filters represent the ``backward" wavelet kernels in Figure \ref{fig: wav_kernel}, while all other operations revert those of the top row. The output is a wavelet noise simulation of PA5 f090, T, first split.}
    \label{fig: wav_schematic}
\end{figure*}
\end{enumerate}

These steps generate the wavelet model square-root covariance $\mathcal{N}_i^{1/2}$ for the purposes of this paper. A schematic depiction of all these steps is given in the top panel of Figure \ref{fig: wav_schematic}.

\bigskip
\bigskip
\subsubsection{Simulations and Covariance Matrix} \label{sec: noise_models_wav_formalism}
Just as the wavelet noise model estimation algorithm followed directly from \S\ref{sec: noise_models_tile_algorithm}, so too does its simulation algorithm carry over from \S\ref{sec: noise_models_tile_formalism}. That is, the procedure for generating a noise simulation from the wavelet model consists of first drawing a realization in wavelet map space, and then reverting the operations that transformed $\bm\nu_i$ into \mbf{u_i} in Equation \ref{eq: wav_uncorrelated_vector}.

\begin{enumerate}
    \item For each wavelet, draw Gaussian white noise in wavelet map space and take the inner product with the square-root covariance:
    \begin{align}
        \mathcal{N}_i^{\frac{1}{2}}\bm\eta_i
    \end{align}
    where $\bm\eta_i$ is real-valued white noise with the same shape as \mbf{u_i}. 
    
    \item Next, transform out of wavelet map space and back to wavelet harmonic space with a per-wavelet SHT:
    \begin{align}
       \mathbf Y^{\dag}\mathbf W \mathcal{N}_i^{\frac{1}{2}}\bm\eta_i.
    \end{align}
    That is, this vector contains the SHT of the $M=26$ independent wavelet noise realizations.  
    
    \item Just as in a tiled noise simulation, we now need to stitch the wavelet realizations back together, only this time the stitching occurs in harmonic space. The same considerations carry over from the tiled model as well: we should avoid naively pasting each wavelet next to one another, which would result in sharp discontinuities in the synthesized harmonic transform. Thus, we first multiply the wavelets by smoothly tapered profiles in harmonic space. For this purpose, we reuse the same profiles, or ``kernels," from the forward transform, as shown in Figure \ref{fig: wav_kernel}. As before, we also require that the sum over wavelets of the per-harmonic-mode noise variance is preserved; equivalently, that the wavelet kernels satisfy Equation \ref{eq: admissibility}. Conveniently, the wavelets of \citet{Wiaux2008, sd-wavelets} do so by construction. We then simply add up the 26 smoothly tapered wavelet realizations in harmonic space, resulting in a single harmonic transform. Together, this constitutes the ``backward tiling transform."
    
    The matrix form of the backward tiling transform is identical to Equation \ref{eq: backward_tile_transform}. Since we have reused the same wavelet profiles as in the forward transform ($\mathcal{T}_b = \mathcal{T}_f)$, we therefore have:
    \begin{align}
        \mathcal{T}_f^{\dag} \mathbf Y^{\dag}\mathbf W \mathcal{N}_i^{\frac{1}{2}}\bm\eta_i
    \end{align}
    where the vector now has the same shape as the harmonic transform of $\bm\nu_i$, and likewise lives in the original, ``untiled" harmonic space.
    
    \item We apply the inverse of the stationarity filter from the model estimation. That is, we reapply the square-root noise auto- and cross-spectra directly in harmonic space:
    \begin{align}
        \mathbf C_{\mathrm{est}}^{\frac{1}{2}} \mathcal{T}_f^{\dag} \mathbf Y^{\dag}\mathbf W \mathcal{N}_i^{\frac{1}{2}}\bm\eta_i.
    \end{align}
    
    \item Finally, transforming the entire simulation back into map space completes the noise simulation:
    \begin{align} \label{eq: wav_sim}
        \bm\xi_i = \mathbf Y \mathbf C_{\mathrm{est}}^{\frac{1}{2}} \mathcal{T}_f^{\dag} \mathbf Y^{\dag}\mathbf W \mathcal{N}_i^{\frac{1}{2}}\bm\eta_i.
    \end{align}
\end{enumerate}

These steps generate a wavelet model noise simulation, $\bm\xi_i$, for the purposes of this paper. A schematic depiction of these steps is given in the bottom panel of Figure \ref{fig: wav_schematic}. As before, it is easy to use Equation \ref{eq: wav_sim} to write the full map-based noise covariance matrix in the wavelet model:
\begin{align} \label{eq: wav_covmat}
    \mathbf N_i = \mathbf Y \mathbf C_{\mathrm{est}}^{\frac{1}{2}} \mathcal{T}_f^{\dag} \mathbf Y^{\dag}\mathbf W \mathcal{N}_i \mathbf W\mathbf Y \mathcal{T}_f \mathbf C_{\mathrm{est}}^{\frac{1}{2}} \mathbf Y^{\dag}
\end{align}
where we have used that $\mathbf C_{\mathrm{est}}^{1/2}$ and $\mathbf W$ are Hermitian.

This completes our discussion of the isotropic wavelet model. The direct parallels to the tiled model construction are apparent. By tiling harmonic space instead of map space, the wavelet model is well-adapted to capture the scale-dependence of the map depth; however, by construction, the isotropic wavelets are circular in map space (Figure \ref{fig: wav_kernel}) --- they are insensitive to any local noise stripiness.

\subsection{Directional Wavelet Noise Model} \label{sec: noise_models_fdw}
Our directional wavelet, or ``directional," noise model in principle captures all enumerated aspects of the ACT map-based noise, including the spatially-varying stripy patterns and scale-dependent map depth, simultaneously. Its key feature is a decomposition using a set of directional wavelets: each wavelet filters for a particular range of angular scales (like the ``isotropic" wavelet model), as well as a particular direction on the sky (unlike the ``isotropic" wavelet model). That is, by assuming the noise is stationary within a ``directional wavelet" with both a well-defined angular scale \textit{and} direction, the decomposition naturally leads to scale- and direction-dependent ``inverse-variance" maps. Moreover, because the main difference between the wavelet model and directional model is just the definition of their wavelet profiles, their model estimation and simulation algorithms are nearly identical. The directional model can therefore be viewed as a generalization of the wavelet model.

The directional model's only algorithmic change is that we build the wavelets in 2D Fourier space instead of harmonic space.\footnote{For harmonic space analogs, see e.g., the ``curvelets" of \citet{curvelet-sd-wavelets} and the ``steerable wavelets" of \citet{directional-sd-wavelets}. We note that the latter wavelets were used in the scale-discretised, directional wavelet ILC (SILC) analysis of \citet{Spin-SILC}. The total data covariance in SILC is the directional generalization of the NILC covariance, just as our directional wavelet model generalizes our isotropic model.} Doing so admits speed improvements, increased flexibility in wavelet design, and generally simpler implementation. One consequence of defining wavelets in Fourier space is that their map space kernels change physical size over declination --- that is, their size in pixels stays fixed as the physical sizes of the pixels themselves change. In principle, then, the wavelets could mix physical scales. In practice, however, the effect of this is limited: the map depth morphology undergoes a smooth transition from small to large scales (again, see Figure \ref{fig: ivar_by_scale}). We should also note that the Fourier space wavelets would encounter regularity issues at the poles; however, this too is not relevant for the ACT footprint, whose largest declination is $-63^\circ$. Fourier space aside, again we can heavily borrow from \S\ref{sec: noise_models_wav_algorithm} and \S\ref{sec: noise_models_wav_formalism} in what follows.

\bigskip
\bigskip
\subsubsection{Model Estimation} \label{sec: noise_models_fdw_algorithm}
Again, our goal is to construct a sparse representation of the noise covariance matrix. We give an abridged version of the model estimation algorithm from the wavelet model in \S\ref{sec: noise_models_wav_algorithm}, since the steps are highly similar. We take care, however, to explain notable differences in the implementation, especially in relation to the design of the directional wavelets.

\begin{enumerate}
    \item \label{enum: fdw_algorithm, item: flattening} We apply the same stationarity filter as in \S\ref{sec: noise_models_wav_algorithm}, Step \ref{enum: wav_algorithm, item: flattening}: flattening the noise power spectra as measured in $\bm\mu_{\mathrm{est}}$. Again, this helps remove the absolute dependence of the noise variance on angular scale, $\ell$. We still apply this filter to the maps in harmonic space:
    \begin{align}
        \mathbf C_{\mathrm{est}}^{-\frac{1}{2}} \mathbf Y^{\dag}\mathbf W \bm\nu_i.
    \end{align}
    
    \item In preparation for the wavelet decomposition, we need to transform the maps from harmonic space to 2D Fourier space. To do so, we first return to map space, and then take the DFT to Fourier space:
    \begin{align}
        \mathbf F \mathbf Y \mathbf C_{\mathrm{est}}^{-\frac{1}{2}} \mathbf Y^{\dag}\mathbf W \bm\nu_i.
    \end{align}
    We note that this Fourier space noise vector still contains $6N_{\mathrm{pix}}$ elements, since the DFT, \mbf{F}, is a square matrix.
    
    \item \label{enum: fdw_algorithm, item: tiling} As in \S\ref{sec: noise_models_wav_algorithm}, Step \ref{enum: wav_algorithm, item: tiling}, our goal is to isolate stationary regions in the transformed noise. Now, we do this by breaking the noise into a set of wavelets defined in Fourier space. We construct the wavelets to be separable radially and azimuthally in the Fourier plane. Modes with comparable radii, $k\equiv|\mathbf k|$, map to comparable angular scales, $\ell$, as well. For instance, had we constructed the isotropic wavelets from \S\ref{sec: noise_models_wav} in 2D Fourier space instead of harmonic space, they would have appeared as concentric annuli. Thus, for the radial part of the directional wavelets we reuse the scale-discrete wavelet profiles of \citet{Wiaux2008, sd-wavelets}, only now with a scaling parameter $\lambda=1.6$ (i.e., we have \textit{increased} the logarithmic wavelet spacing), $J_0=5$, and $J=19$, yielding 16 profiles (see the top panel of Figure \ref{fig: fdw_kernel_1d}). Compared to the isotropic model, the directional model has fewer and wider radial wavelet kernels. Also note, since we are in Fourier space, we consider the wavelets to be functions of $k$ instead of $\ell$.
    \begin{figure}
        \centering
        \includegraphics[width=\columnwidth]{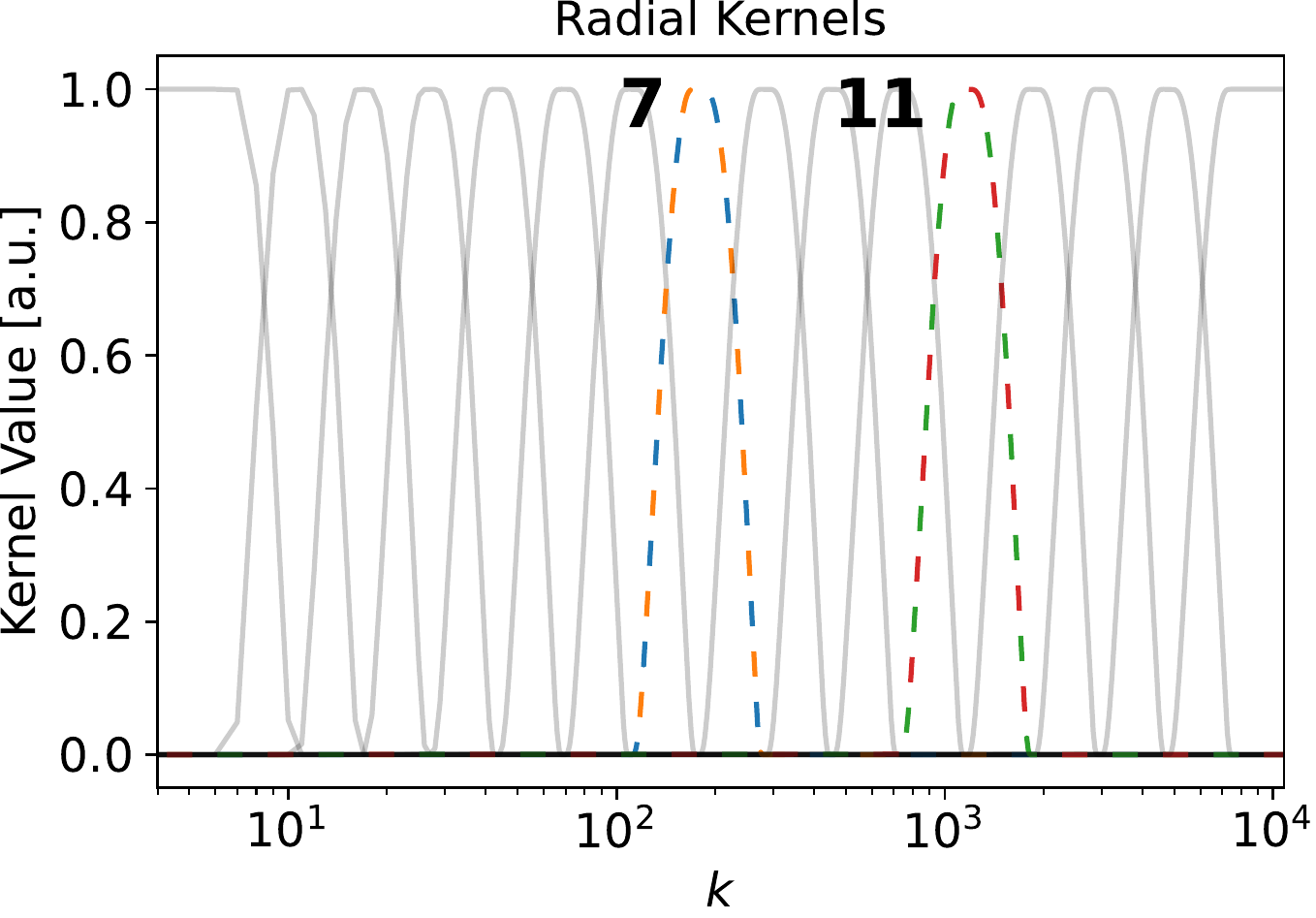}
        \includegraphics[width=\columnwidth]{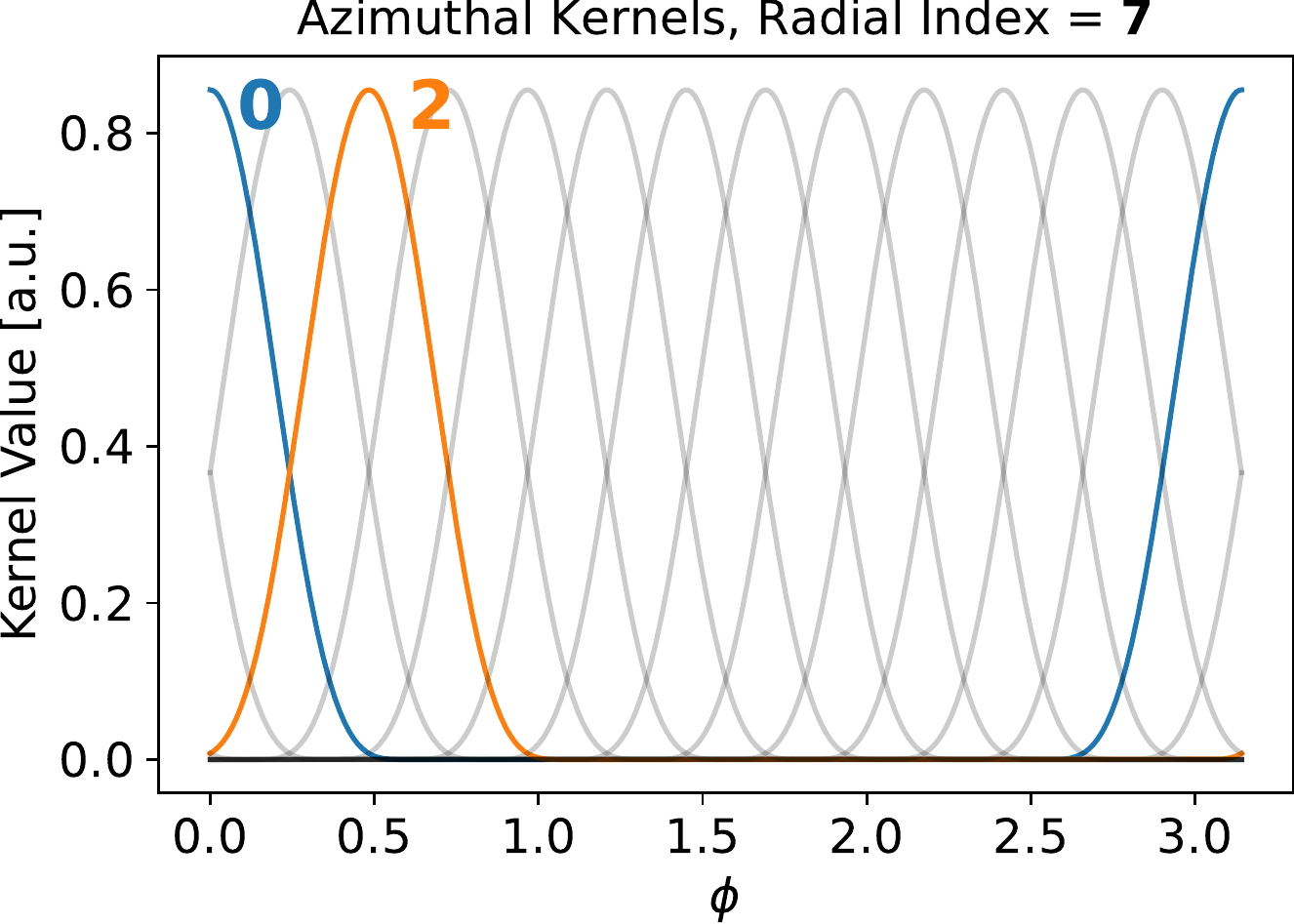}
        \includegraphics[width=\columnwidth]{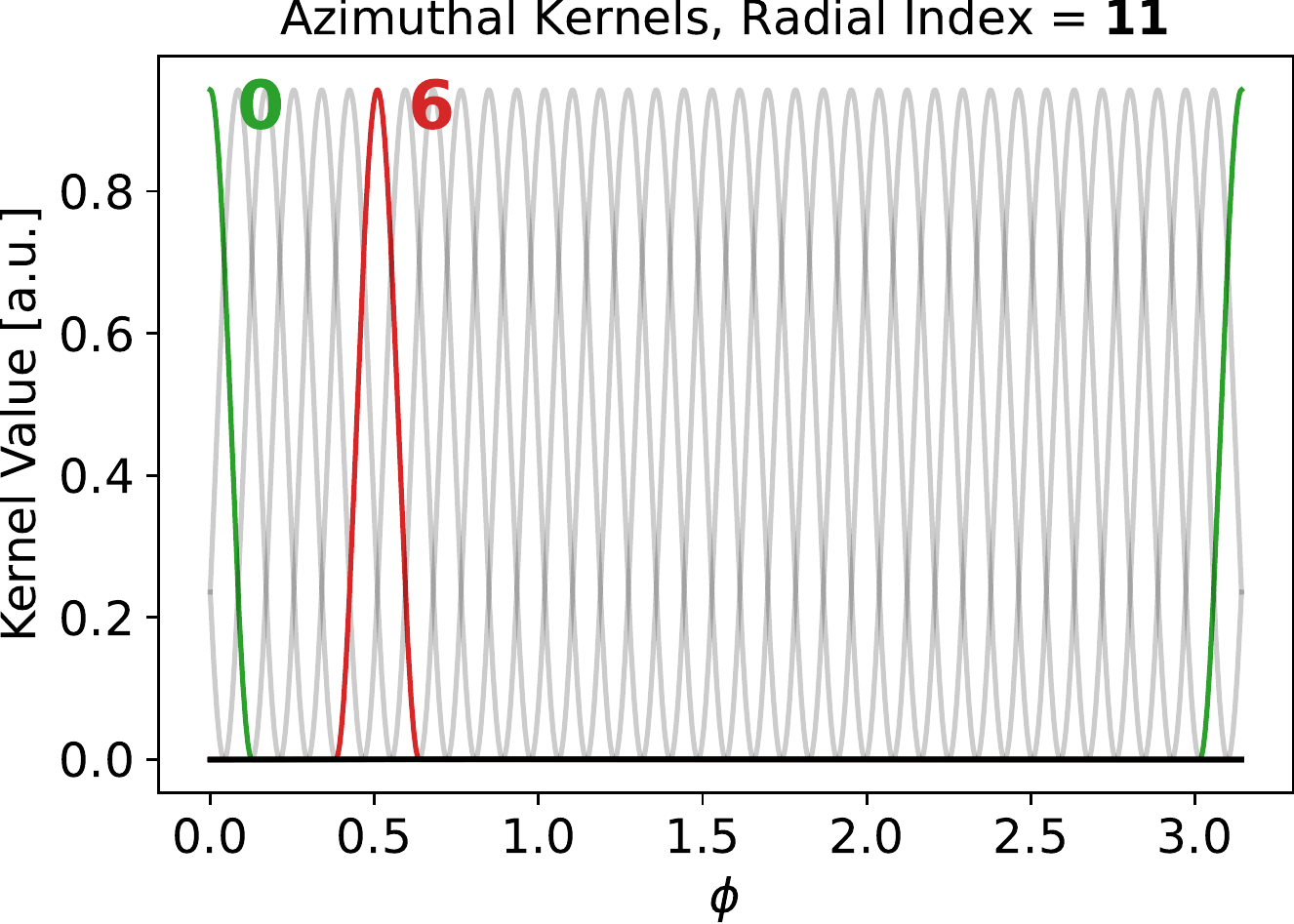}
        \caption{\textit{Top:} Radial slices of the Fourier directional wavelet kernels over angular scale. These profiles are again from the scale-discrete wavelet family \citep{Wiaux2008, sd-wavelets}. The blue and orange dashed wavelet around $k\sim100$, labeled by a ``radial" index of 7, corresponds to the azimuthal slices shown in the middle panel; the green and red dashed wavelet around $\ell\sim1,000$, labeled by a ``radial" index of 11, corresponds to the azimuthal slices shown in the bottom panel. \textit{Middle and Bottom:} We construct a wavelet set with increasing directional locality as a function of $k$. For example, 13 wavelets tile the azimuthal axis for the lower-$k$ wavelet, while 37 kernels tile the azimuthal axis at the higher-$k$ wavelet. A full 2D wavelet is given by the outer product of the radial and azimuthal slices. Numeric labels are for comparison with Figure \ref{fig: fdw_kernel_2d}.} 
        \label{fig: fdw_kernel_1d}
    \end{figure}
    By augmenting the wavelets with an azimuthal part, we create wavelets local not only in their angular scale, $k$, but in their direction on the sky as well. The azimuthal profile shapes (middle and bottom panels of Figure \ref{fig: fdw_kernel_1d}) are defined in Appendix \ref{apx: az_kernels}; their outer product with the radial profiles yields $M\sim\mathcal{O}(300)$ 2D Fourier space wavelets. The full 2D wavelets are shown in Figure \ref{fig: fdw_kernel_2d}. 
    \begin{figure*}
        \centering
        \includegraphics[width=\textwidth]{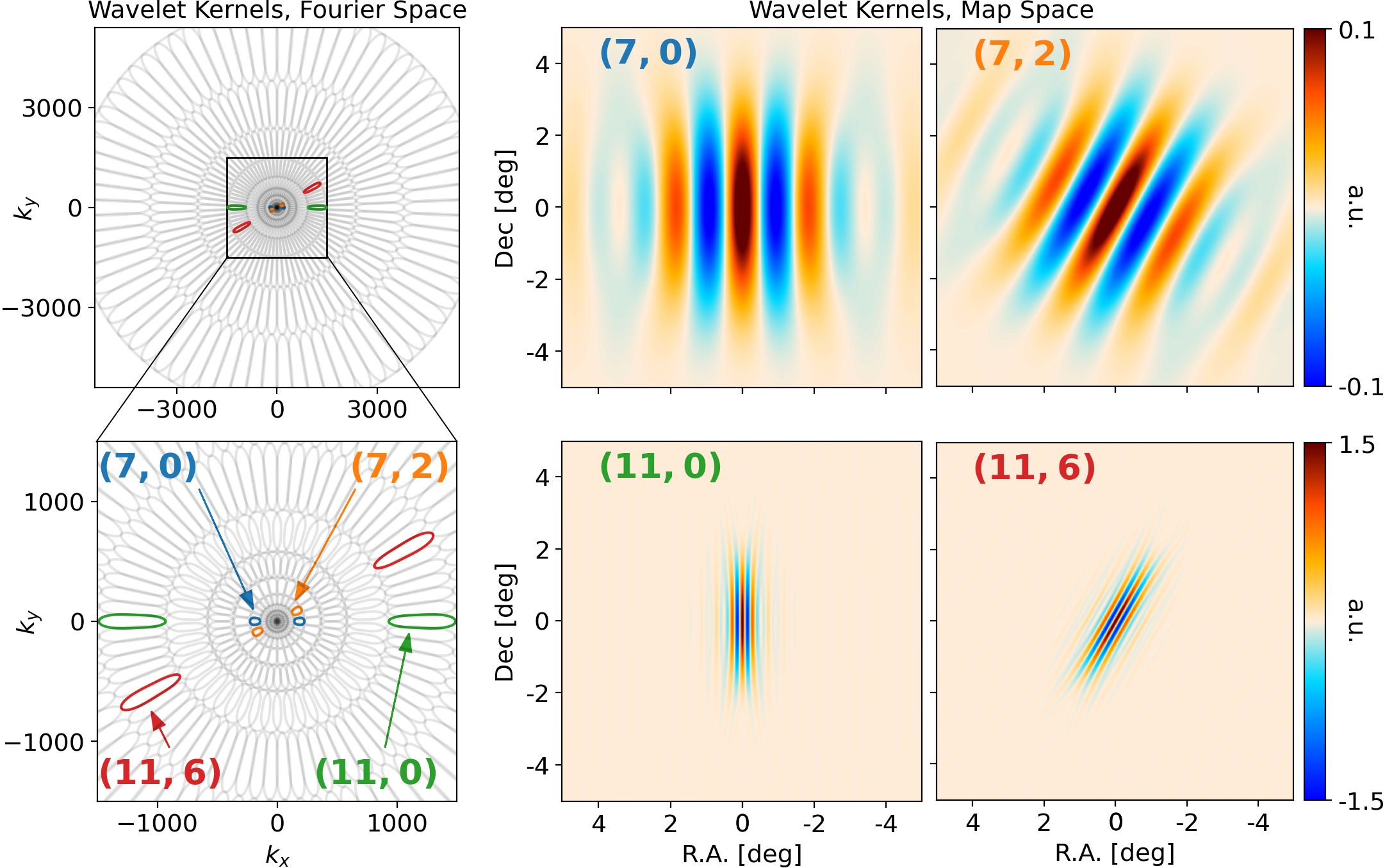}
        \caption{\textit{Left:} Contour-levels of the full 2D wavelet kernel set. The blue (7, 0), orange (7, 2), green (11, 0), and red (11, 6) 2D wavelets, formed by the product of the 1D profiles in Figure \ref{fig: fdw_kernel_1d}, are highlighted (numeric labels match those of Figure \ref{fig: fdw_kernel_1d}). The contours are at the half-maxima of the squared wavelets. We see the wavelets are attuned to ``radially oriented" features in 2D Fourier space. \textit{Right:} The map space wavelet kernels for the same blue, orange, green, and red wavelets. Wavelets (7, 0) and (7, 2) pick out low-frequency features by design and are spatially nonlocal. Wavelets (11, 0) and (11, 6) pick out high-frequency features by design and are spatially local overall. The opposite rotation sense in map space versus Fourier space results from the plotting convention (in map space, the R.A. decreases to the right).}
        \label{fig: fdw_kernel_2d}
    \end{figure*}
    As before, these 2D Fourier-space wavelet profiles smoothly overlap to prevent ringing and increase their map-space resolution.
    
    The forward tiling transform associated with the directional wavelets is similar to that of \S\ref{sec: noise_models_wav}. We note that the action of ``cropping" out each wavelet from the full 2D Fourier space is akin to the $\ell_{\mathrm{max}}$-bandlimiting from the isotropic wavelet model, except that we are restricted to a rectangular geometry (as in the tiled model). That is, rather than retaining all harmonic modes up to the wavelet $\ell_{\mathrm{max}}$, we retain all Fourier modes up the wavelet $|k_{y, \mathrm{max}}|$ and $|k_{x, \mathrm{max}}|$. Thus, each projection matrix \mbf{P_j} is a $N_{\mathrm{pix,wav}}\times N_{\mathrm{pix}}$ cropping matrix, and $\bm\tau_{j,f}$ is a $N_{\mathrm{pix,wav}}\times N_{\mathrm{pix,wav}}$ is a square diagonal matrix with the $j^{\mathrm{th}}$ wavelet kernel values along the diagonal. The forward tiling transform matrix form is again identical to Equation \ref{eq: forward_tile_transform}, yielding the following weighted, transformed, tiled noise vectors:
    \begin{align}
        \mathcal{T}_f \mathbf F \mathbf Y \mathbf C_{\mathrm{est}}^{-\frac{1}{2}} \mathbf Y^{\dag}\mathbf W \bm\nu_i.
    \end{align}
\end{enumerate}

In Figures \ref{fig: fdw_kernel_1d} and \ref{fig: fdw_kernel_2d}, we clearly observe what makes these wavelets directional: by isolating a compact range of azimuthal angles in Fourier space, their map space kernels are stripy with a well-defined orientation on the sky. Additionally, those figures demonstrate a critical design principle: the ``zero-sum" tradeoff in scale, azimuthal (i.e., directional), and spatial locality. In Figure \ref{fig: fdw_kernel_2d}, for instance, the left panel shows that at large scales (near the ``center" of Fourier space), each wavelet exhibits a high degree of scale locality and low azimuthal locality. Correspondingly, their map space kernels (in the top row of the right panel) are spatially extended overall, but are \textit{proportionally} compact along their orientation. At small scales (far from the ``center" of Fourier space), each wavelet exhibits successively less scale locality, and successively more azimuthal locality. Correspondingly, their map space kernels (in the bottom row of the right panel) are spatially compact overall, but are proportionally \textit{extended} along their orientation. This behavior is reminiscent of the ``parabolic scaling" in \citet{curvelet-sd-wavelets}, wherein the smaller-scale wavelets become progressively more extended. We discuss the engineering of this tradeoff in more detail in Appendix \ref{apx: az_kernels}.

\begin{enumerate}[resume]
    \item We again transform from the space in which the noise is assumed to be stationary (Fourier space) into the space where we take the noise to be diagonal (map space), this time with an inverse DFT. Thus, we assume each pixel in the following vector:
    \begin{align}
        \mathbf u_i = \mathbf F^{\dag} \mathcal{T}_f \mathbf F \mathbf Y \mathbf C_{\mathrm{est}}^{-\frac{1}{2}} \mathbf Y^{\dag}\mathbf W \bm\nu_i
    \end{align}
    is uncorrelated with all other pixels.
    
    \item We evaluate the same pixel-diagonal wavelet power maps, $\mathcal{N}(\mathbf x)_{i,j}^{ab}$, as in Equation \ref{eq: wav_ps}. The only change is due to the rectangular geometry, the normalization factor $f_{j, 2}$ is the same form as the tiled model (Equation \ref{eq: mask_correction}), with $N_{\mathrm{pix, wav}}$ replacing $N_{\mathrm{pix, tile}}$. The highly diagonal wavelet covariance matrix, $\mathcal{N}_i$, is likewise the same form as in the isotropic wavelet model.
    
    \item \label
    {enum: fdw_algorithm, item: smoothing} We apply the same smoothing method with the same definition of $f$ as in the isotropic wavelet model (\S\ref{sec: noise_models_wav_algorithm}, Step \ref{enum: wav_algorithm, item: smoothing}) to reduce sample variance.
    
    \item We save to disk the square-root matrices $\mathcal{N}_i^{1/2}$ and $\mathbf C_{\mathrm{est}}^{1/2}$. 
\end{enumerate}

These steps generate the wavelet model square-root covariance $\mathcal{N}_i^{1/2}$ for the purposes of this paper. We do not show the directional wavelet model maps, but they are qualitatively similar to Figure \ref{fig: wav_cov}. The top panel of Figure \ref{fig: wav_schematic}, showing the schematic model estimation steps, is likewise similar for the directional wavelets after substituting spherical harmonic transforms for Fourier transforms.

\subsubsection{Simulations and Covariance Matrix} \label{sec: noise_models_fdw_formalism}
Apart from the wavelet construction, the process of drawing a noise simulation from the directional model is, again, almost identical to the wavelet model. We give an abridged version of the simulation algorithm here.

\begin{enumerate}
    \item For each wavelet, draw Gaussian white noise in wavelet map space and take the inner product with the square-root covariance:
    \begin{align}
        \mathcal{N}_i^{\frac{1}{2}}\bm\eta_i.
    \end{align}

    \item Transform out of wavelet map space and back to wavelet Fourier space with a per-wavelet DFT:
    \begin{align}
       \mathbf F \mathcal{N}_i^{\frac{1}{2}}\bm\eta_i.
    \end{align}
    
    \item Apply a ``backward tiling transform" in Fourier space. That is, again, we need to stitch the wavelets back together to form a single noise Fourier transform. For the same reasons as the isotropic wavelet model, we choose to reuse the forward tiling transform --- $\mathcal{T}_b \equiv \mathcal{T}_f$ --- noting that we construct our wavelets (Appendix \ref{apx: az_kernels}) to satisfy Equation \ref{eq: admissibility}. We therefore have:
    \begin{align}
        \mathcal{T}_f^{\dag} \mathbf F \mathcal{N}_i^{\frac{1}{2}}\bm\eta_i
    \end{align}
    where the vector now has the same shape as the Fourier transform of $\bm\nu_i$, and likewise lives in the original, ``untiled" Fourier space.
    
    \item We reapply the noise auto- and cross-spectra in harmonic space. To do so, we first transform out of Fourier space and into map space, and then into harmonic space, and then multiply by the square-root noise pseudospectra:
    \begin{align}
        \mathbf C_{\mathrm{est}}^{\frac{1}{2}} \mathbf Y^{\dag} \mathbf W \mathbf F^{\dag} \mathcal{T}_f^{\dag} \mathbf F \mathcal{N}_i^{\frac{1}{2}}\bm\eta_i.
    \end{align}
    
    \item Transforming the entire simulation back into map space completes the noise simulation:
    \begin{align} \label{eq: fdw_sim}
        \bm\xi_i = \mathbf Y \mathbf C_{\mathrm{est}}^{\frac{1}{2}} \mathbf Y^{\dag}\mathbf W \mathbf F^{\dag} \mathcal{T}_f^{\dag} \mathbf F \mathcal{N}_i^{\frac{1}{2}}\bm\eta_i.
    \end{align}
\end{enumerate}

These steps generate a directional model noise simulation, $\bm\xi_i$, for the purposes of this paper. The schematic depiction in the bottom panel of Figure \ref{fig: wav_schematic} is highly representative of the simulation algorithm. As before, we use Equation \ref{eq: fdw_sim} to write the full map-based noise covariance matrix in the directional model:
\begin{align} \label{eq: fdw_covmat}
    \mathbf N_i = \mathbf Y \mathbf C_{\mathrm{est}}^{\frac{1}{2}} \mathbf Y^{\dag}\mathbf W \mathbf F^{\dag} \mathcal{T}_f^{\dag} \mathbf F \mathcal{N}_i \mathbf F^{\dag} \mathcal{T}_f \mathbf F \mathbf W\mathbf Y \mathbf C_{\mathrm{est}}^{\frac{1}{2}} \mathbf Y^{\dag}
\end{align}
where we have used that $\mathbf C_{\mathrm{est}}^{1/2}$ and $\mathbf W$ are Hermitian.

This completes our discussion of the directional wavelet model. By generalizing the wavelets to isolate a particular direction on the sky in addition to a range of angular scales, the directional model naturally captures the spatially-varying noise anisotropy and the scale-dependent map depth simultaneously.

\begin{figure*}
    \centering
    \includegraphics[width=\textwidth]{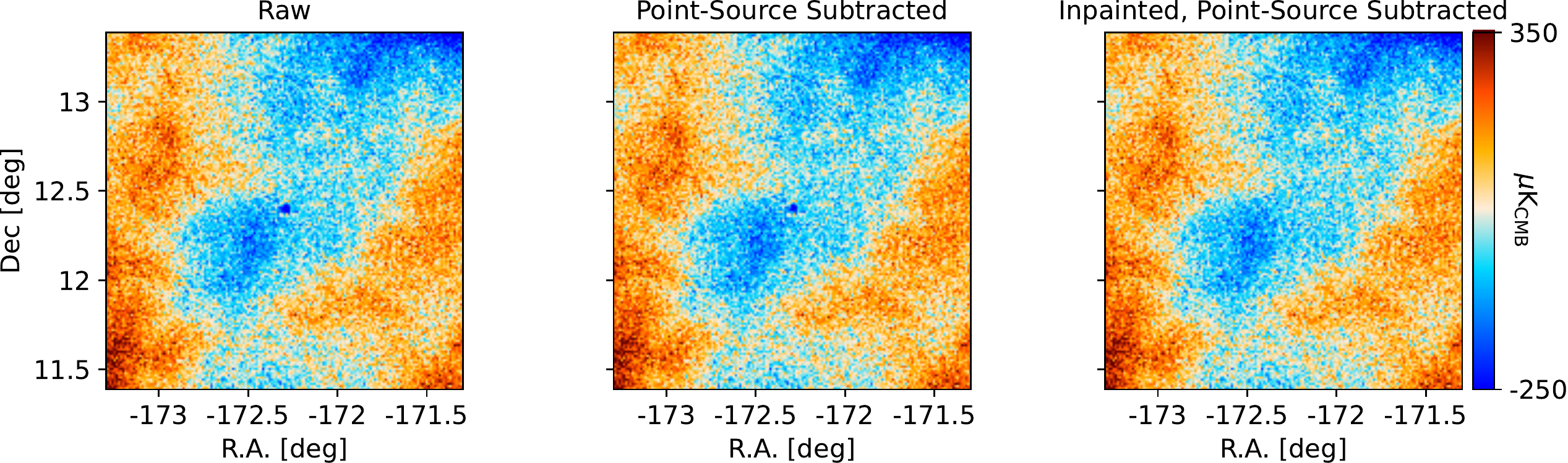}
    \caption{$2^\circ\times2^\circ$ region centered on M87 in the PA5 f090, first split, T difference map. \textit{Left:} Raw difference map (not point-source subtracted). \textit{Center:} Using the point-source subtracted maps described in the text partially mitigates the residual point source but does not eliminate it. \textit{Right:} The inpainted difference map.}
    \label{fig: inpainting}
\end{figure*}

\subsection{Preprocessing and Implementation} \label{sec: implementation}
Now that we have defined the core features and formalism of the models, we complete their discussion with some final implementation details. In Sections \S\ref{sec: implementation_ssi} and \S\ref{sec: implementation_d}, we describe the data preprocessing we perform on the noise maps, $\bm\nu_i$, prior to using them as inputs to the noise models. This is necessary to mitigate residual map signal that does not cancel in Equation \ref{eq: diff_def}. In Section \S\ref{sec: implementation_masks}, we define the ``pseudospectrum estimate mask," and other masks, used throughout the analysis. Sections \S\ref{sec: implementation_ssi}, \S\ref{sec: implementation_d}, and \S\ref{sec: implementation_masks} are common to all three models; section \S\ref{sec: implementation_smoothing} gives further details on the model smoothing step for the wavelet and directional models only.  

\subsubsection{Source Subtraction and Inpainting} \label{sec: implementation_ssi}
Each of the preceding models sacrifices some degree of map space locality in order to simultaneously resolve features in harmonic/Fourier space. The raw ACT data, however, contain spatially compact point sources; importantly, the brightest of these sources do not exactly cancel when we construct difference maps in Equation \ref{eq: diff_def}. This is likely due to a variety of underlying causes: many sources are themselves time-variable, or the telescope pointing varies slightly between splits, both of which would leak their signal into the map-based noise as defined in Equation \ref{eq: mapmaking_model}. They may be bright enough that small errors in gain calibration between splits can also induce significant signal-to-noise leakage in Equation \ref{eq: diff_def}. The upshot is that raw difference maps contain spatially compact point-source residuals, which, when left untreated, can bias our noise models. 

We mitigate these point-source residuals in two ways. First, we use ``point-source subtracted" maps \citep{A20, N20}. These are maps in which both time-domain and map-space models of point sources, either from a catalog or matched filter applied directly to the ACT DR6 data, have been fit and subtracted from the raw map products \citep{A20, N20}. Use of the point-source subtracted maps as inputs eliminates most of the difference map residuals, but some particularly bright sources remain (see left and center panels of Figure \ref{fig: inpainting}). 

\begin{figure*}
    \centering
    \includegraphics[width=\textwidth]{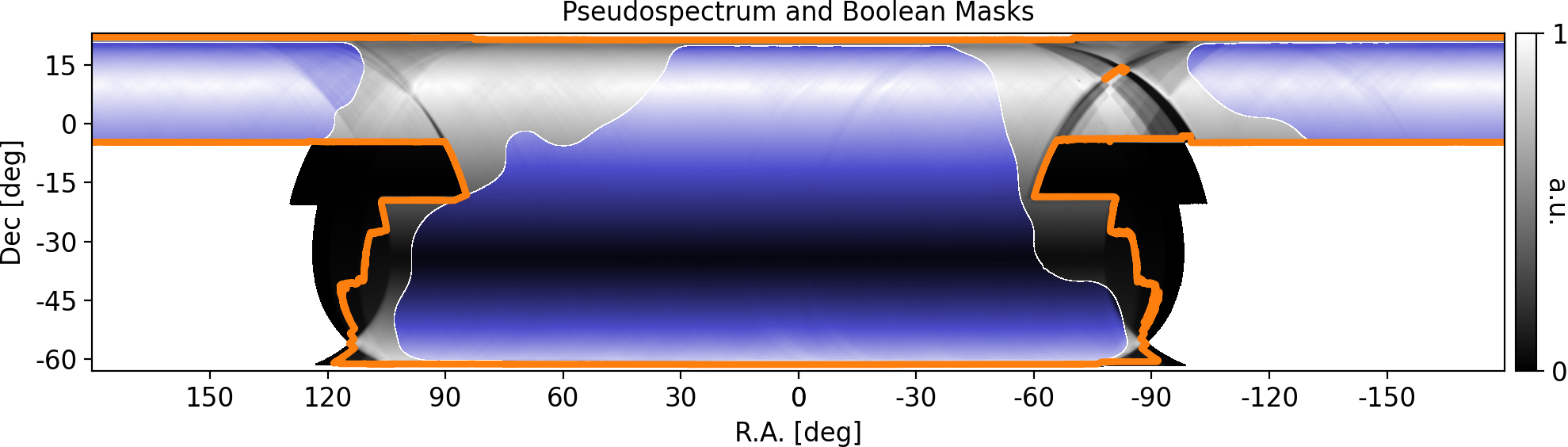}
    \caption{\textit{Grey:} The PA5 f090 cross-linking (square-root of the same data as Figure \ref{fig: ivar_xlink}. The square-root improves contrast in small values). \textit{Blue:} The region preserved by the apodized pseudospectrum estimate mask, $\bm\mu_{\mathrm{est}}$, overplotted on the cross-linking. \textit{Orange:} The outline of the boolean mask, $\bm\mu_{\mathrm{bool}}$, for PA5.}
    \label{fig: masks}
\end{figure*}

To further increase robustness against point-source residuals, we also inpaint the difference maps in small regions around the sources. For computational simplicity, we do not use a constrained-realization \citep[e.g.,][]{Bucher2012} inpainting routine. Rather, we develop a simple, sufficient procedure which leverages the overall smoothness of the noise and compactness of the point-source residuals. For a given point-source location in the difference map, we inpaint only the pixels within a six\,arcmin radius --- the ``inpainted region" --- as follows:
\begin{enumerate}
    \item Set the value of all pixels within the inpainted region equal to the mean value of pixels in the annulus between six and nine\,arcmin. This removes the point-source residual and replaces it with a reasonable constant value. While this is better than having the large spike from the point-source residual, it will leave a discontinuity in the noise properties at the border of the inpainted region. 
    \item We would like the large-scale noise features to be continuous across the inpainted region. To achieve this, we first copy the 120\,arcmin square region centered on the point source from the \textit{updated} difference map --- that is, the map resulting from the first step. We then smooth this copied cutout with a six\,arcmin FWHM Gaussian kernel. Finally, we adopt the values of this smoothed cutout only in the inpainted region of the difference map. This adds some large-scale modes from outside the inpainted region on top of the constant value from the first step.
    \item We would also like the small-scale noise features to be continuous across the inpainted region. For this, we simply add a white-noise realization drawn from the per-pixel difference map variance ($\mathbf w_i^{-1}$) to the difference map inpainted region. 
\end{enumerate}
These three steps ensure that in a small region around inpainted point-source residuals, the noise features are reasonably continuous. The results of this procedure for a sample point source are shown in the right panel of Figure \ref{fig: inpainting}. We inpaint around all sources in the compact source catalog described in \citet{Qu23}, which contains 1,779 point sources. While only a fraction of this catalog is likely bright enough to cause a noticeable residual, our inpainting procedure is benign and we only inpaint 0.12\% (0.28\%) of the sky (ACT footprint).

\subsubsection{Map Downgrading} \label{sec: implementation_d}

We typically complete our preprocessing of the difference maps by downgrading their resolution by a factor of two or four, which also reduces their bandlimit proportionately. There is no fundamental reason for this step; rather, we do so only to save disk space and computation time of noise realization ensembles. We resample in Fourier space to perform the downgrading. This has several benefits over pixel-block-averaging. First, the pixel centers in the downgraded pixelization align with pixel centers in the raw pixelization. This is necessary for ensuring SHTs of the downgraded maps are lossless (i.e., the downgraded maps still follow the Clenshaw-Curtis quadrature rule, see \citet{libsharp}). Second, we do not introduce any additional pixel window or aliasing beyond what is already present in the full-resolution maps. Lastly, unlike harmonic space resampling, Fourier space resampling preserves the entire Fourier plane, which is necessary for the directional wavelet decomposition to work. In principle, bandlimiting the difference maps could introduce ringing around bright features, but in practice these features are already suppressed when we take map differences in Equation \ref{eq: diff_def} and inpaint the difference maps.

\subsubsection{Pseudospectrum and Boolean Masks} \label{sec: implementation_masks}

Our models also rely on two input masks. The first is an apodized mask that leaves a restricted portion of the sky used to measure the noise pseudospectrum and construct the isotropic filter for the noise models. The second is a boolean mask defining the larger area we model and simulate.

This first mask is only used to measure the noise pseudospectra which enter the isotropic filter in \S\ref{sec: noise_models_tile_algorithm} Step \ref{enum: tile_algorithm, item: decoupling_filter}, \S\ref{sec: noise_models_wav_algorithm} Step \ref{enum: wav_algorithm, item: flattening}, and \S\ref{sec: noise_models_fdw_algorithm} Step \ref{enum: fdw_algorithm, item: flattening}. The goal of this mask is produce ``representative" pseudospectra: downweighting or excluding the noisiest data which would otherwise dominate the spectra, while retaining a reasonable fraction of the ACT footprint. Thus we use a slightly modified version of the nominal ACT DR6 CMB lensing analysis sky mask \citep{Qu23}. This is the intersection of two other masks: a root-mean-square threshold cut on the ACT DR6 f090 and f150 coadd maps of 70\,$\mu$K-arcmin, and a \textit{Planck} 60\% Galactic mask \citep{Planck-overview:2018}. We modify it by then cutting pixels not observed by all splits of all detector arrays. This ensures we can use this one mask version for all of our noise models. Finally, we apodize the mask by three degrees with a cosine profile. The outline of the resulting pseudospectrum mask, $\bm\mu_{\mathrm{est}}$, is shown as the blue region in Figure \ref{fig: masks}, overlayed on top of a cross-linking map. To maximize noise model accuracy one might be tempted to select the same mask here as is used in any downstream analysis, but the ensuing proliferation of noise models could be cumbersome from a data management perspective. Instead, we note that our noise models are reasonably robust to $\mathcal{O}(1)$ variations about the noise spectra as measured in our pseudospectrum mask (see Appendix \ref{apx: Nl_benefits}).\footnote{Furthermore, matching the pseudospectrum estimate mask to some desired analysis region is not guaranteed to achieve optimal agreement between the noise power spectra of simulations and data in that region.}

\begin{figure*}
    \centering
    \includegraphics[width=\textwidth]{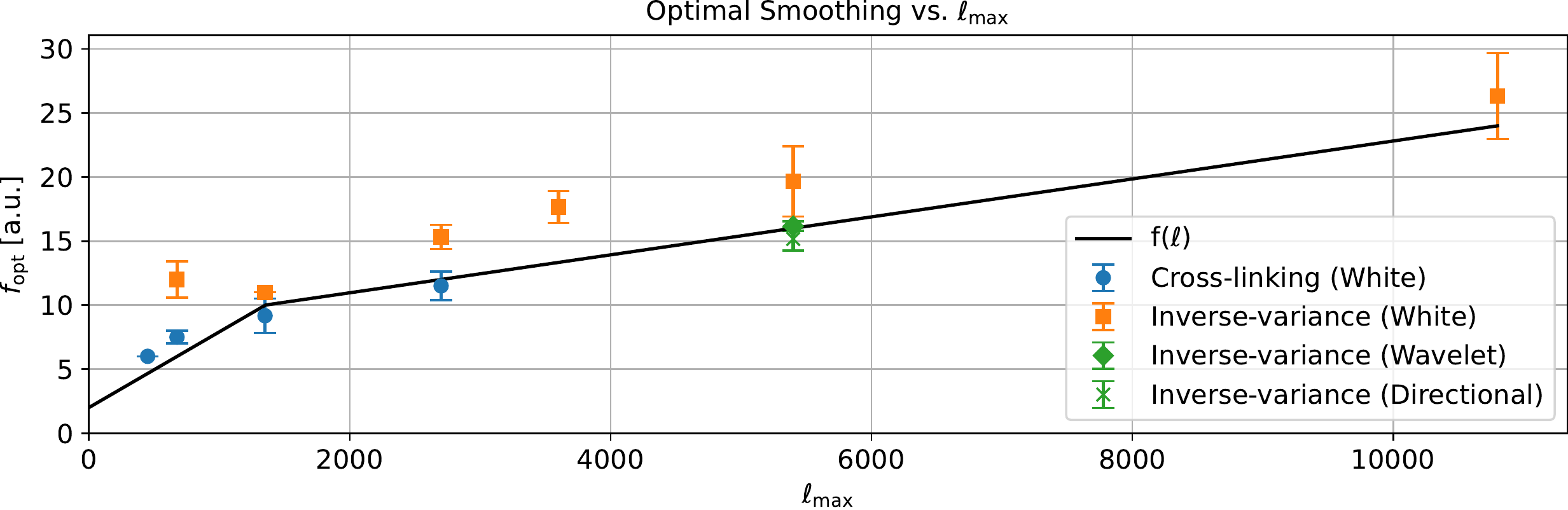}
    \caption{Optimal smoothing factors, $f_{\mathrm{opt}}$, as a function of bandlimit, using the procedure in the text (also see Appendix \ref{apx: smoothing}). The blue circles are obtained by comparing a simple white-noise model to a cross-linking plus inverse-variance-like noise model. The orange circles correspond to a white-noise model following the inverse-variance only. The green diamond (cross) are obtained by comparing the wavelet (directional) noise model to an inverse-variance-like noise covariance. Each data point shows the mean and $1\sigma$ range over the six ACT DR6 arrays/frequencies (first split of each). The solid black line shows our chosen form of $f(\ell)$ for use in the actual model estimation of \S\ref{sec: noise_models_wav_algorithm} and \S\ref{sec: noise_models_fdw_algorithm}.}
    \label{fig: optimal_smoothing}
\end{figure*}

The boolean mask is used to define which data enter our noise models and are simulated. Ideally, we would include all observed pixels in a given difference map, such that we model and simulate all the available data. In practice, however, we found it improved performance to prohibit pixels with the lowest cross-linking from entering the noise models. We detail the motivation and construction of this boolean mask, $\bm\mu_{\mathrm{bool}}$, in Appendix \ref{apx: boolean_mask}. Once defined, we apply it directly to the difference maps (via $\bm\mu_{\mathrm{bool}}*\bm\nu_i$) before the model estimation algorithms, and directly to the simulations (via $\bm\mu_{\mathrm{bool}}*\bm\xi_i$) after the simulation algorithms. The outline of this boolean mask is shown as the orange curve in Figure \ref{fig: masks}. We see it excludes the regions at the edge of the ACT footprint with practically no cross-linking. Importantly, the boolean mask does not cut any data used in the main DR6 cosmology analyses (the CMB power spectrum and CMB lensing).

\subsubsection{Model Smoothing} \label{sec: implementation_smoothing}

The preceding noise models are \textit{themselves} noisy approximations of an underlying, true covariance matrix. This statistical ``model scatter" is made especially acute by the fact that we build models using a single realization of the input data --- a model for each split map. Absent any suppression of the statistical scatter in the estimated covariance matrices, the models will be biased by over-fitting to the data: random fluctuations in the estimated covariance matrix will be recorded as part of the true covariance matrix. Of course, the preceding algorithms do address this issue by smoothing the tiled 2D Fourier power spectra (\S\ref{sec: noise_models_tile_algorithm}, Step \ref{enum: tile_algorithm, item: smoothing}) or wavelet power maps (\S\ref{sec: noise_models_wav_algorithm}, Step \ref{enum: wav_algorithm, item: smoothing} and \S\ref{sec: noise_models_fdw_algorithm}, Step \ref{enum: fdw_algorithm, item: smoothing}). These smoothing steps are essentially averages over adjacent modes or pixels in each tiled or wavelet covariance (we discuss whether also averaging noise models over data splits is a viable alternative in \S\ref{sec: disc_conc}). While necessary to avoid over-fitting to the data, however, over-smoothing the noise models may remove structure in the true noise covariance. For example, in the wavelet power maps of Figure \ref{fig: wav_cov}, a given ``blob" of high variance might be a statistical fluctuation, such that we \textit{would} want to smooth it away, or might be real feature in the true noise covariance, such that we \textit{would not} want to smooth it away. Here, we attempt to balance these competing potential biases.

Starting with the wavelet models, we need to specify the function $f(\ell)$ used in \S\ref{sec: noise_models_wav_algorithm}, Step \ref{enum: wav_algorithm, item: smoothing} and \S\ref{sec: noise_models_fdw_algorithm}, Step \ref{enum: fdw_algorithm, item: smoothing}. Recall, in those smoothing steps we take each wavelet power map, corresponding to a wavelet with a given $\ell_{\mathrm{max}}$, and smooth it with a Gaussian kernel whose FWHM in map pixels is given by $f(\ell_{\mathrm{max}})$. Thus, we want $f(\ell)$ to lie in the ``Goldilocks zone" of not too little, but not too much, smoothing.

Fortunately, we have \textit{a priori} insight into what the noise power maps should resemble: at high-$\ell$, the mapmaker inverse-variance maps, and at low-$\ell$, the mapmaker cross-linking maps (recall Figures \ref{fig: ivar_xlink} and \ref{fig: ivar_by_scale}).\footnote{To be precise, these two mapmaker products trace the spatial variation of the map depth. It is their inverse that traces the map noise power.} We use this knowledge to estimate the optimal smoothing width, $f_{\mathrm{opt}}$, at a given $\ell_{\mathrm{max}}$, according to the following procedure:
\begin{enumerate}
    \item Take the raw mapmaker inverse-variance, initially with a Nyquist bandlimit $\ell_{\mathrm{max}}=21,600$, to be the ``true" noise covariance. Draw a white-noise realization from this covariance to represent a single realization of the map-based noise in the data. 
    \item With varying amounts of smoothing, construct white-noise models using this realization of the map-based noise as the input. This is done by simply squaring the observed noise in each pixel, and smoothing with a Gaussian kernel whose FWHM is $f$ pixels. 
    \item Compare the agreement between the ``estimated" white-noise covariance in the models with the original, ``true" noise covariance. The optimal smoothing, $f_{\mathrm{opt}}$, corresponds to the model that maximizes this agreement. 
    \item Repeat these steps by varying what is taken to be the true noise covariance, by downgrading the inverse-variance map to a lower $\ell_{\mathrm{max}}$ and/or including the cross-linking map. Find $f_{\mathrm{opt}}$ in each case.
\end{enumerate}

The fourth step above is important to ensure the fiducial input covariance is as realistic as possible. First, the wavelet power maps probe noise structure at different resolutions, as set by their $\ell_{\mathrm{max}}$. Consider the limiting case of a white-noise covariance defined by the mapmaker inverse-variance map. In this case, even though there is a ``true" noise covariance at full resolution, such a covariance has different spatial structure when downgraded to lower resolutions. Thus, the optimal smoothing level at those downgraded resolutions may not be the same as at full resolution. Second, we observed in Figure \ref{fig: ivar_by_scale} that the spatial noise structure resembles the mapmaker cross-linking at large scales. Thus, by including the (downgraded) cross-linking in our input covariance at lower values of $\ell_{\mathrm{max}}$, the resulting optimal smoothing level is appropriate for models of the actual data. 

This procedure balances the tension between preserving the true covariance structure and suppressing random fluctuations in realistic wavelet power maps. We show the results in Figure \ref{fig: optimal_smoothing}.\footnote{We give the detailed version of this approach and define a quantitative measure of ``model-to-truth" agreement in Appendix \ref{apx: smoothing}.} In general, the optimal smoothing factor, $f_{\mathrm{opt}}$, increases with $\ell_{\mathrm{max}}$. At large scales --- that is, scales dominated by correlated noise --- the cross-linking governs the spatial noise structure, and $f_{\mathrm{opt}} \sim 5-10$. At small scales, where the inverse-variance dominates the spatial noise structure, $f_{\mathrm{opt}}$ increases toward $\sim 20-25$. The open green diamond and cross in Figure \ref{fig: optimal_smoothing} correct for the fact that our wavelet noise models are not simple white-noise models (also explained further in Appendix \ref{apx: smoothing}). They indicate the optimal smoothing for the wavelet noise models is slightly less than implied by our simple white-noise procedure.

In light of these results, we opt for a simple ad-hoc form of $f(\ell_{\mathrm{max}})$ in all arrays and frequencies, using the model shown in Figure \ref{fig: optimal_smoothing}.\footnote{The lines from (0, 2) to (1,350, 10), and then through (5,400, 16) in ($\ell, f(\ell)$) space.} This respects all arrays' preference for low smoothing factors at low $\ell$, with a shallow increase toward larger smoothing factors at large $\ell$. This curve also remains $\sim 10-30\%$ below the white-noise results for compatibility with our wavelet models. We also see no significant effect in the results of \S\ref{sec: eval_heuristics} when varying $f(\ell_{\mathrm{max}})$ by $\sim 5$ at all but the largest scales. Therefore, in setting the form of $f(\ell_{\mathrm{max}})$ we are primarily concerned with capturing the overall trend than specific, minor variations between arrays. We plan to improve our smoothing procedure in future work.

For the tiled model, there is no mapmaker product that gives an \textit{a priori} understanding of what the 2D Fourier power spectra should look like. Thus, as we discussed in \S\ref{sec: noise_models_tile_algorithm}, Step \ref{enum: tile_algorithm, item: smoothing}, we simply adopt the same smoothing scale as was used in \citet{Mv20} and \citet{N20}. 

\begin{figure*}
    \centering
    \includegraphics[width=\textwidth]{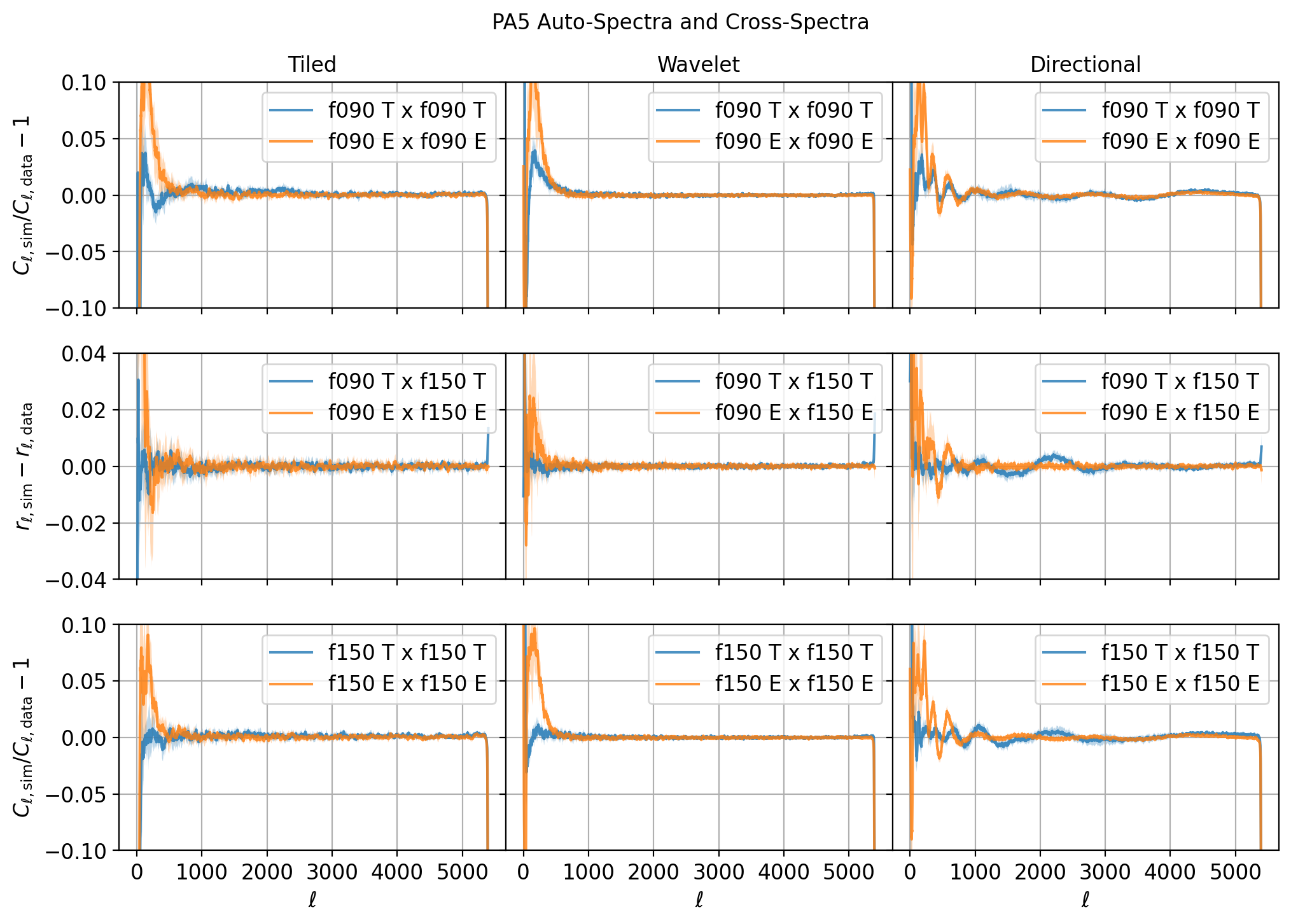}
    \caption{Comparisons of PA5 noise auto- and cross-spectra between the simulations and the data. The solid lines (shaded bands) are the means ($1\sigma$ range) over the eight splits. The top row shows f090 power spectrum ratios. The bottom row shows f150 power spectrum ratios. The middle row shows correlation spectra differences: the correlation is calculated by considering the ratio of the f090 $\times$ f150 cross-spectra to the geometric mean of the auto-spectra (Equation \ref{eq: correlation_spectra}). For all three rows, the nominal result is a value of 0. All spectra are measured in the same apodized pseudospectrum mask from \S\ref{sec: implementation_masks}. Recall the models are bandlimited to $\ell_{\mathrm{max}}=5,400$. Temperature spectra perform well at all scales, exhibiting no more than few-percent deviations at low-$\ell$. As discussed in \S\ref{sec: disc_conc}
    , polarization spectra appear robust for $\ell\gtrsim300$, but exhibit $\sim$10\% deviations for $\ell\lesssim300$.}
    \label{fig: results_ps_ratios}
\end{figure*}

\section{Model Evaluation and Results} \label{sec: eval}
Having constructed the tiled, isotropic wavelet, and directional wavelet noise models, we examine their ability to reproduce ACT map-based noise statistics. As discussed in the previous section, we do so by averaging over large ensembles of noise simulations. A data difference map and example noise simulations for each model are shown in Appendix \ref{apx: example_sims}. For ease of computation, we have chosen to show results from models bandlimited to $\ell_{\mathrm{max}}=5,400$, and averages of 30 simulations.

Because of the density of the data and model noise covariance in map space, a direct comparison of those matrices is not possible. As an informative substitute, we instead examine the noise properties of \S\ref{sec: noise_properties} as captured in simulations. Thus, in \S\ref{sec: eval_heuristics}, we compare simulations to the data by: (a) isotropic noise auto- and cross-spectra (large-scale noise correlations), (b) 2D Fourier power spectra (spatially-varying noise anisotropy), and (c) effective ``inverse-variance'' at different angular scales (scale-dependent map depth). Each comparison is analogous to viewing the full map-based noise covariance matrix along a different ``slice," or through a different ``lens:" it visualizes a particular salient aspect of the matrix structure, since doing so for the entire matrix at once is difficult. In \S\ref{sec: ps_covmat}, we apply the noise simulations to a realistic analysis pipeline. In particular, we examine the covariance matrix of the ACT DR6 power spectrum bandpowers --- that is, the determination of the errors on the power spectrum. We compare a common analytic prescription for the covariance matrix and a Monte Carlo estimate from simulations alone. 

\begin{figure*}
    \centering
    \includegraphics[width=\textwidth]{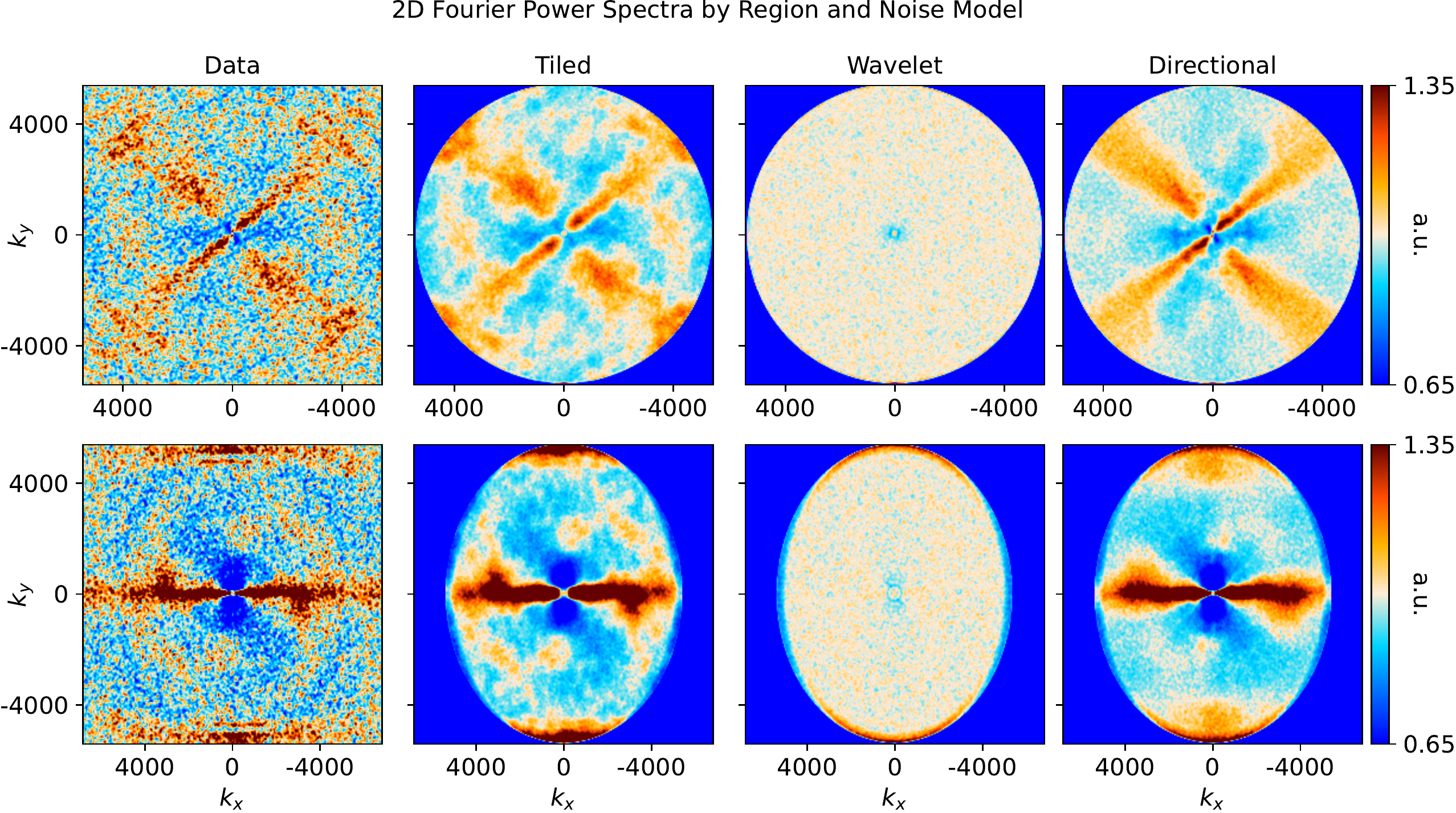}
    \caption{The same 2D Fourier power spectra treatment as Figure \ref{fig: 2d_ps_by_region}, for both data and noise simulations. The top (bottom) row corresponds to the same regions as the left (right) panel of Figure \ref{fig: 2d_ps_by_region}. The first/second/third/fourth columns correspond to the data/tiled/wavelet/directional models. The model bandlimit ($\ell_{\mathrm{max}}=5,400$) causes the oval power loss in the simulations. The bandlimit is nearly circular in 2D Fourier space in the top row, since this is from a region near the equator. The bandlimit becomes ``squished" in the x-direction in the bottom row, since this region is at a higher declination. That is, the pixels in the x-direction are physically closer together at these latitudes, and so the 2D Fourier space samples higher frequencies. Note we account for this in our definition of the 2D Fourier coordinates. The tiled and directional models capture the main anisotropies observed in the data, with the tiled model resolving more fine features. The wavelet model has no ability to measure anisotropies by construction.}
    \label{fig: results_2d_ps_by_region}
\end{figure*}

\subsection{Noise Covariance Model-Data Comparisons} \label{sec: eval_heuristics}
We examine the noise properties from \S\ref{sec: noise_properties} as captured by our simulations and compare to the data. Thus, we directly probe the extent to which each model achieves its design goals. As discussed, each property visualizes a slice of the full covariance matrix, either constructed from the \textit{single} available realization of the data, or averaged over an \textit{ensemble} of simulations. If we have built a good noise model, the covariance slice measured from the data should be consistent with a single noisy realization of the ensemble average covariance slice. Said another way, the ensemble average slice is each model's postulated long-run average of the data slice, if it were possible to rerun the experiment many times. Thus, for a successful model, we expect the data slice to ``scatter" around the smoother long-run average provided by the simulations, as is the case in a typical goodness-of-fit analysis. Indeed, in future work, we aim to add a full $\chi^2$ consistency test of data and simulations, which would require application of the full inverse noise covariance matrix, $\mathbf{N}_i^{-1}$, to map vectors (see \S\ref{sec: disc_conc} for a discussion of the inverse covariance matrix). At any rate, we should keep this signature of good model performance in mind in what follows. 

\subsubsection{Large-Scale Noise Correlations and Correlated Frequencies}
As discussed, extended atmospheric signals are what produce most of the non-trivial map-based noise properties, including pixel-pixel correlations, and correlations between frequencies on the same detector array. Figure \ref{fig: results_ps_ratios} demonstrates the noise models' performance at reconstructing the noise auto-spectra and cross-spectra between frequencies. Generally, all models perform similarly well in reconstructing the noise TT, EE auto- and TE cross-spectra. Their mean values are within 1\% of nominal (zero for the correlation spectra differences and the auto-spectra ratios, since in Figure \ref{fig: results_ps_ratios} we have subtracted 1 from the ratios). Departures from nominal are mainly confined to large scales ($\ell\lesssim 300$), especially in polarization. One possible explanation for this behavior is the pseudospectrum filter, which is applied to all of our models. This filter places a lower bound on the map space locality of features in the simulated noise. As we show in \S\ref{sec: disc_conc}, this lower bound happens to be about three times higher in polarization than temperature. In other words, it is more difficult to contain locally noisier regions of the sky in polarization, and we would interpret Figure \ref{fig: results_ps_ratios} as showing some noisy, large-scale polarized regions bleeding into the analysis mask. We note, however, that ACT DR4 power spectra were limited to $\ell_{\mathrm{min}}=350$ in polarization and $\ell_{\mathrm{min}}=600$ in temperature \citep{C20}. Thus, we take the results of Figure \ref{fig: results_ps_ratios} as a marker of good noise model performance across all three model types.

\subsubsection{Spatially-Varying Noise Anisotropy}
While the tiled and directional models are designed to capture the location-dependent stripiness in the ACT noise, the wavelet model is not. To evaluate their performance in this regard, we simply repeat the analysis in Figure \ref{fig: 2d_ps_by_region}, substituting an average over simulations for the data. The success, or lack thereof, in each model is visually apparent in Figure \ref{fig: results_2d_ps_by_region}. Both the tiled and directional models succeed at reconstructing the primary noise stripiness in each region: the ``x-like" pattern near the equator, and the horizontal ``bar" in the poorly cross-linked region. The tiled model appears to capture some secondary features in the data noise that are not oriented radially; recall the directional kernels (Figure \ref{fig: fdw_kernel_2d}) are primarily designed around radial features. The wavelet model stands out: by construction, it has no ability to measure local noise anisotropy. We also observe that the high frequency ``covariance scatter" in the \textit{data} 2D power spectrum is substantially suppressed in the \textit{simulation} spectra. This is expected: as discussed, in the simulation spectra we visualize the true, underlying covariance structure postulated by each model, of which the data spectrum is a single realization. This also inspires confidence in the model smoothing from \S\ref{sec: implementation_smoothing}: the models are not obviously over-fit to the random fluctuations in the data covariance. In sum, the tiled model and directional models succeed at capturing the primary features of the spatially-varying noise anisotropy in the data, while the wavelet model captures no such features.

\subsubsection{Scale-Dependent Map Depth}
\begin{figure*}
    \centering
    \includegraphics[width=\textwidth]{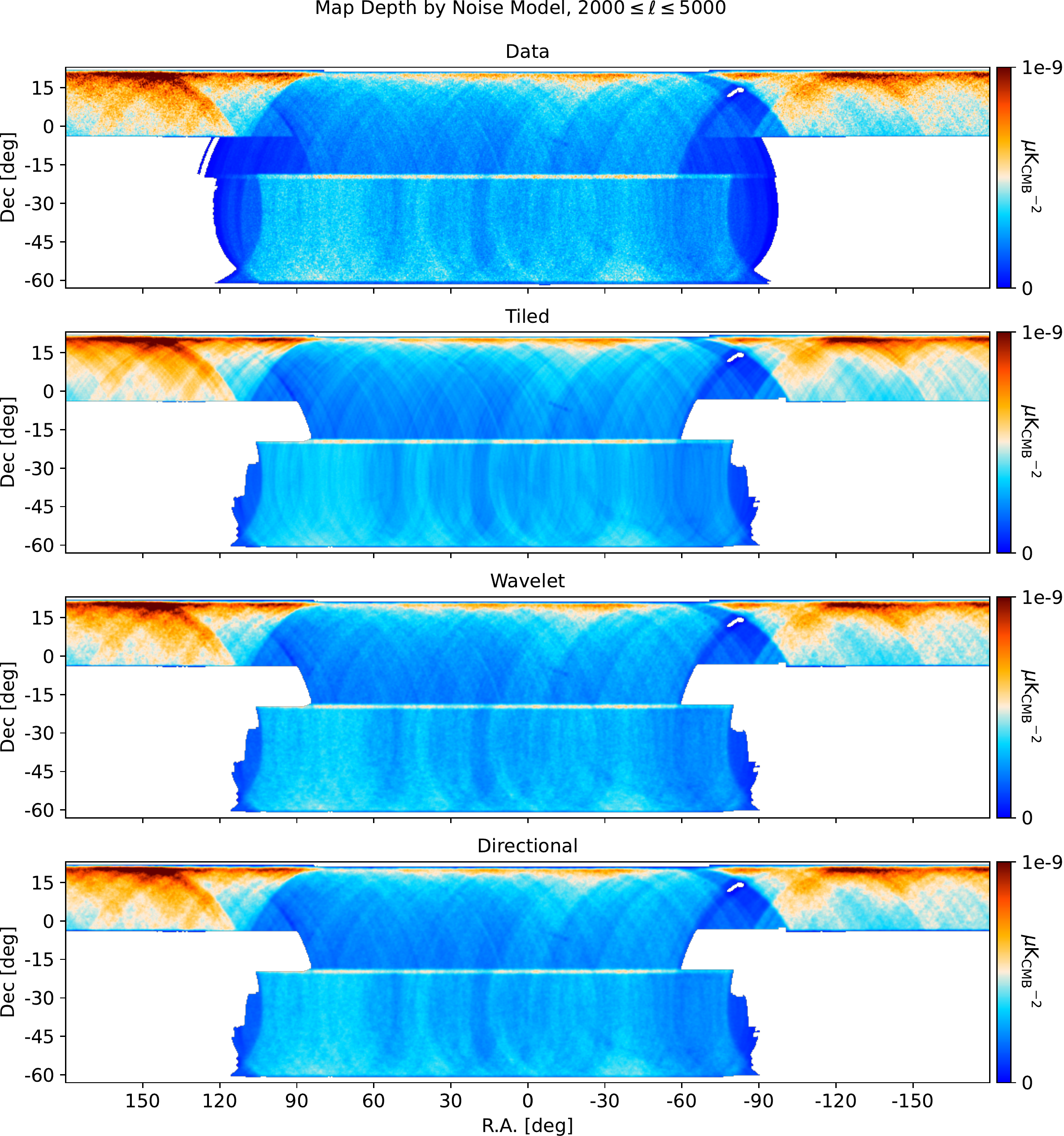}
    \caption{The same small-scale ``inverse-variance map" as the top panel of Figure \ref{fig: ivar_by_scale}, for both data and noise simulations. The first/second/third/fourth rows depict the data/tiled/wavelet/directional models. As is discussed in \S\ref{sec: implementation_masks}, we see that simulations are masked with the boolean mask that removes sky regions with the lowest cross-linking. All models do a good job reproducing the high-$\ell$ map depth in the data.} 
    \label{fig: results_ivar_by_scale_highell}
\end{figure*}

\begin{figure*}
    \centering
    \includegraphics[width=\textwidth]{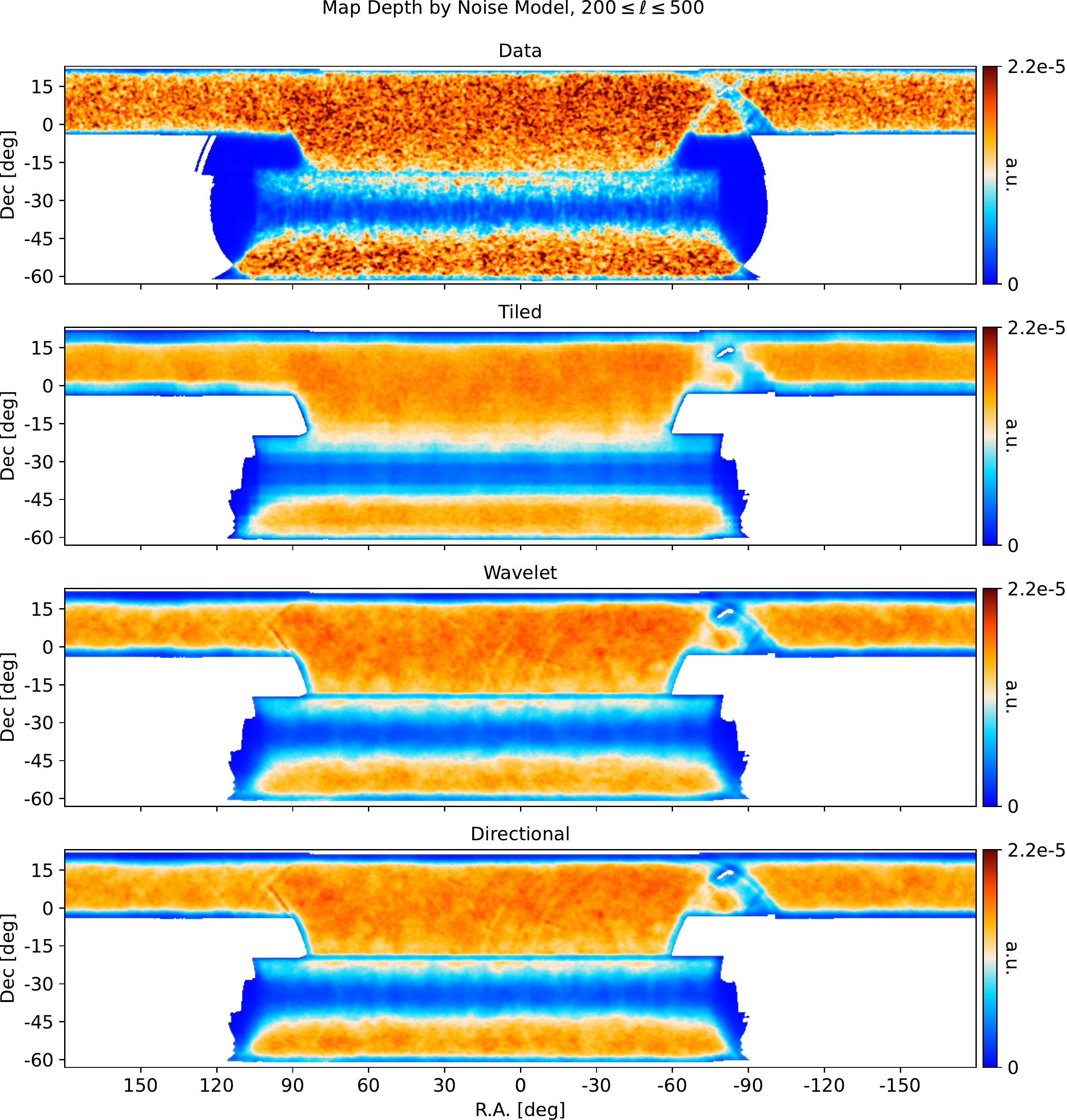}
    \caption{The same large-scale ``inverse-variance map" as the bottom panel of Figure \ref{fig: ivar_by_scale}, for both data and noise simulations. The first/second/third/fourth rows depict the data/tiled/wavelet/directional models. All three models exhibit slightly less inverse-variance (more noise power) than the data, as is expected from the elevated polarized power spectra ratios in Figure \ref{fig: results_ps_ratios} for $\ell\lesssim 300$. The wavelet and directional models capture fine structures in the data; the tiled model resolution at large scales is limited to the tile size --- map features are ``squared-off" or unresolved.}
    \label{fig: results_ivar_by_scale_lowell}
\end{figure*}

In addition to spatial correlations and stripiness, we showed how the atmosphere and scan strategy together induce a scale dependence to the map depth in \S\ref{sec: noise_properties}. In particular, we showed that the small-scale noise power appears to follow the morphology of the mapmaker inverse-variance maps, while the large-scale noise power is sensitive to the scan cross-linking. We compare the ability of the noise models to reproduce these effects in Figures \ref{fig: results_ivar_by_scale_highell} and \ref{fig: results_ivar_by_scale_lowell}. When filtered for small scales (Figure \ref{fig: results_ivar_by_scale_highell}), all three models succeed at reproducing the mapmaker inverse-variance-like morphology in the data. When filtered for large scales (Figure \ref{fig: results_ivar_by_scale_lowell}), the two wavelet-based models succeed at reconstructing the mapmaker cross-linking-like morphology, for which they were nominally designed.

The tiled model, while not explicitly designed for this purpose, also performs moderately well. This is because a scale-dependent map depth is equivalent to a spatially-dependent noise power spectrum. The tiles do model a spatially dependent (2D Fourier) power spectrum, but limited to a resolution of the tile size ($\sim4^{\circ}$). As a result, some features in the tiled large-scale spatial noise power are not resolved --- for instance, the fine structures near a declination of $-20^\circ$. Other structures are ``squared-off," such as the transitions between high- and low-cross-linking regions. 

For the large-scale maps, we also note that all three models show slightly less inverse-variance (more noise power) than the data. This is a restatement of the result in Figure \ref{fig: results_ps_ratios} that the polarized noise power spectra ratios are elevated for $\ell\lesssim 300$. Finally, as for the 2D Fourier spectra of Figure \ref{fig: results_2d_ps_by_region}, we observe the same suppression of high-frequency scatter in the simulated averages as compared to the data. This is again expected, and further supports the model smoothing of \S\ref{sec: implementation_smoothing}. In sum, the two wavelet models perform somewhat better than the tiled model at capturing the scale-dependent map depth in the data.

In short, the models behave as designed. All three models reconstruct the overall noise auto- and cross-spectra. The tiled model is best at capturing spatially-varying noise stripiness; it performs less well at capturing scale-dependent map depth. The wavelet model cannot capture stripiness; however, it is best at capturing scale-dependent map depth. Finally, the directional model performs well in both analyses.

\begin{figure*}
    \centering
    \includegraphics[width=\textwidth]{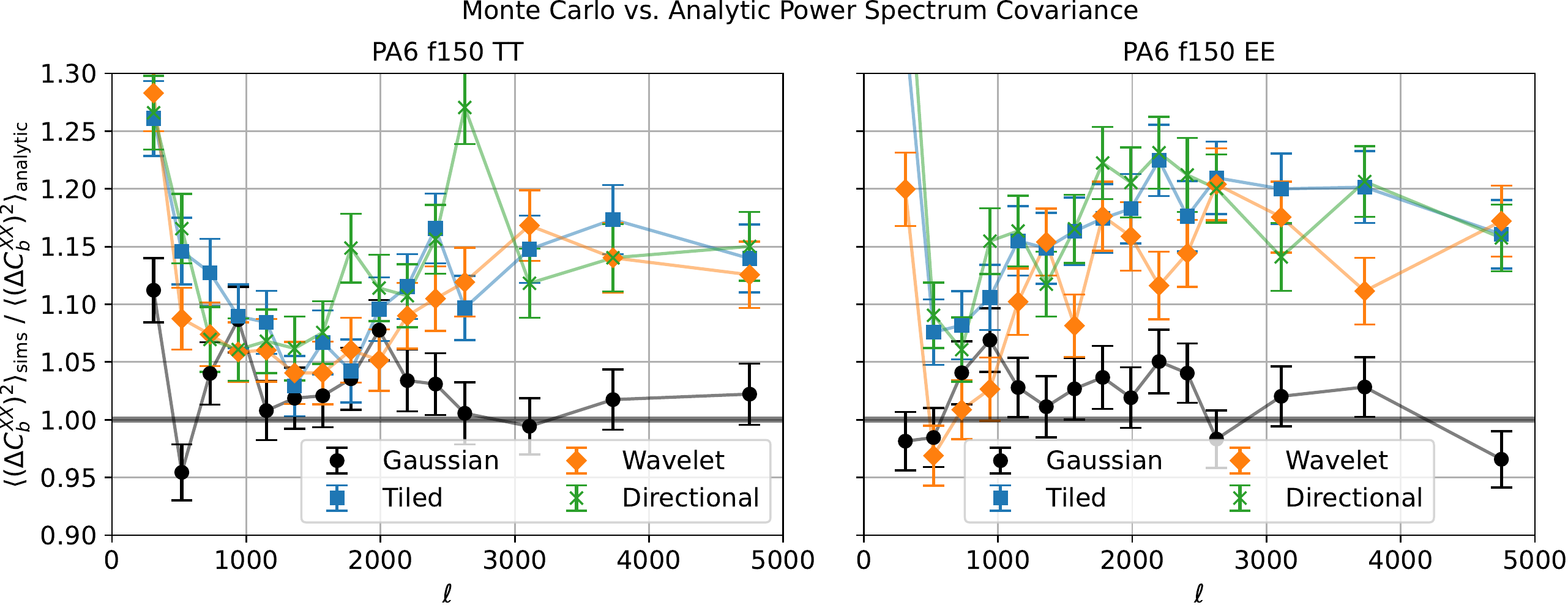}
    \caption{Ratios of the power spectrum covariance matrix diagonal elements between a Monte Carlo construction using 300 noise simulations, and an analytic approach common in the literature. The nominal result is a value of 1. TT (EE) ratios exhibit an excess of up to $\sim 15\%$ ($\sim 20\%)$. The ``Gaussian" simulations are a sanity check: they are drawn directly from the measured noise power spectra and should match the analytic covariance. Significant deviations of the ``Gaussian" simulations from nominal (i.e., in TT for $\ell\lesssim300$) suggest discounting this comparison at those scales. Data points are plotted at the center of the multipole bins, and show the per-bin mean and $1\sigma$ error in the mean coming from the 300 simulations.}
    \label{fig: ps_cov_diag_ratios}
\end{figure*}

\subsection{ACT DR6 Power Spectrum Covariance} \label{sec: ps_covmat}

In this section, we examine the implications of our noise simulations on the estimation of the covariance matrix for the CMB angular power spectrum. Specifically, we examine the matrix, $\Xi$, given by:
\begin{align} \label{eq: ps_covmat}
    \Xi_{b,b'}^{WXYZ} \equiv \langle \Delta C_b^{WX} \Delta C_{b'}^{YZ} \rangle
\end{align}
where $\Delta C_b^{WX}$ is the \textit{residual} of the recovered power spectrum in $\ell$-bin $b$, between fields $W$ and $X$ (one of T, E, or B), with respect to a fiducial model. An accurate covariance matrix is a required deliverable of any CMB experiment: inaccuracies would directly propagate into cosmological parameter uncertainties and could also skew power spectrum pipeline null tests.

While the exact calculation of Equation \ref{eq: ps_covmat} is difficult \citep[see e.g.][]{planckxi15}, several approximations exist. These incorporate knowledge of a fiducial sky model, instrumental noise, and incomplete sky coverage into computationally tractable, \textit{analytic} expressions of the covariance matrix \citep[e.g.,][]{Brown2005, Efstathiou2006, Couchot2017, namaster2019, Louis2020}. These analytic expressions may also be augmented with simulations to correct for some of their assumptions, for example, by adding variance at low-$\ell$ to account for masking point sources with small holes \citep{planckv18, Li2021}. As far as their treatment of instrumental noise, these semi-analytic methods adopt one of two approaches. The first approach assumes the noise is diagonal in the harmonic domain --- that is, completely described by the noise power spectrum. This is the default approach in the \texttt{NaMaster}\footnote{\url{https://github.com/LSSTDESC/NaMaster}} \citep{namaster2019} and \texttt{pspy}\footnote{\url{https://github.com/simonsobs/pspy}} \citep{Louis2020} codes, which together form the SO power spectrum pipeline, \texttt{PSpipe}\footnote{\url{https://github.com/simonsobs/PSpipe}}. The second approach generalizes the first by combining knowledge of the noise power spectrum and the inhomogeneous white-noise depth \citep[e.g.,][]{planckxi15, Efstathiou2021}.

In the following subsections we compare Equation \ref{eq: ps_covmat} estimated from an analytic expression following the former approach (assuming diagonal noise in the harmonic domain), to a Monte Carlo estimate from simulations alone. We will examine analytic expressions following the latter approach in future work. We focus on the PA6 f150 band, whose results are representative of the full DR6 data. 

\subsubsection{Analytic and Monte Carlo Covariances}
We construct an analytic covariance matrix assuming the signal and noise are completely described by their respective power spectra --- that is, assuming uniform white-noise depth. In particular, we use the methods of \texttt{pspy} available in \texttt{PSpipe}. Our implementation adopts the best-fit $\Lambda$CDM cosmology from \citet{Planck-cosmology:2018} as the primary CMB power spectrum signal, and adds a foreground model power spectrum consistent with \citet{Dunkley2013}, derived from maps with a 15 mJy source flux cut. We incorporate the noise pseudospectra estimated directly from the DR6 data (i.e., the unnormalized version of the spectra in Figure \ref{fig: N_ell_r_ell}), following Equation 8 of \citet{Louis2020} (see also Equation 2.55 of \citet{namaster2019} and Equation 10 of \citet{Li2021}). To keep our analysis straightforward and avoid issues with point source holes, we reuse the pseudospectrum estimate mask, $\bm\mu_{\mathrm{est}}$, from the previous sections (shown in Figure \ref{fig: masks}). 

Next, we prescribe the simulations from which we build a Monte Carlo covariance matrix. We first draw signal map realizations from the same CMB-plus-foreground signal model that entered the analytic matrix, and convolve them with the PA6 f150 beam. We then add noise realizations in map space using the methods of \S\ref{sec: noise_models}. We evaluate the power spectra of 300 such simulations, for each of the eight PA6 f150 splits, and directly compute the Monte Carlo covariance matrix (Equation \ref{eq: ps_covmat}) by averaging the residuals of the estimated power spectra compared to the input signal power spectrum. As a sanity check we also generate a set of ``Gaussian'' Monte Carlo simulations whose noise realizations are simply drawn from the noise power spectra, and estimate their covariance matrix following the same method. Both the analytic and Monte Carlo matrices account for the mode-coupling induced by the analysis mask,\footnote{The noise power spectra entering the analytic matrix were measured after applying the same Fourier space ground pickup filter as in ACT DR4 \citep{C20} and Figure \ref{fig: bb_ratios}. Likewise, we applied this filter to the simulations. The final matrices then deconvolve the transfer function induced by this pickup filter.} and are binned in $\ell$ to improve their conditioning.

\subsubsection{Comparison}
We compare the diagonals of the Monte Carlo and analytic matrices in Figure \ref{fig: ps_cov_diag_ratios}. In terms of Equation \ref{eq: ps_covmat}, that is, we only consider the elements of $\Xi$ with $b=b'$ and $W=X=Y=Z$. As expected, the ``Gaussian'' simulations produce a Monte Carlo matrix whose diagonal agrees with the analytic expression to a few percent, demonstrating that our pipeline is not significantly biased. We then note that in spite of their different methods, our three noise models deliver remarkably similar results. In particular, the TT (EE) block of the analytic covariance appears to underestimate its diagonal by up to $\sim 15\%$ ($\sim 20\%$) for $\ell\gtrsim2,000$ ($1,000$). Such discrepancies exceed what has been found in comparable analyses of \textit{Planck} data. For instance, \citet{Couchot2017} compared the diagonals of a Monte Carlo covariance estimate and an analytic expression assuming uniform white-noise depth and found errors of at most a few percent. Meanwhile, \citet{Li2021} performed a similar comparison and suggest errors of order $\sim 10\%$. At a minimum, our noise simulations suggest increasing the ACT DR6 power spectra covariance by up to $\sim 20\%$ in polarization.

The agreement of our three noise models in Figure \ref{fig: ps_cov_diag_ratios} hints at a common underlying mechanism. As discussed, one ACT noise property not accounted for in the analytic covariance of this section is the spatially-varying map depth (regardless of whether the map depth is simultaneously scale-dependent). Thus, the natural next step would be to evaluate an analytic expression that includes the inhomogeneous ACT white-noise depth \citep{Efstathiou2021}, as was used for \textit{Planck} \citep{planckxi15, planckv18}, assessing whether it could sufficiently capture the additional variance observed in our simulations. These results will inform the approach to constructing a power spectrum covariance matrix for ACT DR6, and will be explored in future work.

\section{Discussion and Conclusions} \label{sec: disc_conc}
For ACT DR6, the advent of three separate noise models allows downstream analysis pipelines to evaluate results against a range of map-based noise properties. For instance, we showed how propagation of noise simulations through the ACT DR6 power spectrum pipeline can help measure the extent to which analytic covariance matrix expressions accurately capture the nontrivial ACT noise properties \citep[e.g.,][]{Li2021, Efstathiou2021}. Likewise, the ACT DR6 lensing analysis \citep{Qu23} utilizes these models to demonstrate that the lensing inference is robust with respect to the complicated ACT noise. While checking pipelines for robustness against different noise models can be useful, it is too early to say whether one model is the most performant for particular applications. Finally, consumption of large noise simulation ensembles by ACT and community users requires efficient code and a simple interface to those products. This code --- \texttt{mnms} --- is made public with this paper, and is out-of-the-box applicable to future ground-based CMB experiments such as SO and CMB-S4. Nevertheless, it is worthwhile to evaluate key model assumptions and suggest directions for future development.

\textit{Measuring noise from split differences:} Throughout this paper, we have assumed $\bm\nu_i$ to be an accurate noise estimate. In \S\ref{sec: splits_math} we demonstrated this is a reasonable assumption when averaged over a large sky fraction. However, in regions of especially compact or bright foregrounds, we encounter residual signal in our difference maps. As discussed in \S\ref{sec: implementation_ssi}, we mitigate point-source residuals with inpainting. However, we do not handle bright, extended source residuals, which can be prominent in the Galactic plane (see Figure \ref{apx: example_sims_I}). The underlying cause of these residuals is likely similar as for point sources: a combination of pointing variation in the telescope and small relative calibration errors between splits. Our noise models will therefore be biased in these regions. Future work could include more sophisticated inpainting than our implementation in \S\ref{sec: implementation_ssi} to handle bright, extended regions.

\textit{Non-identically distributed splits:} As noted, unlike ACT DR4, we do not assume that the noise in separate splits is identically distributed. Therefore, we do not average noise models over splits in their sparse basis. Whether to do so is tightly coupled to the model smoothing: averaging over the eight splits of ACT DR6 would reduce the amount of required model smoothing. However, if splits contain different noise structure, averaging over them would bias every model. In fact, we know \textit{a priori} that the noise is not identically distributed over splits because their inverse-variance maps vary considerably. We indeed observe percent-level improvement in low-$\ell$ power spectra ratios when treating splits separately versus when averaging over splits. In principle, including inverse-variance maps as a \textit{mode-decoupling} filter in the wavelet models (as we discussed, and forewent, in \S\ref{sec: noise_models_wav_algorithm} and \S\ref{sec: noise_models_fdw_algorithm}) would eliminate the barrier they present to averaging over splits.

\begin{figure}
    \centering
    \includegraphics[width=\columnwidth]{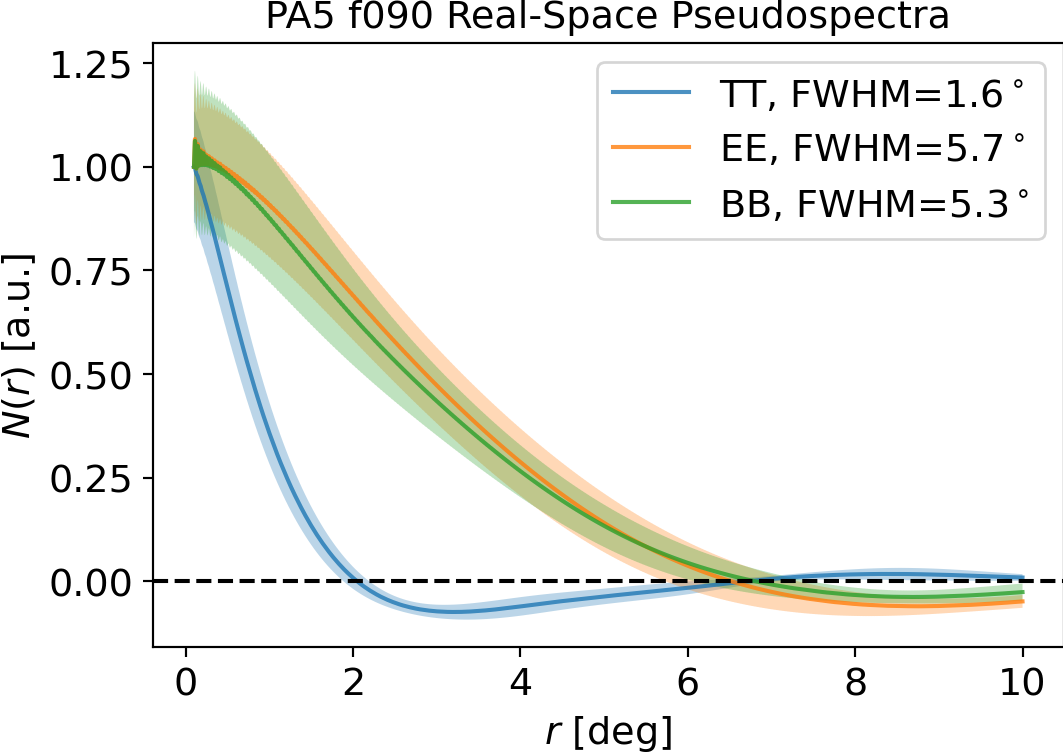}
    \caption{``Beam-like" real-space profiles corresponding to the transform of the noise autospectra from the top panel of Figure \ref{fig: N_ell_r_ell} for PA5 f090. The profiles have been peak-normalized. Solid lines (shaded bands) denote the mean ($1\sigma$ range) over splits.}
    \label{fig: Nl_realspace}
\end{figure}

\textit{Noise pseudospectrum filtering}: Each of our noise models includes the same noise pseudospectrum filter. The benefits of this filter were discussed in \S\ref{sec: noise_models} and are explored further in Appendix \ref{apx: Nl_benefits}. However, this filter has a downside: it limits our ability to model high-resolution noise features in map space, especially in polarization. To see how, consider that the noise pseudospectra in the top panel of Figure \ref{fig: N_ell_r_ell} downweight small scales relative to large scales. Thus, the noise pseudospectrum filter has a similar effect in the noise models as the instrumental beam has on a given sky signal: by suppressing small scales it ``blurs" compact features. We can see this by examining the ``beam-like" real-space profiles corresponding to the transform of noise pseudospectra. Results are shown in Figure \ref{fig: Nl_realspace} for the PA5 f090 band. The FWHM of the EE and BB real-space profiles are both $\sim 5^\circ$, whereas the TT profile is more spatially compact, with a FWHM of $\sim 1^\circ$. In other words, given a point-like noise feature in both temperature and polarization, we would expect our simulations to resolve it to $\sim 1^\circ$ in temperature and $\sim 5^\circ$ in polarization. This may explain, for instance, the greater deviations observed in the low-$\ell$ power spectra ratios (Figure \ref{fig: results_ps_ratios}) for EE than TT: noisy regions from farther outside the analysis mask in polarization than temperature are smoothed into the mask, skewing the ratios. The different real-space widths in Figure \ref{fig: Nl_realspace} are driven mainly by the low-$\ell$ shape of the noise pseudospectra in Figure \ref{fig: N_ell_r_ell}: in some bands, the TT noise power at the largest scales ($\ell\lesssim 100$) level off and even decrease. This does not occur in polarization. We will seek to better understand the shape of the low-$\ell$ noise power spectra in future work.

\textit{Powers of noise covariance matrix:} In this paper, we only interact with the full noise covariance matrix by drawing samples from the Gaussian distribution it defines. Downstream analyses may wish to raise the matrix to an arbitrary power and then apply it to a map vector. In particular, Monte Carlo methods, inverse-covariance filtering, and Wiener filtering require the full inverse noise covariance matrix. Such an implementation is beyond the scope of this work; nevertheless, we can comment on its feasibility. Unfortunately, due to the overlap of tiles in map space (likewise, of wavelets in harmonic/Fourier space) the class of noise covariance matrices prescribed here are not analytically invertible. An exception would be the case of non-overlapping (i.e., tophat) tiles or wavelets, but with a detrimental impact to their performance as noise models (as noted in the text) from their lack of smoothness. In principle, the exact application of the inverse noise covariance matrix to a map vector is still possible with a conjugate gradient method, but may be slow. We will explore this and other alternatives in future work. 

\textit{Prospective:} These results offer concrete insights for the design of next-generation, ground-based CMB experiments. Foremost among them is the central role of the map cross-linking in modulating large-scale noise power. Combined with the tight relationship between cross-linking and the noise anisotropy pattern, spatial uniformity of cross-linking is as, if not more, important to realizing a successful map-based noise model than uniformity of detector hits. We acknowledge that achieving uniform cross-linking may encounter more observational constraints than uniform detector hits (e.g., the dynamics of the telescope and receiver). In a similar vein, projects seeking to use these noise modeling methods should design scan strategies that facilitate identically distributed map splits. While formally difficult, maximizing the uniformity of heuristics across splits (again, e.g., cross-linking, inverse-variance) would go a long way toward allowing noise models to be averaged over splits. Projects should also seek to avoid unnecessary ``hard edges," in their scan strategies. This includes not only the observation footprint boundary, but also adjacent regions of disparate cross-linking or inverse-variance levels. Such boundaries within the ACT DR6 footprint proved especially challenging to handle from a noise modeling perspective. 

Finally, we emphasize the utility of a fast and flexible noise model implementation, as is released in this paper with \texttt{mnms}. Using ACT DR6 data with \texttt{mnms}, this paper has constructed empirical noise models and drawn large ensembles of ACT noise simulations. For future experiments undergoing design and forecast activities, such as SO, \texttt{mnms} can also serve as an emulator of otherwise expensive TOD simulations. Thus, \texttt{mnms} could relieve a bottleneck in the simulation-to-forecast feedback loop, enabling iterative optimization of experiment design. 

\section*{Acknowledgments}
We thank Reijo Keskitalo and Bruno R\'{e}galdo-Saint Blancard for insightful conversations. ZA and JD acknowledge support from NSF grant AST-2108126. CS acknowledges support from the Agencia Nacional de Investigaci\'on y Desarrollo (ANID) through FONDECYT grant no.\ 11191125 and BASAL project FB210003. EC acknowledges support from the European Research Council (ERC) under the European Union’s Horizon 2020 research and innovation programme (Grant agreement No. 849169). KM acknowledges support from the National Research Foundation of South Africa. SKC acknowledges support from NSF award AST-2001866. This work was supported by a grant from the Simons Foundation (CCA 918271, PBL). The Flatiron Institute is supported by the Simons Foundation. 

Support for ACT was through the U.S.~National Science Foundation through awards AST-0408698, AST-0965625, and AST-1440226 for the ACT project, as well as awards PHY-0355328, PHY-0855887 and PHY-1214379. Funding was also provided by Princeton University, the University of Pennsylvania, and a Canada Foundation for Innovation (CFI) award to UBC. ACT operated in the Parque Astron\'omico Atacama in northern Chile under the auspices of the Agencia Nacional de Investigaci\'on y Desarrollo (ANID). The development of multichroic detectors and lenses was supported by NASA grants NNX13AE56G and NNX14AB58G. Detector research at NIST was supported by the NIST Innovations in Measurement Science program. 

The authors are pleased to acknowledge that the work reported on in this paper was substantially performed using the Princeton Research Computing resources at Princeton University which is consortium of groups led by the Princeton Institute for Computational Science and Engineering (PICSciE) and Office of Information Technology's Research Computing. This research used resources of the National Energy Research Scientific Computing Center (NERSC), a U.S. Department of Energy Office of Science User Facility located at Lawrence Berkeley National Laboratory, operated under Contract No. DE-AC02-05CH11231 using NERSC award HEP-ERCAPmp107 for 2022.

Some of the results in this paper have been derived using the \texttt{healpy} and \texttt{HEALPix} package. Other software used in this paper include the following: \software{\texttt{astropy} \citep{astropy:2013, astropy:2018, astropy:2022}, \texttt{cython} \citep{cython}, \texttt{h5py} \citep{h5py}, \texttt{libsharp} \citep{libsharp}, \texttt{matplotlib} \citep{matplotlib}, \texttt{numba} \citep{numba}, \texttt{numpy} \citep{numpy}, \texttt{pyfftw} \citep{pyfftw}, \texttt{pyyaml} (\url{https://pyyaml.org/}), and \texttt{scipy} \citep{scipy}.} 

\bibliography{main}

\begin{thebibliography}{}
\expandafter\ifx\csname natexlab\endcsname\relax\def\natexlab#1{#1}\fi
\providecommand{\url}[1]{\href{#1}{#1}}
\providecommand{\dodoi}[1]{doi:~\href{http://doi.org/#1}{\nolinkurl{#1}}}
\providecommand{\doeprint}[1]{\href{http://ascl.net/#1}{\nolinkurl{http://ascl.net/#1}}}
\providecommand{\doarXiv}[1]{\href{https://arxiv.org/abs/#1}{\nolinkurl{https://arxiv.org/abs/#1}}}

\bibitem[{Abitbol {et~al.}(2017)Abitbol, Ahmed, Barron, Thakur, Bender, Benson,
  Bischoff, Bryan, Carlstrom, Chang, Chuss, Crowley, Cukierman, de~Haan, Dobbs,
  Essinger-Hileman, Filippini, Ganga, Gudmundsson, Halverson, Hanany,
  Henderson, Hill, Ho, Hubmayr, Irwin, Jeong, Johnson, Kernasovskiy, Kovac,
  Kusaka, Lee, Maria, Mauskopf, McMahon, Moncelsi, Nadolski, Nagy, Niemack,
  O'Brient, Padin, Parshley, Pryke, Roe, Rostem, Ruhl, Simon, Staggs, Suzuki,
  Switzer, Tajima, Thompson, Timbie, Tucker, Vieira, Vieregg, Westbrook,
  Wollack, Yoon, Young, \& Young}]{S4}
Abitbol, M.~H., Ahmed, Z., Barron, D., {et~al.} 2017,
  \dodoi{10.48550/arxiv.1706.02464}

\bibitem[{Aiola {et~al.}(2020)Aiola, Calabrese, Maurin, Naess, Schmitt,
  Abitbol, Addison, Ade, Alonso, Amiri, Amodeo, Angile, Austermann, Baildon,
  Battaglia, Beall, Bean, Becker, Bond, Bruno, Calafut, Campusano, Carrero,
  Chesmore, Cho, Choi, Clark, Cothard, Crichton, Crowley, Darwish, Datta,
  Denison, Devlin, Duell, Duff, Duivenvoorden, Dunkley, D{\"{u}}nner,
  Essinger-Hileman, Fankhanel, Ferraro, Fox, Fuzia, Gallardo, Gluscevic, Golec,
  Grace, Gralla, Guan, Hall, Halpern, Han, Hargrave, Hasselfield, Helton,
  Henderson, Hensley, Hill, Hilton, Hilton, Hincks, Hlo{\v{z}}ek, Ho, Hubmayr,
  Huffenberger, Hughes, Infante, Irwin, Jackson, Klein, Knowles, Koopman,
  Kosowsky, Lakey, Li, Li, Li, Lokken, Louis, Lungu, MacInnis, Madhavacheril,
  Maldonado, Mallaby-Kay, Marsden, McMahon, Menanteau, Moodley, Morton,
  Namikawa, Nati, Newburgh, Nibarger, Nicola, Niemack, Nolta, Orlowski-Sherer,
  Page, Pappas, Partridge, Phakathi, Pisano, Prince, Puddu, Qu, Rivera,
  Robertson, Rojas, Salatino, Schaan, Schillaci, Sehgal, Sherwin, Sierra,
  Sievers, Sifon, Sikhosana, Simon, Spergel, Staggs, Stevens, Storer, Sunder,
  Switzer, Thorne, Thornton, Trac, Treu, Tucker, Vale, Engelen, Lanen,
  Vavagiakis, Wagoner, Wang, Ward, Wollack, Xu, Zago, \& Zhu}]{A20}
Aiola, S., Calabrese, E., Maurin, L., {et~al.} 2020, Journal of Cosmology and
  Astroparticle Physics, 2020, 47, \dodoi{10.1088/1475-7516/2020/12/047}

\bibitem[{Akrami {et~al.}(2020)Akrami, Andersen, Ashdown, Baccigalupi,
  Ballardini, Banday, Barreiro, Bartolo, Basak, Benabed, Bernard, Bersanelli,
  Bielewicz, Bond, Borrill, Burigana, Butler, Calabrese, Casaponsa, Chiang,
  Colombo, Combet, Crill, Cuttaia, {De Bernardis}, {De Rosa}, {De Zotti},
  Delabrouille, {DI Valentino}, DIego, Dor{\'{e}}, Douspis, Dupac, Eriksen,
  Fernandez-Cobos, Finelli, Frailis, Fraisse, Franceschi, Frolov, Galeotta,
  Galli, Ganga, Gerbino, Ghosh, Gonz{\'{a}}lez-Nuevo, G{\'{o}}rski, Gruppuso,
  Gudmundsson, Handley, Helou, Herranz, Hildebrandt, Hivon, Huang, Jaffe,
  Jones, Keih{\"{a}}nen, Keskitalo, Kiiveri, Kim, Kisner, Krachmalnicoff, Kunz,
  Kurki-Suonio, Lasenby, Lattanzi, Lawrence, {Le Jeune}, Levrier, Liguori,
  Lilje, Lilley, Lindholm, L{\'{o}}pez-Caniego, Lubin,
  Mac{\'{I}}as-P{\'{e}}rez, Maino, Mandolesi, Marcos-Caballero, Maris, Martin,
  Mart{\'{i}}nez-Gonz{\'{a}}lez, Matarrese, Mauri, McEwen, Meinhold, Mennella,
  Migliaccio, Mitra, Molinari, Montier, Morgante, Moss, Natoli, Paoletti,
  Partridge, Patanchon, Pearson, Pearson, Perrotta, Piacentini, Polenta,
  Rachen, Reinecke, Remazeilles, Renzi, Rocha, Rosset, Roudier,
  Rubi{\~{n}}o-Mart{\'{i}}n, Ruiz-Granados, Salvati, Savelainen, Scott,
  Sirignano, Sirri, Spencer, Suur-Uski, Svalheim, Tauber, Tavagnacco, Tenti,
  Terenzi, Thommesen, Toffolatti, Tomasi, Tristram, Trombetti, Valiviita, {Van
  Tent}, Vielva, Villa, Vittorio, Wandelt, Wehus, Zacchei, \& Zonca}]{NPIPE}
Akrami, Y., Andersen, K.~J., Ashdown, M., {et~al.} 2020, Astronomy \&
  Astrophysics, 643, A42, \dodoi{10.1051/0004-6361/202038073}

\bibitem[{{Allys} {et~al.}(2020){Allys}, {Marchand}, {Cardoso},
  {Villaescusa-Navarro}, {Ho}, \& {Mallat}}]{Allys2020}
{Allys}, E., {Marchand}, T., {Cardoso}, J.~F., {et~al.} 2020, \prd, 102,
  103506, \dodoi{10.1103/PhysRevD.102.103506}

\bibitem[{Anderson {et~al.}(2018)Anderson, Ade, Ahmed, Austermann, Avva, Barry,
  Thakur, Bender, Benson, Bleem, Byrum, Carlstrom, Carter, Cecil, Chang, Cho,
  Cliche, Crawford, Cukierman, Denison, de~Haan, Ding, Dobbs, Dutcher, Everett,
  Foster, Gannon, Gilbert, Groh, Halverson, Harke-Hosemann, Harrington,
  Henning, Hilton, Holder, Holzapfel, Huang, Irwin, Jeong, Jonas, Khaire, Knox,
  Kofman, Korman, Kubik, Kuhlmann, Kuklev, Kuo, Lee, Leitch, Lowitz, Meyer,
  Michalik, Montgomery, Nadolski, Natoli, Nguyen, Noble, Novosad, Padin, Pan,
  Pearson, Posada, Rahlin, Reichardt, Ruhl, Saunders, Sayre, Shirley,
  Shirokoff, Smecher, Sobrin, Stark, Story, Suzuki, Tang, Thompson, Tucker,
  Vale, Vanderlinde, Vieira, Wang, Whitehorn, Yefremenko, Yoon, \&
  Young}]{SPT_3G_2}
Anderson, A.~J., Ade, P.~A., Ahmed, Z., {et~al.} 2018, Journal of Low
  Temperature Physics, 193, 1057, \dodoi{10.1007/s10909-018-2007-z}

\bibitem[{{Astropy Collaboration} {et~al.}(2013){Astropy Collaboration},
  {Robitaille}, {Tollerud}, {Greenfield}, {Droettboom}, {Bray}, {Aldcroft},
  {Davis}, {Ginsburg}, {Price-Whelan}, {Kerzendorf}, {Conley}, {Crighton},
  {Barbary}, {Muna}, {Ferguson}, {Grollier}, {Parikh}, {Nair}, {Unther},
  {Deil}, {Woillez}, {Conseil}, {Kramer}, {Turner}, {Singer}, {Fox}, {Weaver},
  {Zabalza}, {Edwards}, {Azalee Bostroem}, {Burke}, {Casey}, {Crawford},
  {Dencheva}, {Ely}, {Jenness}, {Labrie}, {Lim}, {Pierfederici}, {Pontzen},
  {Ptak}, {Refsdal}, {Servillat}, \& {Streicher}}]{astropy:2013}
{Astropy Collaboration}, {Robitaille}, T.~P., {Tollerud}, E.~J., {et~al.} 2013,
  \aap, 558, A33, \dodoi{10.1051/0004-6361/201322068}

\bibitem[{{Astropy Collaboration} {et~al.}(2018){Astropy Collaboration},
  {Price-Whelan}, {Sip{\H{o}}cz}, {G{\"u}nther}, {Lim}, {Crawford}, {Conseil},
  {Shupe}, {Craig}, {Dencheva}, {Ginsburg}, {Vand erPlas}, {Bradley},
  {P{\'e}rez-Su{\'a}rez}, {de Val-Borro}, {Aldcroft}, {Cruz}, {Robitaille},
  {Tollerud}, {Ardelean}, {Babej}, {Bach}, {Bachetti}, {Bakanov}, {Bamford},
  {Barentsen}, {Barmby}, {Baumbach}, {Berry}, {Biscani}, {Boquien}, {Bostroem},
  {Bouma}, {Brammer}, {Bray}, {Breytenbach}, {Buddelmeijer}, {Burke},
  {Calderone}, {Cano Rodr{\'\i}guez}, {Cara}, {Cardoso}, {Cheedella}, {Copin},
  {Corrales}, {Crichton}, {D'Avella}, {Deil}, {Depagne}, {Dietrich}, {Donath},
  {Droettboom}, {Earl}, {Erben}, {Fabbro}, {Ferreira}, {Finethy}, {Fox},
  {Garrison}, {Gibbons}, {Goldstein}, {Gommers}, {Greco}, {Greenfield},
  {Groener}, {Grollier}, {Hagen}, {Hirst}, {Homeier}, {Horton}, {Hosseinzadeh},
  {Hu}, {Hunkeler}, {Ivezi{\'c}}, {Jain}, {Jenness}, {Kanarek}, {Kendrew},
  {Kern}, {Kerzendorf}, {Khvalko}, {King}, {Kirkby}, {Kulkarni}, {Kumar},
  {Lee}, {Lenz}, {Littlefair}, {Ma}, {Macleod}, {Mastropietro}, {McCully},
  {Montagnac}, {Morris}, {Mueller}, {Mumford}, {Muna}, {Murphy}, {Nelson},
  {Nguyen}, {Ninan}, {N{\"o}the}, {Ogaz}, {Oh}, {Parejko}, {Parley}, {Pascual},
  {Patil}, {Patil}, {Plunkett}, {Prochaska}, {Rastogi}, {Reddy Janga},
  {Sabater}, {Sakurikar}, {Seifert}, {Sherbert}, {Sherwood-Taylor}, {Shih},
  {Sick}, {Silbiger}, {Singanamalla}, {Singer}, {Sladen}, {Sooley},
  {Sornarajah}, {Streicher}, {Teuben}, {Thomas}, {Tremblay}, {Turner},
  {Terr{\'o}n}, {van Kerkwijk}, {de la Vega}, {Watkins}, {Weaver}, {Whitmore},
  {Woillez}, {Zabalza}, \& {Astropy Contributors}}]{astropy:2018}
{Astropy Collaboration}, {Price-Whelan}, A.~M., {Sip{\H{o}}cz}, B.~M., {et~al.}
  2018, \aj, 156, 123, \dodoi{10.3847/1538-3881/aabc4f}

\bibitem[{{Astropy Collaboration} {et~al.}(2022){Astropy Collaboration},
  {Price-Whelan}, {Lim}, {Earl}, {Starkman}, {Bradley}, {Shupe}, {Patil},
  {Corrales}, {Brasseur}, {N{"o}the}, {Donath}, {Tollerud}, {Morris},
  {Ginsburg}, {Vaher}, {Weaver}, {Tocknell}, {Jamieson}, {van Kerkwijk},
  {Robitaille}, {Merry}, {Bachetti}, {G{"u}nther}, {Aldcroft},
  {Alvarado-Montes}, {Archibald}, {B{'o}di}, {Bapat}, {Barentsen}, {Baz{'a}n},
  {Biswas}, {Boquien}, {Burke}, {Cara}, {Cara}, {Conroy}, {Conseil}, {Craig},
  {Cross}, {Cruz}, {D'Eugenio}, {Dencheva}, {Devillepoix}, {Dietrich},
  {Eigenbrot}, {Erben}, {Ferreira}, {Foreman-Mackey}, {Fox}, {Freij}, {Garg},
  {Geda}, {Glattly}, {Gondhalekar}, {Gordon}, {Grant}, {Greenfield}, {Groener},
  {Guest}, {Gurovich}, {Handberg}, {Hart}, {Hatfield-Dodds}, {Homeier},
  {Hosseinzadeh}, {Jenness}, {Jones}, {Joseph}, {Kalmbach}, {Karamehmetoglu},
  {Ka{l}uszy{'n}ski}, {Kelley}, {Kern}, {Kerzendorf}, {Koch}, {Kulumani},
  {Lee}, {Ly}, {Ma}, {MacBride}, {Maljaars}, {Muna}, {Murphy}, {Norman},
  {O'Steen}, {Oman}, {Pacifici}, {Pascual}, {Pascual-Granado}, {Patil},
  {Perren}, {Pickering}, {Rastogi}, {Roulston}, {Ryan}, {Rykoff}, {Sabater},
  {Sakurikar}, {Salgado}, {Sanghi}, {Saunders}, {Savchenko}, {Schwardt},
  {Seifert-Eckert}, {Shih}, {Jain}, {Shukla}, {Sick}, {Simpson},
  {Singanamalla}, {Singer}, {Singhal}, {Sinha}, {Sip{H{o}}cz}, {Spitler},
  {Stansby}, {Streicher}, {{{S}}umak}, {Swinbank}, {Taranu}, {Tewary},
  {Tremblay}, {Val-Borro}, {Van Kooten}, {Vasovi{'c}}, {Verma}, {de Miranda
  Cardoso}, {Williams}, {Wilson}, {Winkel}, {Wood-Vasey}, {Xue}, {Yoachim},
  {Zhang}, {Zonca}, \& {Astropy Project Contributors}}]{astropy:2022}
{Astropy Collaboration}, {Price-Whelan}, A.~M., {Lim}, P.~L., {et~al.} 2022,
  apj, 935, 167, \dodoi{10.3847/1538-4357/ac7c74}

\bibitem[{Barron {et~al.}(2022)Barron, Ahmed, Aguilar, Anderson, Baker, Barry,
  Beall, Bender, Benson, Besuner, Cecil, Chang, Chapman, Chesmore, Derylo,
  Doriese, Duff, Elleflot, Filippini, Flaugher, Gomez, Grimes, Gualtieri,
  Gullett, Haller, Henderson, Henke, Herbst, Huber, Hubmayr, Jonas, Joseph,
  King, Kovac, Kubik, Lisovenko, McMahon, Moncelsi, Nagy, Osherson, Reese,
  Ruhl, Sapozhnikov, Schillaci, Simon, Suzuki, Wang, Westbrook, Yefremenko, \&
  Zhang}]{Barron2022}
Barron, D.~R., Ahmed, Z., Aguilar, J., {et~al.} 2022,
  \dodoi{10.48550/arxiv.2208.02284}

\bibitem[{Behnel {et~al.}(2011)Behnel, Bradshaw, Citro, Dalcin, Seljebotn, \&
  Smith}]{cython}
Behnel, S., Bradshaw, R., Citro, C., {et~al.} 2011, Computing in Science
  Engineering, 13, 31, \dodoi{10.1109/MCSE.2010.118}

\bibitem[{Benson {et~al.}(2014)Benson, Ade, Ahmed, Allen, Arnold, Austermann,
  Bender, Bleem, Carlstrom, Chang, Cho, Cliche, Crawford, Cukierman, de~Haan,
  Dobbs, Dutcher, Everett, Gilbert, Halverson, Hanson, Harrington, Hattori,
  Henning, Hilton, Holder, Holzapfel, Irwin, Keisler, Knox, Kubik, Kuo, Lee,
  Leitch, Li, McDonald, Meyer, Montgomery, Myers, Natoli, Nguyen, Novosad,
  Padin, Pan, Pearson, Reichardt, Ruhl, Saliwanchik, Simard, Smecher, Sayre,
  Shirokoff, Stark, Story, Suzuki, Thompson, Tucker, Vanderlinde, Vieira,
  Vikhlinin, Wang, Yefremenko, \& Yoon}]{SPT_3G_1}
Benson, B.~A., Ade, P. A.~R., Ahmed, Z., {et~al.} 2014, in Millimeter,
  Submillimeter, and Far-Infrared Detectors and Instrumentation for Astronomy
  VII, ed. W.~S. Holland \& J.~Zmuidzinas, Vol. 9153, International Society for
  Optics and Photonics (SPIE), 552 -- 572, \dodoi{10.1117/12.2057305}

\bibitem[{Bleem {et~al.}(2022)Bleem, Crawford, Ansarinejad, Benson, Bocquet,
  Carlstrom, Chang, Chown, Crites, de~Haan, Dobbs, Everett, George, Gualtieri,
  Halverson, Holder, Holzapfel, Hrubes, Knox, Lee, Luong-Van, Marrone, McMahon,
  Meyer, Millea, Mocanu, Mohr, Natoli, Omori, Padin, Pryke, Raghunathan,
  Reichardt, Ruhl, Schaffer, Shirokoff, Staniszewski, Stark, Vieira, \&
  Williamson}]{Bleem2022}
Bleem, L.~E., Crawford, T.~M., Ansarinejad, B., {et~al.} 2022, The
  Astrophysical Journal Supplement Series, 258, 36,
  \dodoi{10.3847/1538-4365/AC35E9}

\bibitem[{Brown {et~al.}(2005)Brown, Castro, \& Taylor}]{Brown2005}
Brown, M.~L., Castro, P.~G., \& Taylor, A.~N. 2005, Monthly Notices of the
  Royal Astronomical Society, 360, 1262,
  \dodoi{10.1111/j.1365-2966.2005.09111.x}

\bibitem[{{Bucher} \& {Louis}(2012)}]{Bucher2012}
{Bucher}, M., \& {Louis}, T. 2012, \mnras, 424, 1694,
  \dodoi{10.1111/j.1365-2966.2012.21138.x}

\bibitem[{Chan {et~al.}(2017)Chan, Leistedt, Kitching, \&
  McEwen}]{curvelet-sd-wavelets}
Chan, J.~Y., Leistedt, B., Kitching, T.~D., \& McEwen, J.~D. 2017, IEEE
  Transactions on Signal Processing, 65, 5, \dodoi{10.1109/TSP.2016.2600506}

\bibitem[{Cheng {et~al.}(2020)Cheng, Ting, Menard, \& Bruna}]{Cheng2020}
Cheng, S., Ting, Y.~S., Menard, B., \& Bruna, J. 2020, Monthly Notices of the
  Royal Astronomical Society, 499, 5902, \dodoi{10.1093/MNRAS/STAA3165}

\bibitem[{{Choi} {et~al.}(2018){Choi}, {Austermann}, {Beall}, {Crowley},
  {Datta}, {Duff}, {Gallardo}, {Ho}, {Hubmayr}, {Koopman}, {Li}, {Nati},
  {Niemack}, {Page}, {Salatino}, {Simon}, {Staggs}, {Stevens}, {Ullom}, \&
  {Wollack}}]{PA56}
{Choi}, S.~K., {Austermann}, J., {Beall}, J.~A., {et~al.} 2018, Journal of Low
  Temperature Physics, 193, 267, \dodoi{10.1007/s10909-018-1982-4}

\bibitem[{Choi {et~al.}(2020)Choi, Hasselfield, Ho, Koopman, Lungu, Abitbol,
  Addison, Ade, Aiola, Alonso, Amiri, Amodeo, Angile, Austermann, Baildon,
  Battaglia, Beall, Bean, Becker, Bond, Bruno, Calabrese, Calafut, Campusano,
  Carrero, Chesmore, Cho, Clark, Cothard, Crichton, Crowley, Darwish, Datta,
  Denison, Devlin, Duell, Duff, Duivenvoorden, Dunkley, D{\"{u}}nner,
  Essinger-Hileman, Fankhanel, Ferraro, Fox, Fuzia, Gallardo, Gluscevic, Golec,
  Grace, Gralla, Guan, Hall, Halpern, Han, Hargrave, Henderson, Hensley, Hill,
  Hilton, Hilton, Hincks, Hlo{\v{z}}ek, Hubmayr, Huffenberger, Hughes, Infante,
  Irwin, Jackson, Klein, Knowles, Kosowsky, Lakey, Li, Li, Li, Lokken, Louis,
  MacInnis, Madhavacheril, Maldonado, Mallaby-Kay, Marsden, Maurin, McMahon,
  Menanteau, Moodley, Morton, Naess, Namikawa, Nati, Newburgh, Nibarger,
  Nicola, Niemack, Nolta, Orlowski-Sherer, Page, Pappas, Partridge, Phakathi,
  Prince, Puddu, Qu, Rivera, Robertson, Rojas, Salatino, Schaan, Schillaci,
  Schmitt, Sehgal, Sherwin, Sierra, Sievers, Sifon, Sikhosana, Simon, Spergel,
  Staggs, Stevens, Storer, Sunder, Switzer, Thorne, Thornton, Trac, Treu,
  Tucker, Vale, Engelen, Lanen, Vavagiakis, Wagoner, Wang, Ward, Wollack, Xu,
  Zago, \& Zhu}]{C20}
Choi, S.~K., Hasselfield, M., Ho, S.-P.~P., {et~al.} 2020, Journal of Cosmology
  and Astroparticle Physics, 2020, 45, \dodoi{10.1088/1475-7516/2020/12/045}

\bibitem[{Chown {et~al.}(2018)Chown, Omori, Aylor, Benson, Simard,
  Staniszewski, Stark, Story, Vanderlinde, Vieira, Williamson, \&
  Wu}]{Chown2018}
Chown, R., Omori, Y., Aylor, K., {et~al.} 2018, The Astrophysical Journal
  Supplement Series, 7, \dodoi{10.3847/1538-4365/aae694}

\bibitem[{Collette(2013)}]{h5py}
Collette, A. 2013, Python and HDF5 (O'Reilly).
\newblock \url{https://www.h5py.org/}

\bibitem[{Couchot {et~al.}(2017)Couchot, Henrot-Versill{\'{e}}, Perdereau,
  Plaszczynski, {Rouill{\'{e}} D'Orfeuil}, Spinelli, \& Tristram}]{Couchot2017}
Couchot, F., Henrot-Versill{\'{e}}, S., Perdereau, O., {et~al.} 2017, Astronomy
  \& Astrophysics, 602, A41, \dodoi{10.1051/0004-6361/201629815}

\bibitem[{Coulton {et~al.}(in prep.)Coulton, Madhavacheril, Hill, \&
  Duivenvoorden}]{Coulton23}
Coulton, W., Madhavacheril, M., Hill, C., \& Duivenvoorden, A. in prep.

\bibitem[{{Dahal} {et~al.}(2022){Dahal}, {Appel}, {Datta}, {Brewer}, {Ali},
  {Bennett}, {Bustos}, {Chan}, {Chuss}, {Cleary}, {Couto}, {Denis},
  {D{\"u}nner}, {Eimer}, {Espinoza}, {Essinger-Hileman}, {Golec}, {Harrington},
  {Helson}, {Iuliano}, {Karakla}, {Li}, {Marriage}, {McMahon}, {Miller},
  {Novack}, {N{\'u}{\~n}ez}, {Osumi}, {Padilla}, {Palma}, {Parker}, {Petroff},
  {Reeves}, {Rhoades}, {Rostem}, {Valle}, {Watts}, {Weiland}, {Wollack}, \&
  {Xu}}]{CLASS_2}
{Dahal}, S., {Appel}, J.~W., {Datta}, R., {et~al.} 2022, \apj, 926, 33,
  \dodoi{10.3847/1538-4357/ac397c}

\bibitem[{{Dahlen} \& {Simons}(2008)}]{Dahlen2008}
{Dahlen}, F.~A., \& {Simons}, F.~J. 2008, Geophysical Journal International,
  174, 774, \dodoi{10.1111/j.1365-246X.2008.03854.x}

\bibitem[{{De Belsunce} {et~al.}(2021){De Belsunce}, Gratton, Coulton, \&
  Efstathiou}]{DeBelsunce2021}
{De Belsunce}, R., Gratton, S., Coulton, W., \& Efstathiou, G. 2021, Monthly
  Notices of the Royal Astronomical Society, 507, 1072,
  \dodoi{10.1093/MNRAS/STAB2215}

\bibitem[{Delabrouille {et~al.}(2009)Delabrouille, Cardoso, {Le Jeune},
  Betoule, Fay, \& Guilloux}]{WMAP_NILC}
Delabrouille, J., Cardoso, J.-F., {Le Jeune}, M., {et~al.} 2009, Astronomy \&
  Astrophysics, 493, 835, \dodoi{10.1051/0004-6361:200810514}

\bibitem[{Dunkley {et~al.}(2009)Dunkley, Spergel, Komatsu, Hinshaw, Larson,
  Nolta, Odegard, Page, Bennett, Gold, Hill, Jarosik, Weiland, Halpern, Kogut,
  Limon, Meyer, Tucker, Wollack, \& Wright}]{Dunkley2009}
Dunkley, J., Spergel, D.~N., Komatsu, E., {et~al.} 2009, The Astrophysical
  Journal, 701, 1804, \dodoi{10.1088/0004-637X/701/2/1804}

\bibitem[{Dunkley {et~al.}(2013)Dunkley, Calabrese, Sievers, Addison,
  Battaglia, Battistelli, Bond, Das, Devlin, Dünner, Fowler, Gralla, Hajian,
  Halpern, Hasselfield, Hincks, Hlozek, Hughes, Irwin, Kosowsky, Louis,
  Marriage, Marsden, Menanteau, Moodley, Niemack, Nolta, Page, Partridge,
  Sehgal, Spergel, Staggs, Switzer, Trac, \& Wollack}]{Dunkley2013}
Dunkley, J., Calabrese, E., Sievers, J., {et~al.} 2013, Journal of Cosmology
  and Astroparticle Physics, 2013, 025, \dodoi{10.1088/1475-7516/2013/07/025}

\bibitem[{{D{\"u}nner} {et~al.}(2013){D{\"u}nner}, {Hasselfield}, {Marriage},
  {Sievers}, {Acquaviva}, {Addison}, {Ade}, {Aguirre}, {Amiri}, {Appel},
  {Barrientos}, {Battistelli}, {Bond}, {Brown}, {Burger}, {Calabrese},
  {Chervenak}, {Das}, {Devlin}, {Dicker}, {Bertrand Doriese}, {Dunkley},
  {Essinger-Hileman}, {Fisher}, {Gralla}, {Fowler}, {Hajian}, {Halpern},
  {Hern{\'a}ndez-Monteagudo}, {Hilton}, {Hilton}, {Hincks}, {Hlozek},
  {Huffenberger}, {Hughes}, {Hughes}, {Infante}, {Irwin}, {Baptiste Juin},
  {Kaul}, {Klein}, {Kosowsky}, {Lau}, {Limon}, {Lin}, {Louis}, {Lupton},
  {Marsden}, {Martocci}, {Mauskopf}, {Menanteau}, {Moodley}, {Moseley},
  {Netterfield}, {Niemack}, {Nolta}, {Page}, {Parker}, {Partridge}, {Quintana},
  {Reid}, {Sehgal}, {Sherwin}, {Spergel}, {Staggs}, {Swetz}, {Switzer},
  {Thornton}, {Trac}, {Tucker}, {Warne}, {Wilson}, {Wollack}, \& {Zhao}}]{D13}
{D{\"u}nner}, R., {Hasselfield}, M., {Marriage}, T.~A., {et~al.} 2013, \apj,
  762, 10, \dodoi{10.1088/0004-637X/762/1/10}

\bibitem[{{Efstathiou}(2006)}]{Efstathiou2006}
{Efstathiou}, G. 2006, \mnras, 370, 343,
  \dodoi{10.1111/j.1365-2966.2006.10486.x}

\bibitem[{Efstathiou \& Gratton(2021)}]{Efstathiou2021}
Efstathiou, G., \& Gratton, S. 2021, The Open Journal of Astrophysics, 4,
  27518, \dodoi{10.21105/ASTRO.1910.00483}

\bibitem[{{Eriksen} {et~al.}(2008){Eriksen}, {Jewell}, {Dickinson}, {Banday},
  {G{\'o}rski}, \& {Lawrence}}]{Eriksen2008}
{Eriksen}, H.~K., {Jewell}, J.~B., {Dickinson}, C., {et~al.} 2008, \apj, 676,
  10, \dodoi{10.1086/525277}

\bibitem[{Errard {et~al.}(2015)Errard, Ade, Akiba, Arnold, Atlas, Baccigalupi,
  Barron, Boettger, Borrill, Chapman, Chinone, Cukierman, Delabrouille, Dobbs,
  Ducout, Elleflot, Fabbian, Feng, Feeney, Gilbert, Goeckner-Wald, Halverson,
  Hasegawa, Hattori, Hazumi, Hill, Holzapfel, Hori, Inoue, Jaehnig, Jaffe,
  Jeong, Katayama, Kaufman, Keating, Kermish, Keskitalo, Kisner, {Le Jeune},
  Lee, Leitch, Leon, Linder, Matsuda, Matsumura, Miller, Myers, Navaroli,
  Nishino, Okamura, Paar, Peloton, Poletti, Puglisi, Rebeiz, Reichardt,
  Richards, Ross, Rotermund, Schenck, Sherwin, Siritanasak, Smecher, Stebor,
  Steinbach, Stompor, Suzuki, Tajima, Takakura, Tikhomirov, Tomaru, Whitehorn,
  Wilson, Yadav, \& Zahn}]{Errard2015}
Errard, J., Ade, P.~A., Akiba, Y., {et~al.} 2015, The Astrophysical Journal,
  809, 63, \dodoi{10.1088/0004-637X/809/1/63}

\bibitem[{Fowler {et~al.}(2007)Fowler, Niemack, Dicker, Aboobaker, Ade,
  Battistelli, Devlin, Fisher, Halpern, Hargrave, Hincks, Kaul, Klein, Lau,
  Limon, Marriage, Mauskopf, Page, Staggs, Swetz, Switzer, Thornton, \&
  Tucker}]{ACT_telescope}
Fowler, J.~W., Niemack, M.~D., Dicker, S.~R., {et~al.} 2007, Appl. Opt., 46,
  3444, \dodoi{10.1364/AO.46.003444}

\bibitem[{Freeman \& Adelson(1991)}]{Freeman1991}
Freeman, W.~T., \& Adelson, E.~H. 1991, IEEE Trans. Patt. Anal. and Machine
  Intell, 13, 891

\bibitem[{Galitzki {et~al.}(2018)Galitzki, Baildon, Barron, Lashner, Lee, Li,
  Limon, Lungu, Matsuda, Mauskopf, May, McCallum, McMahon, Nati, Niemack,
  Orlowski-Scherer, Parshley, Piccirillo, Rao, Salatino, Seibert, Sierra,
  Silva-Feaver, Simon, Staggs, Stevens, Suzuki, Teply, Thornton, Tsai, Ullom,
  Vavagiakis, Vissers, Westbrook, Wollack, Xu, Zhu, Raum, Beckman, Jeong, Ali,
  Arnold, Ashton, Austermann, Baccigalupi, Beall, Bruno, Bryan, Calisse,
  Chesmore, Chinone, Choi, Coppi, Crowley, Crowley, Cukierman, Devlin, Dicker,
  Dober, Duff, Dunkley, Fabbian, Gallardo, Gerbino, Goeckner-Wald, Golec,
  Gudmundsson, Healy, Henderson, Hill, Hilton, Ho, Howe, Hubmayr, Keating,
  Koopman, Kuichi, \& Kusaka}]{SO_instrument}
Galitzki, N., Baildon, T., Barron, D., {et~al.} 2018, in Millimeter,
  Submillimeter, and Far-Infrared Detectors and Instrumentation for Astronomy
  IX, ed. J.~Zmuidzinas \& J.-R. Gao, Vol. 10708 (SPIE), 3,
  \dodoi{10.1117/12.2312985}

\bibitem[{Garc{\'{i}}a-Garc{\'{i}}a {et~al.}(2019)Garc{\'{i}}a-Garc{\'{i}}a,
  Alonso, \& Bellini}]{namaster2019}
Garc{\'{i}}a-Garc{\'{i}}a, C., Alonso, D., \& Bellini, E. 2019, Journal of
  Cosmology and Astroparticle Physics, 2019, 043,
  \dodoi{10.1088/1475-7516/2019/11/043}

\bibitem[{Gomersall(2016)}]{pyfftw}
Gomersall, H. 2016.
\newblock \url{http://dx.doi.org/10.5281/zenodo.59508}

\bibitem[{{G{\'o}rski} {et~al.}(2005){G{\'o}rski}, {Hivon}, {Banday},
  {Wandelt}, {Hansen}, {Reinecke}, \& {Bartelmann}}]{HEALPix}
{G{\'o}rski}, K.~M., {Hivon}, E., {Banday}, A.~J., {et~al.} 2005, \apj, 622,
  759, \dodoi{10.1086/427976}

\bibitem[{Hanson {et~al.}(2009)Hanson, Rocha, \& G{\'{o}}rski}]{Hanson2009}
Hanson, D., Rocha, G., \& G{\'{o}}rski, K. 2009, Mon. Not. R. Astron. Soc, 400,
  2169, \dodoi{10.1111/j.1365-2966.2009.15614.x}

\bibitem[{Harris {et~al.}(2020)Harris, Millman, van~der Walt, Gommers,
  Virtanen, Cournapeau, Wieser, Taylor, Berg, Smith, Kern, Picus, Hoyer, van
  Kerkwijk, Brett, Haldane, del R{\'{i}}o, Wiebe, Peterson,
  G{\'{e}}rard-Marchant, Sheppard, Reddy, Weckesser, Abbasi, Gohlke, \&
  Oliphant}]{numpy}
Harris, C.~R., Millman, K.~J., van~der Walt, S.~J., {et~al.} 2020, Nature, 585,
  357, \dodoi{10.1038/s41586-020-2649-2}

\bibitem[{Henderson {et~al.}(2016)Henderson, Allison, Austermann, Baildon,
  Battaglia, Beall, Becker, {De Bernardis}, Bond, Calabrese, Choi, Coughlin,
  Crowley, Datta, Devlin, Duff, Dunkley, D{\"{u}}nner, van Engelen, Gallardo,
  Grace, Hasselfield, Hills, Hilton, Hincks, Hloẑek, Ho, Hubmayr,
  Huffenberger, Hughes, Irwin, Koopman, Kosowsky, Li, McMahon, Munson, Nati,
  Newburgh, Niemack, Niraula, Page, Pappas, Salatino, Schillaci, Schmitt,
  Sehgal, Sherwin, Sievers, Simon, Spergel, Staggs, Stevens, Thornton, {Van
  Lanen}, Vavagiakis, Ward, \& Wollack}]{AdvACT}
Henderson, S.~W., Allison, R., Austermann, J., {et~al.} 2016, Journal of Low
  Temperature Physics, 184, 772, \dodoi{10.1007/s10909-016-1575-z}

\bibitem[{Henning {et~al.}(2018)Henning, Sayre, Reichardt, Ade, Anderson,
  Austermann, Beall, Bender, Benson, Bleem, Carlstrom, Chang, Chiang, Cho,
  Citron, {Corbett Moran}, Crawford, Crites, {De Haan}, Dobbs, Everett,
  Gallicchio, George, Gilbert, Halverson, Harrington, Hilton, Holder,
  Holzapfel, Hoover, Hou, Hrubes, Huang, Hubmayr, Irwin, Keisler, Knox, Lee,
  Leitch, Li, Lowitz, Manzotti, Mcmahon, Meyer, Mocanu, Montgomery, Nadolski,
  Natoli, Nibarger, Novosad, Padin, Pryke, Ruhl, Saliwanchik, Schaffer,
  Sievers, Smecher, Stark, Story, Tucker, Vanderlinde, Veach, Vieira, Wang,
  Whitehorn, Wu, \& Yefremenko}]{Henning2018}
Henning, J.~W., Sayre, J.~T., Reichardt, C.~L., {et~al.} 2018, The
  Astrophysical Journal, 852, 97, \dodoi{10.3847/1538-4357/aa9ff4}

\bibitem[{Hivon {et~al.}(2002)Hivon, Gorski, Netterfield, Crill, Prunet, \&
  Hansen}]{MASTER}
Hivon, E., Gorski, K.~M., Netterfield, C.~B., {et~al.} 2002, The Astrophysical
  Journal, 567, 2–17, \dodoi{10.1086/338126}

\bibitem[{Ho {et~al.}(2017)Ho, Austermann, Beall, Choi, Cothard, Crowley,
  Datta, Devlin, Duff, Gallardo, Hasselfield, Henderson, Hilton, Hubmayr,
  Koopman, Li, McMahon, Niemack, Salatino, Simon, Staggs, Ward, Ullom,
  Vavagiakis, \& Wollack}]{PA4}
Ho, S.-P.~P., Austermann, J., Beall, J.~A., {et~al.} 2017, in Millimeter,
  Submillimeter, and Far-Infrared Detectors and Instrumentation for Astronomy
  VIII, ed. W.~S. Holland \& J.~Zmuidzinas, Vol. 9914, International Society
  for Optics and Photonics (SPIE), 991418, \dodoi{10.1117/12.2233113}

\bibitem[{Hunter(2007)}]{matplotlib}
Hunter, J.~D. 2007, Computing in Science \& Engineering, 9, 90,
  \dodoi{10.1109/MCSE.2007.55}

\bibitem[{{Jarosik} {et~al.}(2007){Jarosik}, {Barnes}, {Greason}, {Hill},
  {Nolta}, {Odegard}, {Weiland}, {Bean}, {Bennett}, {Dor{\'e}}, {Halpern},
  {Hinshaw}, {Kogut}, {Komatsu}, {Limon}, {Meyer}, {Page}, {Spergel}, {Tucker},
  {Wollack}, \& {Wright}}]{Jarosik2007}
{Jarosik}, N., {Barnes}, C., {Greason}, M.~R., {et~al.} 2007, \apjs, 170, 263,
  \dodoi{10.1086/513697}

\bibitem[{{Kang} {et~al.}(2018){Kang}, {Ade}, {Ahmed}, {Aikin}, {Alexander},
  {Barkats}, {Benton}, {Bischoff}, {Bock}, {Boenish}, {Bowens-Rubin}, {Brevik},
  {Buder}, {Bullock}, {Buza}, {Connors}, {Cornelison}, {Crill}, {Crumrine},
  {Dierickx}, {Duband}, {Dvorkin}, {Filippini}, {Fliescher}, {Grayson}, {Hall},
  {Halpern}, {Harrison}, {Hildebrandt}, {Hilton}, {Hui}, {Irwin}, {Karkare},
  {Karpel}, {Kaufman}, {Keating}, {Kefeli}, {Kernasovskiy}, {Kovac}, {Kuo},
  {Larsen}, {Lau}, {Leitch}, {Lueker}, {Megerian}, {Moncelsi}, {Namikawa},
  {Netterfield}, {Nguyen}, {O'Brient}, {Ogburn}, {Palladino}, {Pryke},
  {Racine}, {Richter}, {Schillaci}, {Schwarz}, {Sheehy}, {Soliman}, {St.
  Germaine}, {Staniszewski}, {Steinbach}, {Sudiwala}, {Teply}, {Thompson},
  {Tolan}, {Tucker}, {Turner}, {Umilt{\`a}}, {Vieregg}, {Wandui}, {Weber},
  {Wiebe}, {Willmert}, {Wong}, {Wu}, {Yang}, {Yoon}, \& {Zhang}}]{BICEP3_2}
{Kang}, J.~H., {Ade}, P.~A.~R., {Ahmed}, Z., {et~al.} 2018, in Society of
  Photo-Optical Instrumentation Engineers (SPIE) Conference Series, Vol. 10708,
  Millimeter, Submillimeter, and Far-Infrared Detectors and Instrumentation for
  Astronomy IX, 107082N, \dodoi{10.1117/12.2313854}

\bibitem[{{Kesden} {et~al.}(2003){Kesden}, {Cooray}, \&
  {Kamionkowski}}]{Kesden2003_N0}
{Kesden}, M., {Cooray}, A., \& {Kamionkowski}, M. 2003, \prd, 67, 123507,
  \dodoi{10.1103/PhysRevD.67.123507}

\bibitem[{Lam {et~al.}(2015)Lam, Pitrou, \& Seibert}]{numba}
Lam, S.~K., Pitrou, A., \& Seibert, S. 2015, in Proceedings of the Second
  Workshop on the LLVM Compiler Infrastructure in HPC, LLVM '15 (New York, NY,
  USA: Association for Computing Machinery), \dodoi{10.1145/2833157.2833162}

\bibitem[{Leistedt {et~al.}(2013)Leistedt, McEwen, Vandergheynst, \&
  Wiaux}]{sd-wavelets}
Leistedt, B., McEwen, J.~D., Vandergheynst, P., \& Wiaux, Y. 2013, Astronomy \&
  Astrophysics, 558, A128, \dodoi{10.1051/0004-6361/201220729}

\bibitem[{Li {et~al.}(2021)Li, Louis, Calabrese, Jense, Alonso, Bond, Choi,
  Dunkley, Fabbian, Garrido, Jaffe, Madhavacheril, Meerburg, Natale, \&
  Qu}]{Li2021}
Li, Z., Louis, T., Calabrese, E., {et~al.} 2021,
  \dodoi{10.48550/arxiv.2112.13839}

\bibitem[{Louis {et~al.}(2020)Louis, Naess, Garrido, \& Challinor}]{Louis2020}
Louis, T., Naess, S., Garrido, X., \& Challinor, A. 2020, Phys. Rev. D, 102,
  123538, \dodoi{10.1103/PhysRevD.102.123538}

\bibitem[{Louis {et~al.}(2017)Louis, Grace, Hasselfield, Lungu, Maurin,
  Addison, Ade, Aiola, Allison, Amiri, Angile, Battaglia, Beall, {De
  Bernardis}, Bond, Britton, Calabrese, Cho, Choi, Coughlin, Crichton, Crowley,
  Datta, Devlin, Dicker, Dunkley, D{\"{u}}nner, Ferraro, Fox, Gallardo, Gralla,
  Halpern, Henderson, Hill, Hilton, Hilton, Hincks, Hlozek, {Patty Ho}, Huang,
  Hubmayr, Huffenberger, Hughes, Infante, Irwin, Kasanda, Klein, Koopman,
  Kosowsky, Li, Madhavacheril, Marriage, McMahon, Menanteau, Moodley, Munson,
  Naess, Nati, Newburgh, Nibarger, Niemack, Nolta, Nu{\~{n}}ez, Page, Pappas,
  Partridge, Rojas, Schaan, Schmitt, Sehgal, Sherwin, Sievers, Simon, Spergel,
  Staggs, Switzer, Thornton, Trac, Treu, Tucker, Engelen, Ward, \&
  Wollack}]{L17}
Louis, T., Grace, E., Hasselfield, M., {et~al.} 2017, Journal of Cosmology and
  Astroparticle Physics, 2017, 031, \dodoi{10.1088/1475-7516/2017/06/031}

\bibitem[{Madhavacheril {et~al.}(2021)Madhavacheril, Smith, Sherwin, \&
  Naess}]{cross-estimator}
Madhavacheril, M.~S., Smith, K.~M., Sherwin, B.~D., \& Naess, S. 2021, Journal
  of Cosmology and Astroparticle Physics, 2021, 028,
  \dodoi{10.1088/1475-7516/2021/05/028}

\bibitem[{Madhavacheril {et~al.}(2020)Madhavacheril, Hill, Naess, Addison,
  Aiola, Baildon, Battaglia, Bean, Bond, Calabrese, Calafut, Choi, Darwish,
  Datta, Devlin, Dunkley, D{\"{u}}nner, Ferraro, Gallardo, Gluscevic, Halpern,
  Han, Hasselfield, Hilton, Hincks, Hlo{\v{z}}ek, Ho, Huffenberger, Hughes,
  Koopman, Kosowsky, Lokken, Louis, Lungu, Macinnis, Maurin, Mcmahon, Moodley,
  Nati, Niemack, Page, Partridge, Robertson, Sehgal, Schaan, Schillaci,
  Sherwin, Sif{\'{o}}n, Simon, Spergel, Staggs, Storer, {Van Engelen},
  Vavagiakis, Wollack, Xu, \& Henry}]{Mv20}
Madhavacheril, M.~S., Hill, J.~C., Naess, S., {et~al.} 2020, Physical Review D,
  102, \dodoi{10.1103/PhysRevD.102.023534}

\bibitem[{Mallaby-Kay {et~al.}(2021)Mallaby-Kay, Atkins, Aiola, Amodeo,
  Austermann, Beall, Becker, Bond, Calabrese, Chesmore, Choi, Crowley, Darwish,
  Denison, Devlin, Duff, Duivenvoorden, Dunkley, Ferraro, Fichman, Gallardo,
  Golec, Guan, Han, Hasselfield, Hill, Hilton, Hilton, Hlo{\v{z}}ek, Hubmayr,
  Huffenberger, Hughes, Koopman, Louis, MacInnis, Madhavacheril, McMahon,
  Moodley, Naess, Namikawa, Nati, Newburgh, Nibarger, Niemack, Page, Salatino,
  Schaan, Schillaci, Sehgal, Sherwin, Sif{\'{o}}n, Simon, Staggs, Storer,
  Ullom, Engelen, Lanen, Vale, Wollack, \& Xu}]{MK21}
Mallaby-Kay, M., Atkins, Z., Aiola, S., {et~al.} 2021, The Astrophysical
  Journal Supplement Series, 255, 11, \dodoi{10.3847/1538-4365/ABFCC4}

\bibitem[{McEwen {et~al.}(2018)McEwen, Durastanti, \&
  Wiaux}]{directional-sd-wavelets}
McEwen, J.~D., Durastanti, C., \& Wiaux, Y. 2018, Applied and Computational
  Harmonic Analysis, 44, 59, \dodoi{10.1016/J.ACHA.2016.03.009}

\bibitem[{Millea {et~al.}(2021)Millea, Daley, Chou, Anderes, Ade, Anderson,
  Austermann, Avva, Beall, Bender, Benson, Bianchini, Bleem, Carlstrom, Chang,
  Chaubal, Chiang, Citron, Moran, Crawford, Crites, de~Haan, Dobbs, Everett,
  Gallicchio, George, Goeckner-Wald, Guns, Gupta, Halverson, Henning, Hilton,
  Holder, Holzapfel, Hrubes, Huang, Hubmayr, Irwin, Knox, Lee, Li, Lowitz,
  McMahon, Meyer, Mocanu, Montgomery, Natoli, Nibarger, Noble, Novosad, Omori,
  Padin, Patil, Pryke, Reichardt, Ruhl, Saliwanchik, Schaffer, Sievers,
  Smecher, Stark, Thorne, Tucker, Veach, Vieira, Wang, Whitehorn, Wu, \&
  Yefremenko}]{Millea2021}
Millea, M., Daley, C.~M., Chou, T.-L., {et~al.} 2021, The Astrophysical
  Journal, 922, 259, \dodoi{10.3847/1538-4357/AC02BB}

\bibitem[{{Morris} {et~al.}(2022){Morris}, {Bustos}, {Calabrese}, {Choi},
  {Duivenvoorden}, {Dunkley}, {D{\"u}nner}, {Gallardo}, {Hasselfield},
  {Hincks}, {Mroczkowski}, {Naess}, {Niemack}, {Page}, {Partridge}, {Salatino},
  {Staggs}, {Treu}, {Wollack}, \& {Xu}}]{Morris2022}
{Morris}, T.~W., {Bustos}, R., {Calabrese}, E., {et~al.} 2022, \prd, 105,
  042004, \dodoi{10.1103/PhysRevD.105.042004}

\bibitem[{Naess {et~al.}(2014)Naess, Hasselfield, McMahon, Niemack, Addison,
  Ade, Allison, Amiri, Battaglia, Beall, {De Bernardis}, Bond, Britton,
  Calabrese, Cho, Coughlin, Crichton, Das, Datta, Devlin, Dicker, Dunkley,
  D{\"{u}}nner, Fowler, Fox, Gallardo, Grace, Gralla, Hajian, Halpern,
  Henderson, Hill, Hilton, Hilton, Hincks, Hlozek, Ho, Hubmayr, Huffenberger,
  Hughes, Infante, Irwin, Jackson, Kasanda, Klein, Koopman, Kosowsky, Li,
  Louis, Lungu, Madhavacheril, Marriage, Maurin, Menanteau, Moodley, Munson,
  Newburgh, Nibarger, Nolta, Page, Pappas, Partridge, Rojas, Schmitt, Sehgal,
  Sherwin, Sievers, Simon, Spergel, Staggs, Switzer, Thornton, Trac, Tucker,
  Uehara, Engelen, Ward, \& Wollack}]{N14}
Naess, S., Hasselfield, M., McMahon, J., {et~al.} 2014, Journal of Cosmology
  and Astroparticle Physics, 2014, 007, \dodoi{10.1088/1475-7516/2014/10/007}

\bibitem[{Naess {et~al.}(2020)Naess, Aiola, Austermann, Battaglia, Beall,
  Becker, Bond, Calabrese, Choi, Cothard, Crowley, Darwish, Datta, Denison,
  Devlin, Duell, Duff, Duivenvoorden, Dunkley, D{\"{u}}nner, Fox, Gallardo,
  Halpern, Han, Hasselfield, Hill, Hilton, Hilton, Hincks, Hlo{\v{z}}ek, Ho,
  Hubmayr, Huffenberger, Hughes, Kosowsky, Louis, Madhavacheril, McMahon,
  Moodley, Nati, Nibarger, Niemack, Page, Partridge, Salatino, Schaan,
  Schillaci, Schmitt, Sherwin, Sehgal, Sif{\'{o}}n, Spergel, Staggs, Stevens,
  Storer, Ullom, Vale, Engelen, Lanen, Vavagiakis, Wollack, \& Xu}]{N20}
Naess, S., Aiola, S., Austermann, J.~E., {et~al.} 2020, Journal of Cosmology
  and Astroparticle Physics, 2020, 46, \dodoi{10.1088/1475-7516/2020/12/046}

\bibitem[{{Planck Collaboration I} {et~al.}(2020){Planck Collaboration I},
  {Aghanim, N.}, {Akrami, Y.}, {Arroja, F.}, {Ashdown, M.}, {Aumont, J.},
  {Baccigalupi, C.}, {Ballardini, M.}, {Banday, A. J.}, {Barreiro, R. B.},
  {Bartolo, N.}, {Basak, S.}, {Battye, R.}, {Benabed, K.}, {Bernard, J.-P.},
  {Bersanelli, M.}, {Bielewicz, P.}, {Bock, J. J.}, {Bond, J. R.}, {Borrill,
  J.}, {Bouchet, F. R.}, {Boulanger, F.}, {Bucher, M.}, {Burigana, C.},
  {Butler, R. C.}, {Calabrese, E.}, {Cardoso, J.-F.}, {Carron, J.}, {Casaponsa,
  B.}, {Challinor, A.}, {Chiang, H. C.}, {Colombo, L. P. L.}, {Combet, C.},
  {Contreras, D.}, {Crill, B. P.}, {Cuttaia, F.}, {de Bernardis, P.}, {de
  Zotti, G.}, {Delabrouille, J.}, {Delouis, J.-M.}, {D{\'{e}}sert, F.-X.}, {Di
  Valentino, E.}, {Dickinson, C.}, {Diego, J. M.}, {Donzelli, S.}, {Dor{\'{e}},
  O.}, {Douspis, M.}, {Ducout, A.}, {Dupac, X.}, {Efstathiou, G.}, {Elsner,
  F.}, {En{\ss}lin, T. A.}, {Eriksen, H. K.}, {Falgarone, E.}, {Fantaye, Y.},
  {Fergusson, J.}, {Fernandez-Cobos, R.}, {Finelli, F.}, {Forastieri, F.},
  {Frailis, M.}, {Franceschi, E.}, {Frolov, A.}, {Galeotta, S.}, {Galli, S.},
  {Ganga, K.}, {G{\'{e}}nova-Santos, R. T.}, {Gerbino, M.}, {Ghosh, T.},
  {Gonz{\'{a}}lez-Nuevo, J.}, {G{\'{o}}rski, K. M.}, {Gratton, S.}, {Gruppuso,
  A.}, {Gudmundsson, J. E.}, {Hamann, J.}, {Handley, W.}, {Hansen, F. K.},
  {Helou, G.}, {Herranz, D.}, {Hildebrandt, S. R.}, {Hivon, E.}, {Huang, Z.},
  {Jaffe, A. H.}, {Jones, W. C.}, {Karakci, A.}, {Keih{\"{a}}nen, E.},
  {Keskitalo, R.}, {Kiiveri, K.}, {Kim, J.}, {Kisner, T. S.}, {Knox, L.},
  {Krachmalnicoff, N.}, {Kunz, M.}, {Kurki-Suonio, H.}, {Lagache, G.},
  {Lamarre, J.-M.}, {Langer, M.}, {Lasenby, A.}, {Lattanzi, M.}, {Lawrence, C.
  R.}, {Le Jeune, M.}, {Leahy, J. P.}, {Lesgourgues, J.}, {Levrier, F.},
  {Lewis, A.}, {Liguori, M.}, {Lilje, P. B.}, {Lilley, M.}, {Lindholm, V.},
  {L{\'{o}}pez-Caniego, M.}, {Lubin, P. M.}, {Ma, Y.-Z.},
  {Mac{\'{i}}as-P{\'{e}}rez, J. F.}, {Maggio, G.}, {Maino, D.}, {Mandolesi,
  N.}, {Mangilli, A.}, {Marcos-Caballero, A.}, {Maris, M.}, {Martin, P. G.},
  {Martinelli, M.}, {Mart{\'{i}}nez-Gonz{\'{a}}lez, E.}, {Matarrese, S.},
  {Mauri, N.}, {McEwen, J. D.}, {Meerburg, P. D.}, {Meinhold, P. R.},
  {Melchiorri, A.}, {Mennella, A.}, {Migliaccio, M.}, {Millea, M.}, {Mitra,
  S.}, {Miville-Desch{\^{e}}nes, M.-A.}, {Molinari, D.}, {Moneti, A.},
  {Montier, L.}, {Morgante, G.}, {Moss, A.}, {Mottet, S.}, {M{\"{u}}nchmeyer,
  M.}, {Natoli, P.}, {N{\o}rgaard-Nielsen, H. U.}, {Oxborrow, C. A.}, {Pagano,
  L.}, {Paoletti, D.}, {Partridge, B.}, {Patanchon, G.}, {Pearson, T. J.},
  {Peel, M.}, {Peiris, H. V.}, {Perrotta, F.}, {Pettorino, V.}, {Piacentini,
  F.}, {Polastri, L.}, {Polenta, G.}, {Puget, J.-L.}, {Rachen, J. P.},
  {Reinecke, M.}, {Remazeilles, M.}, {Renault, C.}, {Renzi, A.}, {Rocha, G.},
  {Rosset, C.}, {Roudier, G.}, {Rubi{\~{n}}o-Mart{\'{i}}n, J. A.},
  {Ruiz-Granados, B.}, {Salvati, L.}, {Sandri, M.}, {Savelainen, M.}, {Scott,
  D.}, {Shellard, E. P. S.}, {Shiraishi, M.}, {Sirignano, C.}, {Sirri, G.},
  {Spencer, L. D.}, {Sunyaev, R.}, {Suur-Uski, A.-S.}, {Tauber, J. A.},
  {Tavagnacco, D.}, {Tenti, M.}, {Terenzi, L.}, {Toffolatti, L.}, {Tomasi, M.},
  {Trombetti, T.}, {Valiviita, J.}, {Van Tent, B.}, {Vibert, L.}, {Vielva, P.},
  {Villa, F.}, {Vittorio, N.}, {Wandelt, B. D.}, {Wehus, I. K.}, {White, M.},
  {White, S. D. M.}, {Zacchei, A.}, \& {Zonca, A.}}]{Planck-overview:2018}
{Planck Collaboration I}, {Aghanim, N.}, {Akrami, Y.}, {et~al.} 2020, Astronomy
  \& Astrophysics, 641, A1, \dodoi{10.1051/0004-6361/201833880}

\bibitem[{{Planck Collaboration III} {et~al.}(2020){Planck Collaboration III},
  Aghanim, Akrami, Ashdown, Aumont, Baccigalupi, Ballardini, Banday, Barreiro,
  Bartolo, Basak, Benabed, Bernard, Bersanelli, Bielewicz, Bond, Borrill,
  Bouchet, Boulanger, Bucher, Burigana, Calabrese, Cardoso, Carron, Challinor,
  Chiang, Colombo, Combet, Couchot, Crill, Cuttaia, {De Bernardis}, {De Rosa},
  {De Zotti}, Delabrouille, Delouis, {Di Valentino}, Diego, Dor{\'{e}},
  Douspis, Ducout, Dupac, Efstathiou, Elsner, En{\ss}lin, Eriksen, Falgarone,
  Fantaye, Finelli, Frailis, Fraisse, Franceschi, Frolov, Galeotta, Galli,
  Ganga, G{\'{e}}nova-Santos, Gerbino, Ghosh, Gonz{\'{a}}lez-Nuevo,
  G{\'{o}}rski, Gratton, Gruppuso, Gudmundsson, Handley, Hansen,
  Henrot-Versill{\'{e}}, Herranz, Hivon, Huang, Jaffe, Jones, Karakci,
  Keih{\"{a}}nen, Keskitalo, Kiiveri, Kim, Kisner, Krachmalnicoff, Kunz,
  Kurki-Suonio, Lagache, Lamarre, Lasenby, Lattanzi, Lawrence, Levrier,
  Liguori, Lilje, Lindholm, L{\'{o}}pez-Caniego, Ma, Maci{\'{a}}s-P{\'{e}}rez,
  Maggio, Maino, Mandolesi, Mangilli, Martin, Mart{\'{i}}nez-Gonz{\'{a}}lez,
  Matarrese, Mauri, McEwen, Melchiorri, Mennella, Migliaccio,
  Miville-Desch{\^{e}}nes, Molinari, Moneti, Montier, Morgante, Moss, Mottet,
  Natoli, Pagano, Paoletti, Partridge, Patanchon, Patrizii, Perdereau,
  Perrotta, Pettorino, Piacentini, Puget, Rachen, Reinecke, Remazeilles, Renzi,
  Rocha, Roudier, Salvati, Sandri, Savelainen, Scott, Sirignano, Sirri,
  Spencer, Sunyaev, Suur-Uski, Tauber, Tavagnacco, Tenti, Toffolatti, Tomasi,
  Tristram, Trombetti, Valiviita, Vansyngel, {Van Tent}, Vibert, Vielva, Villa,
  Vittorio, Wandelt, Wehus, \& Zonca}]{FFP10}
{Planck Collaboration III}, Aghanim, N., Akrami, Y., {et~al.} 2020, Astronomy
  \& Astrophysics, 641, A3, \dodoi{10.1051/0004-6361/201832909}

\bibitem[{{Planck Collaboration V} {et~al.}(2020){Planck Collaboration V},
  Aghanim, Akrami, Ashdown, Aumont, Baccigalupi, Ballardini, Banday, Barreiro,
  Bartolo, Basak, Benabed, Bernard, Bersanelli, Bielewicz, Bock, Bond, Borrill,
  Bouchet, Boulanger, Bucher, Burigana, Butler, Calabrese, Cardoso, Carron,
  Casaponsa, Challinor, Chiang, Colombo, Combet, Crill, Cuttaia, {De
  Bernardis}, {De Rosa}, {De Zotti}, Delabrouille, Delouis, {Di Valentino},
  Diego, Dor{\'{e}}, Douspis, Ducout, Dupac, Dusini, Efstathiou, Elsner,
  En{\ss}lin, Eriksen, Fantaye, Fernandez-Cobos, Finelli, Frailis, Fraisse,
  Franceschi, Frolov, Galeotta, Galli, Ganga, G{\'{e}}nova-Santos, Gerbino,
  Ghosh, Giraud-H{\'{e}}raud, Gonz{\'{a}}lez-Nuevo, G{\'{o}}rski, Gratton,
  Gruppuso, Gudmundsson, Hamann, Handley, Hansen, Herranz, Hivon, Huang, Jaffe,
  Jones, Keih{\"{a}}nen, Keskitalo, Kiiveri, Kim, Kisner, Krachmalnicoff, Kunz,
  Kurki-Suonio, Lagache, Lamarre, Lasenby, Lattanzi, Lawrence, {Le Jeune},
  Levrier, Lewis, Liguori, Lilje, Lilley, Lindholm, L{\'{o}}pez-Caniego, Lubin,
  Ma, Maci{\'{a}}s-P{\'{e}}rez, Maggio, Maino, Mandolesi, Mangilli,
  Marcos-Caballero, Maris, Martin, Mart{\'{i}}nez-Gonz{\'{a}}lez, Matarrese,
  Mauri, McEwen, Meinhold, Melchiorri, Mennella, Migliaccio, Millea,
  Miville-Desch{\^{e}}nes, Molinari, Moneti, Montier, Morgante, Moss, Natoli,
  N{\o}rgaard-Nielsen, Pagano, Paoletti, Partridge, Patanchon, Peiris,
  Perrotta, Pettorino, Piacentini, Polenta, Puget, Rachen, Reinecke,
  Remazeilles, Renzi, Rocha, Rosset, Roudier, Rubi{\~{n}}o-Mart{\'{i}}n,
  Ruiz-Granados, Salvati, Sandri, Savelainen, Scott, Shellard, Sirignano,
  Sirri, Spencer, Sunyaev, Suur-Uski, Tauber, Tavagnacco, Tenti, Toffolatti,
  Tomasi, Trombetti, Valiviita, {Van Tent}, Vielva, Villa, Vittorio, Wandelt,
  Wehus, Zacchei, \& Zonca}]{planckv18}
{Planck Collaboration V}, Aghanim, N., Akrami, Y., {et~al.} 2020, Astronomy \&
  Astrophysics, 641, A5, \dodoi{10.1051/0004-6361/201936386}

\bibitem[{{Planck Collaboration VI} {et~al.}(2020){Planck Collaboration VI},
  Aghanim, Akrami, Ashdown, Aumont, Baccigalupi, Ballardini, Banday, Barreiro,
  Bartolo, Basak, Battye, Benabed, Bernard, Bersanelli, Bielewicz, Bock, Bond,
  Borrill, Bouchet, Boulanger, Bucher, Burigana, Butler, Calabrese, Cardoso,
  Carron, Challinor, Chiang, Chluba, Colombo, Combet, Contreras, Crill,
  Cuttaia, de~Bernardis, de~Zotti, Delabrouille, Delouis, {Di Valentino},
  Diego, Dor{\'{e}}, Douspis, Ducout, Dupac, Dusini, Efstathiou, Elsner,
  Ensslin, Eriksen, Fantaye, Farhang, Fergusson, Fernandez-Cobos, Finelli,
  Forastieri, Frailis, Fraisse, Franceschi, Frolov, Galeotta, Galli, Ganga,
  G{\'{e}}nova-Santos, Gerbino, Ghosh, Gonz{\'{a}}lez-Nuevo, G{\'{o}}rski,
  Gratton, Gruppuso, Gudmundsson, Hamann, Handley, Hansen, Herranz,
  Hildebrandt, Hivon, Huang, Jaffe, Jones, Karakci, Keih{\"{a}}nen, Keskitalo,
  Kiiveri, Kim, Kisner, Knox, Krachmalnicoff, Kunz, Kurki-Suonio, Lagache,
  Lamarre, Lasenby, Lattanzi, Lawrence, {Le Jeune}, Lemos, Lesgourgues,
  Levrier, Lewis, Liguori, Lilje, Lilley, Lindholm, L{\'{o}}pez-Caniego, Lubin,
  Ma, Mac\'ias-P{\'{e}}rez, Maggio, Maino, Mandolesi, Mangilli,
  Marcos-Caballero, Maris, Martin, Martinelli, Mart\'inez-Gonz{\'{a}}lez,
  Matarrese, Mauri, McEwen, Meinhold, Melchiorri, Mennella, Migliaccio, Millea,
  Mitra, Miville-Desch{\^{e}}nes, Molinari, Montier, Morgante, Moss, Natoli,
  Norgaard-Nielsen, Pagano, Paoletti, Partridge, Patanchon, Peiris, Perrotta,
  Pettorino, Piacentini, Polastri, Polenta, Puget, Rachen, Reinecke,
  Remazeilles, Renzi, Rocha, Rosset, Roudier, Rubi{\~{n}}o-Mart\'\in,
  Ruiz-Granados, Salvati, Sandri, Savelainen, Scott, Shellard, Sirignano,
  Sirri, Spencer, Sunyaev, Suur-Uski, Tauber, Tavagnacco, Tenti, Toffolatti,
  Tomasi, Trombetti, Valenziano, Valiviita, {Van Tent}, Vibert, Vielva, Villa,
  Vittorio, Wandelt, Wehus, White, White, Zacchei, \&
  Zonca}]{Planck-cosmology:2018}
{Planck Collaboration VI}, Aghanim, N., Akrami, Y., {et~al.} 2020, Astronomy \&
  Astrophysics, 641, A6, \dodoi{10.1051/0004-6361/201833910}

\bibitem[{{Planck Collaboration XI} {et~al.}(2016){Planck Collaboration XI},
  Aghanim, Arnaud, Ashdown, Aumont, Baccigalupi, Banday, Barreiro, Bartlett,
  Bartolo, Battaner, Benabed, Beno{\^{i}}t, Benoit-L{\'{e}}vy, Bernard,
  Bersanelli, Bielewicz, Bock, Bonaldi, Bonavera, Bond, Borrill, Bouchet,
  Boulanger, Bucher, Burigana, Butler, Calabrese, Cardoso, Catalano, Challinor,
  Chiang, Christensen, Clements, Colombo, Combet, Coulais, Crill, Curto,
  Cuttaia, Danese, Davies, Davis, {De Bernardis}, {De Rosa}, {De Zotti},
  Delabrouille, D{\'{e}}sert, {Di Valentino}, Dickinson, Diego, Dolag, Dole,
  Donzelli, Dor{\'{e}}, Douspis, Ducout, Dunkley, Dupac, Efstathiou, Elsner,
  En{\ss}lin, Eriksen, Fergusson, Finelli, Forni, Frailis, Fraisse, Franceschi,
  Frejsel, Galeotta, Galli, Ganga, Gauthier, Gerbino, Giard, Gjerl{\o}w,
  Gonz{\'{a}}lez-Nuevo, G{\'{o}}rski, Gratton, Gregorio, Gruppuso, Gudmundsson,
  Hamann, Hansen, Harrison, Helou, Henrot-Versill{\'{e}},
  Hern{\'{a}}ndez-Monteagudo, Herranz, Hildebrandt, Hivon, Holmes, Hornstrup,
  Huffenberger, Hurier, Jaffe, Jones, Juvela, Keih{\"{a}}nen, Keskitalo,
  Kiiveri, Knoche, Knox, Kunz, Kurki-Suonio, Lagache, L{\"{a}}hteenm{\"{a}}ki,
  Lamarre, Lasenby, Lattanzi, Lawrence, {Le Jeune}, Leonardi, Lesgourgues,
  Levrier, Lewis, Liguori, Lilje, Lilley, Linden-V{\o}rnle, Lindholm,
  L{\'{o}}pez-Caniego, Maci{\'{a}}s-P{\'{e}}rez, Maffei, Maggio, Maino,
  Mandolesi, Mangilli, Maris, Martin, Mart{\'{i}}nez-Gonz{\'{a}}lez, Masi,
  Matarrese, Meinhold, Melchiorri, Migliaccio, Millea, Mitra,
  Miville-Desch{\^{e}}nes, Moneti, Montier, Morgante, Mortlock, Mottet, Munshi,
  Murphy, Narimani, Naselsky, Nati, Natoli, Noviello, Novikov, Novikov,
  Oxborrow, Paci, Pagano, Pajot, Paoletti, Partridge, Pasian, Patanchon,
  Pearson, Perdereau, Perotto, Pettorino, Piacentini, Piat, Pierpaoli,
  Pietrobon, Plaszczynski, Pointecouteau, Polenta, Ponthieu, Pratt, Prunet,
  Puget, Rachen, Reinecke, Remazeilles, Renault, Renzi, Ristorcelli, Rocha,
  Rossetti, Roudier, {Rouill{\'{e}} D'orfeuil}, Rubin{\~{o}}-Mart{\'{i}}n,
  Rusholme, Salvati, Sandri, Santos, Savelainen, Savini, Scott, Serra, Spencer,
  Spinelli, Stolyarov, Stompor, Sunyaev, Sutton, Suur-Uski, Sygnet, Tauber,
  Terenzi, Toffolatti, Tomasi, Tristram, Trombetti, Tucci, Tuovinen, Umana,
  Valenziano, Valiviita, {Van Tent}, Vielva, Villa, Wade, Wandelt, Wehus, Yvon,
  Zacchei, \& Zonca}]{planckxi15}
{Planck Collaboration XI}, Aghanim, N., Arnaud, M., {et~al.} 2016, Astronomy \&
  Astrophysics, 594, A11, \dodoi{10.1051/0004-6361/201526926}

\bibitem[{{Planck Collaboration XII} {et~al.}(2016){Planck Collaboration XII},
  Ade, Aghanim, Arnaud, Ashdown, Aumont, Baccigalupi, Banday, Barreiro,
  Bartlett, Bartolo, Battaner, Benabed, Beno{\^{i}}t, Benoit-L{\'{e}}vy,
  Bernard, Bersanelli, Bielewicz, Bock, Bonaldi, Bonavera, Bond, Borrill,
  Bouchet, Boulanger, Bucher, Burigana, Butler, Calabrese, Cardoso, Castex,
  Catalano, Challinor, Chamballu, Chiang, Christensen, Clements, Colombi,
  Colombo, Combet, Couchot, Coulais, Crill, Curto, Cuttaia, Danese, Davies,
  Davis, {De Bernardis}, {De Rosa}, {De Zotti}, Delabrouille, Delouis,
  D{\'{e}}sert, Dickinson, Diego, Dolag, Dole, Donzelli, Dor{\'{e}}, Douspis,
  Ducout, Dupac, Efstathiou, Elsner, En{\ss}lin, Eriksen, Fergusson, Finelli,
  Forni, Frailis, Fraisse, Franceschi, Frejsel, Galeotta, Galli, Ganga, Ghosh,
  Giard, Giraud-H{\'{e}}raud, Gjerl{\o}w, Gonz{\'{a}}lez-Nuevo, G{\'{o}}rski,
  Gratton, Gregorio, Gruppuso, Gudmundsson, Hansen, Hanson, Harrison,
  Henrot-Versill{\'{e}}, Hern{\'{a}}ndez-Monteagudo, Herranz, Hildebrandt,
  Hivon, Hobson, Holmes, Hornstrup, Hovest, Huffenberger, Hurier, Jaffe, Jaffe,
  Jones, Juvela, Karakci, Keih{\"{a}}nen, Keskitalo, Kiiveri, Kisner, Kneissl,
  Knoche, Kunz, Kurki-Suonio, Lagache, Lamarre, Lasenby, Lattanzi, Lawrence,
  Leonardi, Lesgourgues, Levrier, Liguori, Lilje, Linden-V{\o}rnle, Lindholm,
  L{\'{o}}pez-Caniego, Lubin, MacI{\'{a}}s-P{\'{e}}rez, Maggio, Maino,
  Mandolesi, Mangilli, Maris, Martin, Mart{\'{i}}nez-Gonz{\'{a}}lez, Masi,
  Matarrese, McGehee, Meinhold, Melchiorri, Melin, Mendes, Mennella,
  Migliaccio, Mitra, Miville-Desch{\^{e}}nes, Moneti, Montier, Morgante,
  Mortlock, Moss, Munshi, Murphy, Naselsky, Nati, Natoli, Netterfield,
  N{\o}rgaard-Nielsen, Noviello, Novikov, Novikov, Oxborrow, Paci, Pagano,
  Pajot, Paoletti, Pasian, Patanchon, Pearson, Perdereau, Perotto, Perrotta,
  Pettorino, Piacentini, Piat, Pierpaoli, Pietrobon, Plaszczynski,
  Pointecouteau, Polenta, Pratt, Pr{\'{e}}zeau, Prunet, Puget, Rachen, Rebolo,
  Reinecke, Remazeilles, Renault, Renzi, Ristorcelli, Rocha, Roman, Rosset,
  Rossetti, Roudier, Rubin{\~{o}}-Mart{\'{i}}n, Rusholme, Sandri, Santos,
  Savelainen, Scott, Seiffert, Shellard, Spencer, Stolyarov, Stompor, Sudiwala,
  Sutton, Suur-Uski, Sygnet, Tauber, Terenzi, Toffolatti, Tomasi, Tristram,
  Tucci, Tuovinen, Valenziano, Valiviita, {Van Tent}, Vielva, Villa, Wade,
  Wandelt, Wehus, Welikala, Yvon, Zacchei, \& Zonca}]{FFP8}
{Planck Collaboration XII}, Ade, P.~A., Aghanim, N., {et~al.} 2016, Astronomy
  \& Astrophysics, 594, A12, \dodoi{10.1051/0004-6361/201527103}

\bibitem[{{Planck Collaboration XLVIII} {et~al.}(2016){Planck Collaboration
  XLVIII}, Aghanim, Ashdown, Aumont, Baccigalupi, Ballardini, Banday, Barreiro,
  Bartolo, Basak, Benabed, Bernard, Bersanelli, Bielewicz, Bonavera, Bond,
  Borrill, Bouchet, Boulanger, Burigana, Calabrese, Cardoso, Carron, Chiang,
  Colombo, Comis, Couchot, Coulais, Crill, Curto, Cuttaia, {De Bernardis}, {De
  Zotti}, Delabrouille, {Di Valentino}, Dickinson, Diego, Dor{\'{e}}, Douspis,
  Ducout, Dupac, Dusini, Elsner, En{\ss}lin, Eriksen, Falgarone, Fantaye,
  Finelli, Forastieri, Frailis, Fraisse, Franceschi, Frolov, Galeotta, Galli,
  Ganga, G{\'{e}}nova-Santos, Gerbino, Ghosh, Giraud-H{\'{e}}raud,
  Gonz{\'{a}}lez-Nuevo, G{\'{o}}rski, Gruppuso, Gudmundsson, Hansen, Helou,
  Henrot-Versill{\'{e}}, Herranz, Hivon, Huang, Jaffe, Jones, Keih{\"{a}}nen,
  Keskitalo, Kiiveri, Kisner, Krachmalnicoff, Kunz, Kurki-Suonio, Lamarre,
  Langer, Lasenby, Lattanzi, Lawrence, {Le Jeune}, Levrier, Lilje, Lilley,
  Lindholm, L{\'{o}}pez-Caniego, Ma, MacI{\'{a}}s-P{\'{e}}rez, Maggio, Maino,
  Mandolesi, Mangilli, Maris, Martin, Mart{\'{i}}nez-Gonz{\'{a}}lez, Matarrese,
  Mauri, McEwen, Melchiorri, Mennella, Migliaccio, Miville-Desch{\^{e}}nes,
  Molinari, Moneti, Montier, Morgante, Moss, Natoli, Oxborrow, Pagano,
  Paoletti, Patanchon, Perdereau, Perotto, Pettorino, Piacentini, Plaszczynski,
  Polastri, Polenta, Puget, Rachen, Racine, Reinecke, Remazeilles, Renzi,
  Rocha, Rosset, Rossetti, Roudier, Rubin{\~{o}}-Mart{\'{i}}n, Ruiz-Granados,
  Salvati, Sandri, Savelainen, Scott, Sirignano, Sirri, Soler, Spencer,
  Suur-Uski, Tauber, Tavagnacco, Tenti, Toffolatti, Tomasi, Tristram,
  Trombetti, Valiviita, {Van Tent}, Vielva, Villa, Vittorio, Wandelt, Wehus,
  Zacchei, \& Zonca}]{Aghanim2016}
{Planck Collaboration XLVIII}, Aghanim, N., Ashdown, M., {et~al.} 2016,
  Astronomy \& Astrophysics, 596, A109, \dodoi{10.1051/0004-6361/201629022}

\bibitem[{Qu {et~al.}(in prep.)Qu, Sherwin, Madhavacheril, Han, \&
  Crowley}]{Qu23}
Qu, F., Sherwin, B., Madhavacheril, M., Han, D., \& Crowley, K.~T. in prep.

\bibitem[{{Regaldo-Saint Blancard} {et~al.}(2021){Regaldo-Saint Blancard},
  Allys, Boulanger, Levrier, \& Jeffrey}]{Regaldo-SaintBlancard2021}
{Regaldo-Saint Blancard}, B., Allys, E., Boulanger, F., Levrier, F., \&
  Jeffrey, N. 2021, Astronomy \& Astrophysics, 649, L18,
  \dodoi{10.1051/0004-6361/202140503}

\bibitem[{{Reinecke, M.} \& {Seljebotn, D. S.}(2013)}]{libsharp}
{Reinecke, M.}, \& {Seljebotn, D. S.} 2013, Astronomy \& Astrophysics, 554,
  A112, \dodoi{10.1051/0004-6361/201321494}

\bibitem[{Rogers {et~al.}(2016)Rogers, Peiris, Leistedt, McEwen, \&
  Pontzen}]{Spin-SILC}
Rogers, K.~K., Peiris, H.~V., Leistedt, B., McEwen, J.~D., \& Pontzen, A. 2016,
  Monthly Notices of the Royal Astronomical Society, 463, 2310,
  \dodoi{10.1093/MNRAS/STW2128}

\bibitem[{Salatino {et~al.}(2020)Salatino, Austermann, Thompson, Ade, Bai,
  Beall, Becker, Cai, Chang, Chen, Connors, Chen, Dober, Delabrouille, Duff,
  Gao, Givhan, Ghosh, Hilton, Hu, Hubmayr, Karpel, Kuo, Li, Li, Li, Li, Link,
  Li, Liu, Liu, Liu, Lu, Lucas, Lu, Mates, Mathewson, Mauskopf, Meinke,
  Montana-Lopez, Shi, Sinclair, Stephenson, Sun, Tseng, Tucker, Ullom, Vale,
  van Lanen, Vissers, Walker, Wang, Wang, Wang, Weeks, Wu, Wu, Xia, Xu, Yao,
  Yao, Yoon, Yue, Zhai, Zhang, Zhang, Zhang, Zhang, Zhang, Zhang, Zhang, Zhang,
  Zhao, \& Zhao}]{AliCPT}
Salatino, M., Austermann, J., Thompson, K.~L., {et~al.} 2020, in Millimeter,
  Submillimeter, and Far-Infrared Detectors and Instrumentation for Astronomy
  X, ed. J.~Zmuidzinas \& J.-R. Gao, Vol. 11453, International Society for
  Optics and Photonics (SPIE), 114532A, \dodoi{10.1117/12.2560709}

\bibitem[{{Schillaci} {et~al.}(2020){Schillaci}, {Ade}, {Ahmed}, {Amiri},
  {Barkats}, {Thakur}, {Bischoff}, {Bock}, {Boenish}, {Bullock}, {Buza},
  {Cheshire}, {Connors}, {Cornelison}, {Crumrine}, {Cukierman}, {Dierickx},
  {Duband}, {Fatigoni}, {Filippini}, {Hall}, {Halpern}, {Harrison},
  {Henderson}, {Hildebrandt}, {Hilton}, {Hui}, {Irwin}, {Kang}, {Karkare},
  {Karpel}, {Kefeli}, {Kovac}, {Kuo}, {Lau}, {Megerian}, {Moncelsi},
  {Namikawa}, {Nguyen}, {O'Brient}, {Palladino}, {Precup}, {Prouve}, {Pryke},
  {Racine}, {Reintsema}, {Richter}, {Schmitt}, {Schwarz}, {Sheehy}, {Soliman},
  {Germaine}, {Steinbach}, {Sudiwala}, {Thompson}, {Tucker}, {Turner},
  {Umilt{\`a}}, {Vieregg}, {Wandui}, {Weber}, {Wiebe}, {Willmert}, {Wu},
  {Yang}, {Yoon}, {Young}, {Yu}, \& {Zhang}}]{BICEP_Array}
{Schillaci}, A., {Ade}, P.~A.~R., {Ahmed}, Z., {et~al.} 2020, Journal of Low
  Temperature Physics, 199, 976, \dodoi{10.1007/s10909-020-02394-6}

\bibitem[{Seljebotn {et~al.}(2019)Seljebotn, Baerland, Eriksen, Mardal, \&
  Wehus}]{Seljebotn2019}
Seljebotn, D.~S., Baerland, T., Eriksen, H.~K., Mardal, K.-A., \& Wehus, I.~K.
  2019, Astronomy \& Astrophysics, 627, \dodoi{10.1051/0004-6361/201732037}

\bibitem[{{Smith} {et~al.}(2007){Smith}, {Zahn}, \& {Dor{\'e}}}]{Smith2007}
{Smith}, K.~M., {Zahn}, O., \& {Dor{\'e}}, O. 2007, \prd, 76, 043510,
  \dodoi{10.1103/PhysRevD.76.043510}

\bibitem[{Smith \& Zaldarriaga(2011)}]{Smith2011}
Smith, K.~M., \& Zaldarriaga, M. 2011, Monthly Notices of the Royal
  Astronomical Society, 417, 2, \dodoi{10.1111/J.1365-2966.2010.18175.X}

\bibitem[{Suzuki {et~al.}(2016)Suzuki, Ade, Akiba, Aleman, Arnold, Baccigalupi,
  Barch, Barron, Bender, Boettger, Borrill, Chapman, Chinone, Cukierman, Dobbs,
  Ducout, Dunner, Elleflot, Errard, Fabbian, Feeney, Feng, Fujino, Fuller,
  Gilbert, Goeckner-Wald, Groh, Haan, Hall, Halverson, Hamada, Hasegawa,
  Hattori, Hazumi, Hill, Holzapfel, Hori, Howe, Inoue, Irie, Jaehnig, Jaffe,
  Jeong, Katayama, Kaufman, Kazemzadeh, Keating, Kermish, Keskitalo, Kisner,
  Kusaka, Jeune, Lee, Leon, Linder, Lowry, Matsuda, Matsumura, Miller,
  Mizukami, Montgomery, Navaroli, Nishino, Peloton, Poletti, Puglisi, Rebeiz,
  Raum, Reichardt, Richards, Ross, Rotermund, Segawa, Sherwin, Shirley,
  Siritanasak, Stebor, Stompor, Suzuki, Tajima, Takada, Takakura, Takatori,
  Tikhomirov, Tomaru, Westbrook, Whitehorn, Yamashita, Zahn, \&
  Zahn}]{Polarbear_2}
Suzuki, A., Ade, P., Akiba, Y., {et~al.} 2016, Journal of Low Temperature
  Physics, 184, 805, \dodoi{10.1007/s10909-015-1425-4}

\bibitem[{{Svalheim} {et~al.}(2020){Svalheim}, {Andersen}, {Aurlien},
  {Banerji}, {Bersanelli}, {Bertocco}, {Brilenkov}, {Carbone}, {Colombo},
  {Eriksen}, {Foss}, {Franceschet}, {Fuskeland}, {Galeotta}, {Galloway},
  {Gerakakis}, {Gjerl{\o}w}, {Hensley}, {Herman}, {Iacobellis}, {Ieronymaki},
  {Ihle}, {Jewell}, {Karakci}, {Keih{\"a}nen}, {Keskitalo}, {Maggio}, {Maino},
  {Maris}, {Paradiso}, {Partridge}, {Reinecke}, {Suur-Uski}, {Tavagnacco},
  {Thommesen}, {Watts}, {Wehus}, \& {Zacchei}}]{Svalheim2020}
{Svalheim}, T.~L., {Andersen}, K.~J., {Aurlien}, R., {et~al.} 2020, arXiv
  e-prints, arXiv:2011.08503, \dodoi{10.48550/arXiv.2011.08503}

\bibitem[{Takakura {et~al.}(2017)Takakura, Aguilar, Akiba, Arnold, Baccigalupi,
  Barron, Beckman, Boettger, Borrill, Chapman, Chinone, Cukierman, Ducout,
  Elleflot, Errard, Fabbian, Fujino, Galitzki, Goeckner-Wald, Halverson,
  Hasegawa, Hattori, Hazumi, Hill, Howe, Inoue, Jaffe, Jeong, Kaneko, Katayama,
  Keating, Keskitalo, Kisner, Krachmalnicoff, Kusaka, Lee, Leon, Lowry,
  Matsuda, Matsumura, Navaroli, Nishino, Paar, Peloton, Poletti, Puglisi,
  Reichardt, Ross, Siritanasak, Suzuki, Tajima, Takatori, \&
  Teply}]{Takakura2017}
Takakura, S., Aguilar, M., Akiba, Y., {et~al.} 2017, Journal of Cosmology and
  Astroparticle Physics, 2017, 008, \dodoi{10.1088/1475-7516/2017/05/008}

\bibitem[{Thornton {et~al.}(2016)Thornton, Ade, Aiola, Angil{\`{e}}, Amiri,
  Beall, Becker, Cho, Choi, Corlies, Coughlin, Datta, Devlin, Dicker,
  D{\"{u}}nner, Fowler, Fox, Gallardo, Gao, Grace, Halpern, Hasselfield,
  Henderson, Hilton, Hincks, Ho, Hubmayr, Irwin, Klein, Koopman, Li, Louis,
  Lungu, Maurin, McMahon, Munson, Naess, Nati, Newburgh, Nibarger, Niemack,
  Niraula, Nolta, Page, Pappas, Schillaci, Schmitt, Sehgal, Sievers, Simon,
  Staggs, Tucker, Uehara, van Lanen, Ward, \& Wollack}]{Thornton2016}
Thornton, R.~J., Ade, P. A.~R., Aiola, S., {et~al.} 2016, The Astrophysical
  Journal Supplement Series, 227, 21, \dodoi{10.3847/1538-4365/227/2/21}

\bibitem[{Virtanen {et~al.}(2020)Virtanen, Gommers, Oliphant, Haberland, Reddy,
  Cournapeau, Burovski, Peterson, Weckesser, Bright, {van der Walt}, Brett,
  Wilson, Millman, Mayorov, Nelson, Jones, Kern, Larson, Carey, Polat, Feng,
  Moore, {VanderPlas}, Laxalde, Perktold, Cimrman, Henriksen, Quintero, Harris,
  Archibald, Ribeiro, Pedregosa, {van Mulbregt}, \& {SciPy 1.0
  Contributors}}]{scipy}
Virtanen, P., Gommers, R., Oliphant, T.~E., {et~al.} 2020, Nature Methods, 17,
  261, \dodoi{10.1038/s41592-019-0686-2}

\bibitem[{{Wiaux} {et~al.}(2008){Wiaux}, {McEwen}, {Vandergheynst}, \&
  {Blanc}}]{Wiaux2008}
{Wiaux}, Y., {McEwen}, J.~D., {Vandergheynst}, P., \& {Blanc}, O. 2008, \mnras,
  388, 770, \dodoi{10.1111/j.1365-2966.2008.13448.x}

\bibitem[{{Zaldarriaga} \& {Seljak}(1997)}]{zaldarriaga_1997}
{Zaldarriaga}, M., \& {Seljak}, U. 1997, \prd, 55, 1830,
  \dodoi{10.1103/PhysRevD.55.1830}

\bibitem[{Zonca {et~al.}(2019)Zonca, Singer, Lenz, Reinecke, Rosset, Hivon, \&
  Gorski}]{healpy}
Zonca, A., Singer, L., Lenz, D., {et~al.} 2019, Journal of Open Source
  Software, 4, 1298, \dodoi{10.21105/joss.01298}

\end{thebibliography}
\bibliographystyle{aasjournal}

\appendix

\section{Example Noise Simulations} \label{apx: example_sims}
\begin{figure*}[h]
    \centering
    \includegraphics[width=\textwidth]{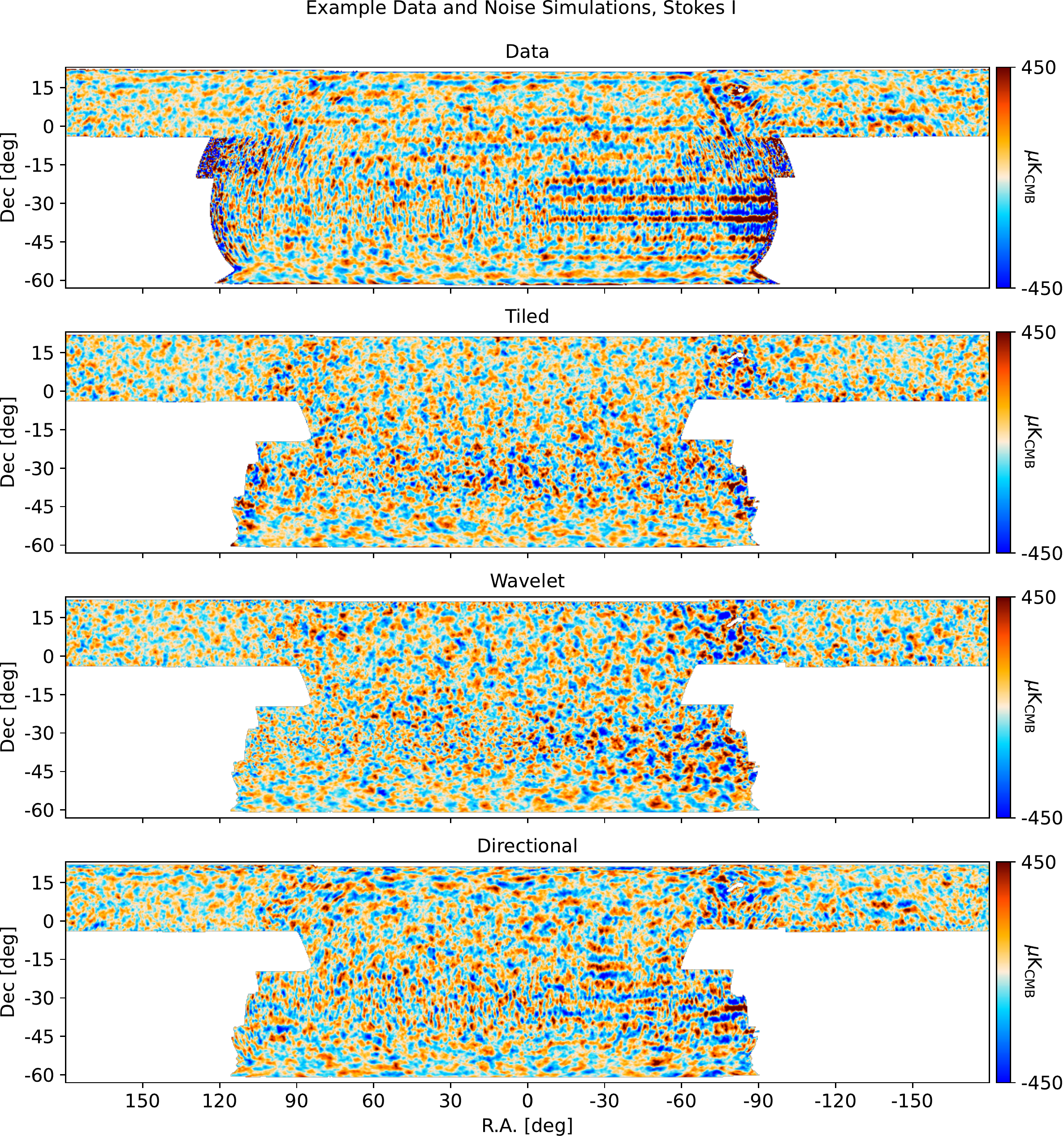}
    \caption{The difference map (Equation \ref{eq: diff_def_3}) for PA5 f090, first split, and simulations from the three noise models (Stokes I).} 
    \label{apx: example_sims_I}
\end{figure*}

\clearpage
\begin{figure*}
    \centering
    \includegraphics[width=\textwidth]{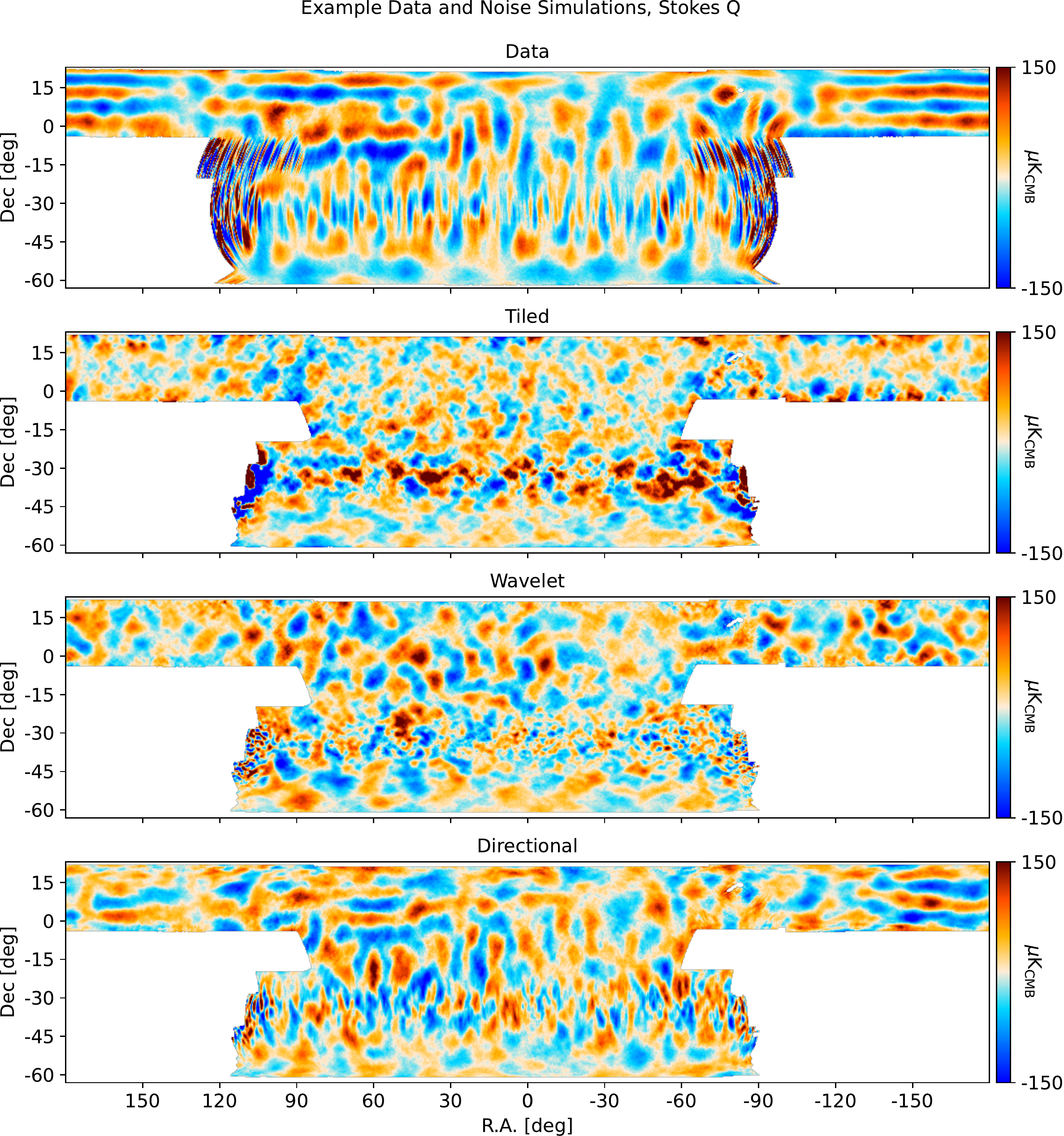}
    \caption{The difference map (Equation \ref{eq: diff_def_3}) for PA5 f090, first split, and simulations from the three noise models (Stokes Q).} 
    \label{apx: example_sims_Q}
\end{figure*}

\clearpage
\section{Difference Map Algebra} \label{apx: diff_maths}
We explicitly show the algebra relating to the construction of the difference maps, $\bm\nu_i$. We first prove Equation \ref{eq: diff_noise_var}. Starting from Equation \ref{eq: diff_def_2}, we have:
\begin{equation}
    \begin{aligned}
        Var(\mathbf d_i) &= \mathbb{E}(\mathbf d_i^2) = \mathbb{E}\left(\left(\mathbf n_i - \frac{\sum_{j}{\mathbf h_j*\mathbf n_j}}{\mathbf\Sigma}\right)^2\right) = \\
        &= \mathbb{E}(\mathbf n_i^2) - 2\mathbb{E}\left(\mathbf n_i * \frac{\sum_{j}{\mathbf h_j*\mathbf n_j}}{\mathbf\Sigma}\right) + \mathbb{E}\left(\frac{\sum_{i}{\mathbf h_i*\sum_{j}{\mathbf h_j*\mathbf n_i*\mathbf n_j}}}{\mathbf\Sigma^2}\right) = \\
        &= \mathbb{E}(\mathbf n_i^2) - \frac{2\sum_{j}{\mathbf h_j*\mathbb{E}(\mathbf n_i*\mathbf n_j)}}{\mathbf\Sigma} + \frac{\sum_{i}{\mathbf h_i*\sum_{j}{\mathbf h_j*\mathbb{E}(\mathbf n_i*\mathbf n_j)}}}{\mathbf\Sigma^2}.
    \end{aligned}
\end{equation}
Now impose the independence of the noise in different splits and the definition of the inverse-variance in split $i$, $\mathbb{E}(\mathbf n_i * \mathbf n_j) = 1/\mathbf{h_i}*\delta_{ij}$. Also substitute the definition $\mathbf\Sigma\equiv\sum_{i}\mathbf h_i$:
\begin{equation}
    \begin{aligned}
        Var(\mathbf d_i) &= \frac{1}{\mathbf h_i} - \frac{2\sum_{j}{\mathbf h_j*1/\mathbf{h_i}*\delta_{ij}}}{\mathbf\Sigma} + \frac{\sum_{i}{\mathbf h_i*\sum_{j}{\mathbf h_j*1/\mathbf{h_i}*\delta_{ij}}}}{\mathbf\Sigma^2} = \\
        &= \frac{1}{\mathbf h_i} - \frac{2}{\mathbf\Sigma} + \frac{\sum_{i}{\mathbf h_i}}{\mathbf\Sigma^2} = \\
        &= \frac{1}{\mathbf h_i} - \frac{1}{\mathbf\Sigma}.
    \end{aligned}
\end{equation}

We then prove Equation \ref{eq: diff_noise_corr}. First, we note that because $\bm\nu_i \propto \mathbf d_i$, the correlation of $\bm\nu_i$ and \mbf{n_i} is the same as that of \mbf{d_i} and \mbf{n_i}. Next, we evaluate the covariance of \mbf{d_i} and \mbf{n_i}:
\begin{equation}
    \begin{aligned}
        Cov(\mathbf d_i, \mathbf n_i) &= \mathbb{E}(\mathbf d_i*\mathbf n_i) = \mathbb{E}\left(\mathbf n_i*\left(\mathbf n_i - \frac{\sum_{j}{\mathbf h_j*\mathbf n_j}}{\mathbf\Sigma}\right)\right) = \\
        &= \mathbb{E}(\mathbf n_i^2) - \frac{\sum_{j}{\mathbf h_j*\mathbb{E}(\mathbf n_i*\mathbf n_j)}}{\mathbf\Sigma} = \\
        &= \frac{1}{\mathbf h_i} - \frac{\sum_{j}{\mathbf h_j*1/\mathbf{h_i}*\delta_{ij}}}{\mathbf\Sigma} = \\
        &= \frac{1}{\mathbf h_i} - \frac{1}{\mathbf\Sigma} = \\
        &= Var(\mathbf d_i).
    \end{aligned}
\end{equation}
Now evaluate the correlation coefficient directly:
\begin{equation}
    \begin{aligned}
        \bm\rho_i = \frac{Cov(\mathbf d_i, \mathbf n_i)}{\sqrt{Var(\mathbf d_i) * Var(\mathbf n_i)}} = \frac{Var(\mathbf d_i)}{\sqrt{Var(\mathbf d_i) * Var(\mathbf n_i)}} = \sqrt{\frac{Var(\mathbf d_i)}{Var(\mathbf n_i)}} = \sqrt{\mathbf h_i*\left(\frac{1}{\mathbf h_i} - \frac{1}{\mathbf\Sigma}\right)} = \sqrt{1 - \frac{\mathbf h_i}{\mathbf\Sigma}}.
    \end{aligned}
\end{equation}

Note that all of these Equations are simply generalizing the familiar ``$N-1$" factors from the unbiased sample variance of a set of equally-weighted observations to the case of unequally-weighted observations.

\section{Derivation of Tiling Constraint} \label{apx: unbiased_covmat}
We offer a derivation on the backward tiling transform constraint (Equation \ref{eq: admissibility}) that each noise model must satisfy. We will show that it emerges from a desire to not bias the diagonal entries of the full noise covariance matrix. We begin by considering a simpler version of the noise simulation equations (Equations \ref{eq: tile_sim}, \ref{eq: wav_sim}, \ref{eq: fdw_sim}). Specifically, we ignore any ``exterior" filtering operations, and we replace any DFTs or SHT matrices with a ``universal" transform matrix \mbf{U} (we also drop the split index $i$ for brevity).  In this way, a noise simulation from all three noise models appears identical:
\begin{equation}
    \begin{aligned}
        \bm\xi_{\mathrm{oper}} &= \mathcal{T}_b^{\dag} \mathbf U \mathcal{N}_{\mathrm{sparse}}^{\frac{1}{2}}\bm\eta \\
        &= \sum_{j=0}^{M-1} \mathbf P_j^{\dag} \bm\tau_{j,\mathrm{b}}^{\dag} \mathbf U \mathcal{N}_{j, \mathrm{sparse}}^{\frac{1}{2}} \bm\eta_j.\\
    \end{aligned}
\end{equation}
where we have generalized the bases of these objects: the simulation is in the ``operational" basis where we define and apply the tiling transform profiles --- for the tiled/wavelet/directional model this is map space/harmonic space/Fourier space --- and the model covariance is in the ``sparse" basis. The matrix \mbf{U} thus projects a vector from the ``sparse" to the ``operational" basis --- it is a DFT or a SHT matrix. 

We want to determine the conditions under which the operational space noise covariance matrix, $\mathbf N_{\mathrm{oper}} = \EV{\bm\xi_{\mathrm{oper}}\bm\xi_{\mathrm{oper}}^\dag}$,
is unbiased:
\begin{equation}
    \begin{aligned}
        \mathbf N_{\mathrm{oper}} = \EV{\bm\xi_{\mathrm{oper}}\bm\xi_{\mathrm{oper}}^\dag} &= \sum_{j=0}^{M-1}\sum_{k=0}^{M-1} \mathbf P_j^{\dag} \bm\tau_{j,\mathrm{b}}^{\dag} \mathbf U \mathcal{N}_{j, \mathrm{sparse}}^{\frac{1}{2}} \EV{\bm\eta_j \bm\eta_k^\dag} \mathcal{N}_{k, \mathrm{sparse}}^{\frac{1}{2},\dag} \mathbf U^\dag \bm\tau_{k,\mathrm{b}} \mathbf P_k \\
        &= \sum_{j=0}^{M-1}\sum_{k=0}^{M-1} \mathbf P_j^{\dag} \bm\tau_{j,\mathrm{b}}^{\dag} \mathbf U \mathcal{N}_{j, \mathrm{sparse}}^{\frac{1}{2}} \mathbf I \delta_{j,k} \mathcal{N}_{k, \mathrm{sparse}}^{\frac{1}{2},\dag} \mathbf U^\dag \bm\tau_{k,\mathrm{b}} \mathbf P_k \\
        &= \sum_{j=0}^{M-1} \mathbf P_j^{\dag} \bm\tau_{j,\mathrm{b}}^{\dag} \mathbf U \mathcal{N}_{j, \mathrm{sparse}} \mathbf U^\dag \bm\tau_{j,\mathrm{b}} \mathbf P_j \\
        &= \sum_{j=0}^{M-1}  \mathbf P_j^{\dag} \bm\tau_{j,\mathrm{b}}^{\dag} \mathbf{N}_{j, \mathrm{oper}} \bm\tau_{j,\mathrm{b}} \mathbf P_j
    \end{aligned}
\end{equation}
where we have identified $\mathbf U \mathcal{N}_{j, \mathrm{sparse}} \mathbf U^\dag$ with $\mathbf N_{j, \mathrm{oper}}$. Now consider a single matrix element indexed by $x$, $x'$ --- that is, the covariance between ``pixels" (or ``modes") in the operational basis, $N_{\mathrm{oper}}^{x,x'}$. Such an element is formed by taking the $x^{\mathrm{th}}$ row of $\mathbf P_j^{\dag} \bm\tau_{j,\mathrm{b}}^{\dag}$ and $x'^{\mathrm{th}}$ column of $\bm\tau_{j,\mathrm{b}} \mathbf P_j$, each of which have at most one nonzero element. Thus, for a given $j$, $N_{\mathrm{oper}}^{x,x'}$ receives a contribution from at most one element of $\mathbf N_{j,\mathrm{oper}}$; let us index that element by $e$ and $e'$:
\begin{equation}
    \begin{aligned} \label{eq: apx_admissibility}
       N_{\mathrm{oper}}^{x,x'} &= \sum_{j=0}^{M-1} (\mathbf P_j^{\dag} \bm\tau_{j,\mathrm{b}}^{\dag})_x (\bm\tau_{j,\mathrm{b}} \mathbf P_j)_{x'} N_{j,\mathrm{oper}}^{e,e'} \\  
       N_{\mathrm{oper}}^{x,x'} &= N_{\mathrm{oper}}^{e,e'} \sum_{j=0}^{M-1} (\mathbf P_j^{\dag} \bm\tau_{j,\mathrm{b}}^{\dag})_x (\bm\tau_{j,\mathrm{b}} \mathbf P_j)_{x'} \\
       1 &= \sum_{j=0}^{M-1} (\mathbf P_j^{\dag} \bm\tau_{j,\mathrm{b}}^{\dag})_x (\bm\tau_{j,\mathrm{b}} \mathbf P_j)_{x'}.
    \end{aligned}
\end{equation}
In the case of $x=x'$ (i.e., the diagonal of the map-based covariance matrix), Equation \ref{eq: apx_admissibility} recovers Equation \ref{eq: admissibility}. We had to use two sleights-of-hand in the second and third lines of Equation \ref{eq: apx_admissibility}. In the second line, we moved $N_{j,\mathrm{oper}}^{e,e'}$ out of the sum, where it became $N_{\mathrm{oper}}^{e,e'}$. In the third line, we asserted that $N_{j,\mathrm{oper}}^{x,x'}=N_{j,\mathrm{oper}}^{e,e'}$. To do so, we assume that any given element of the covariance matrix on the left-hand-side is \textit{equal to} each of its occurrences in each tile or wavelet, $j$. In reality, each tile or wavelet covariance will disagree on the exact value of a given shared element, but if we assume that these separate estimates are unbiased, then satisfying Equation \ref{eq: apx_admissibility} will likewise yield an unbiased full covariance matrix. 

There are a few more caveats of note. Working backwards, it is difficult to construct useful tiles or wavelets that satisfy Equation \ref{eq: apx_admissibility} for $x \neq x'$, and indeed for the kernels in this paper this sum approaches zero as the difference between $x$ and $x'$ increases. The upshot is that we cannot model correlations at scales larger than the tile or wavelet size within the tiling scheme itself. Such ``long-distance" correlations instead are modeled by the filtering steps that we excluded from the preceding argument. Likewise, we assume that our sparse-basis covariance blocks $\mathcal{N}_{j, \mathrm{sparse}}$ are unbiased estimates, but that is likely not true. For instance, Equation \ref{eq: mask_correction} is the zeroth-order correction to the mode-coupling induced by the forward tiling transform. Lastly, we must assume that the sparse-basis covariance blocks are themselves uncorrelated, but we know that is not true, since the forward tiling kernels from which the blocks are measured also overlap. In sum, while satisfying Equation \ref{eq: admissibility} is necessary for a performant noise model, it alone is not sufficient. 

\clearpage
\section{Directional Wavelet Kernels} \label{apx: az_kernels}
As discussed in \S\ref{sec: noise_models_fdw}, the directional model utilizes radially- and azimuthally-separable wavelets in the Fourier plane:
\begin{align}
    \tau_{j,\mathrm{f}}(\mathbf k) \rightarrow \tau_{u,v,\mathrm{f}}(\mathbf k) = \kappa_u(k)\eta_{u,v}(\phi)
\end{align}
where \mbf{k} is the 2D Fourier mode, $k=|\mathbf k|$, $\phi=\mathrm{arg}(\mathbf k)$, $\kappa$ are the wavelet kernels of \citet{Wiaux2008, sd-wavelets} with the parameters given in \S\ref{sec: noise_models_wav_algorithm}, $\eta$ are the azimuthal kernels, and we have expanded the wavelet index $j$ into a ``radial" index $u$ and ``azimuthal" index $v$. Evaluated at each discrete \mbf{k} in the map Fourier transform, $\tau_{j,\mathrm{f}}$ becomes the diagonal of the forward wavelet kernel matrix $\bm\tau_{j,\mathrm{f}}$. The fact that $\eta$ also carries the ``radial" index denotes that each radial kernel has its own distinct set of associated azimuthal kernels.

\begin{figure}
    \centering
    \includegraphics[width=0.49\textwidth]{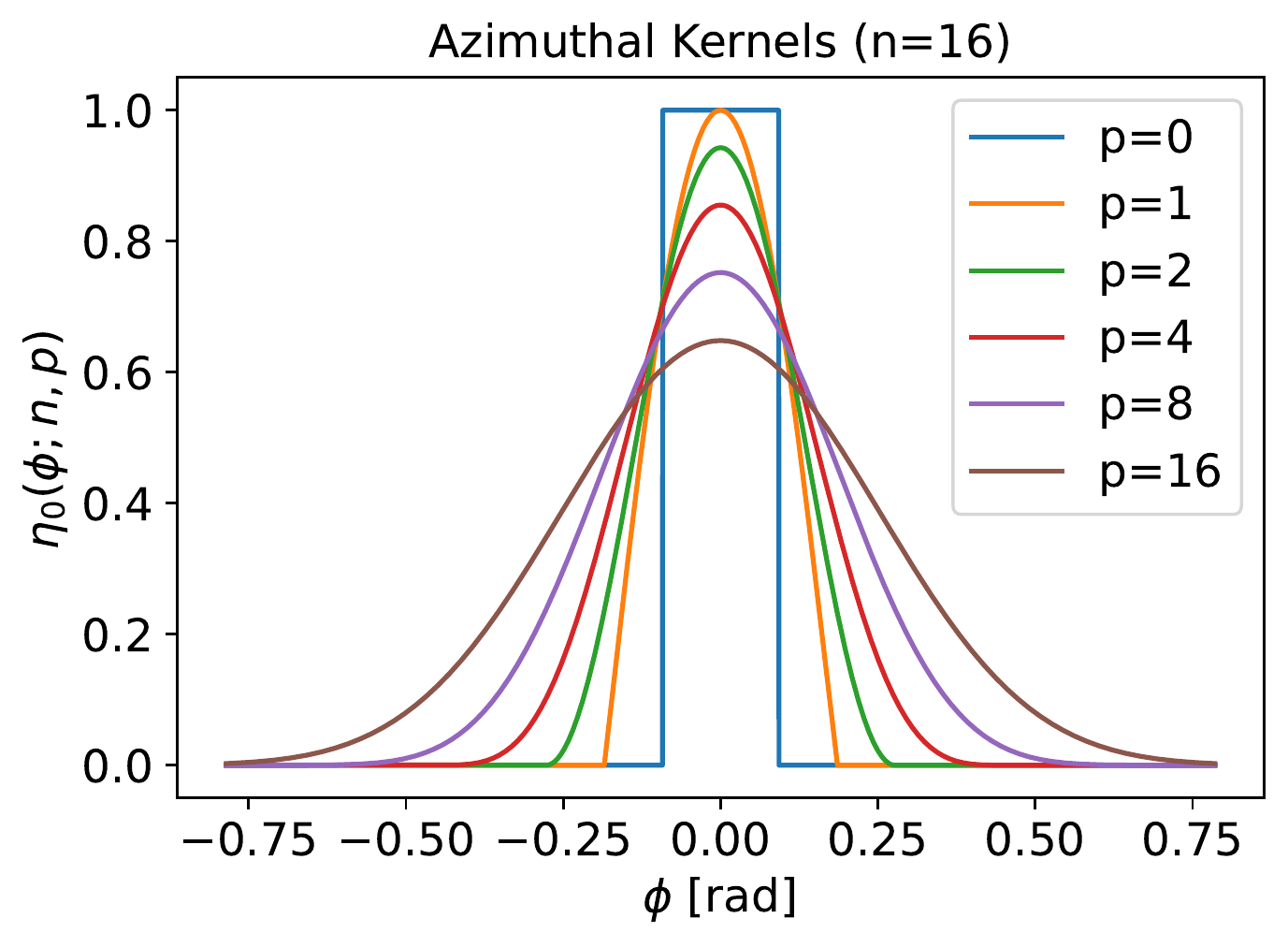}
    \includegraphics[width=0.49\textwidth]{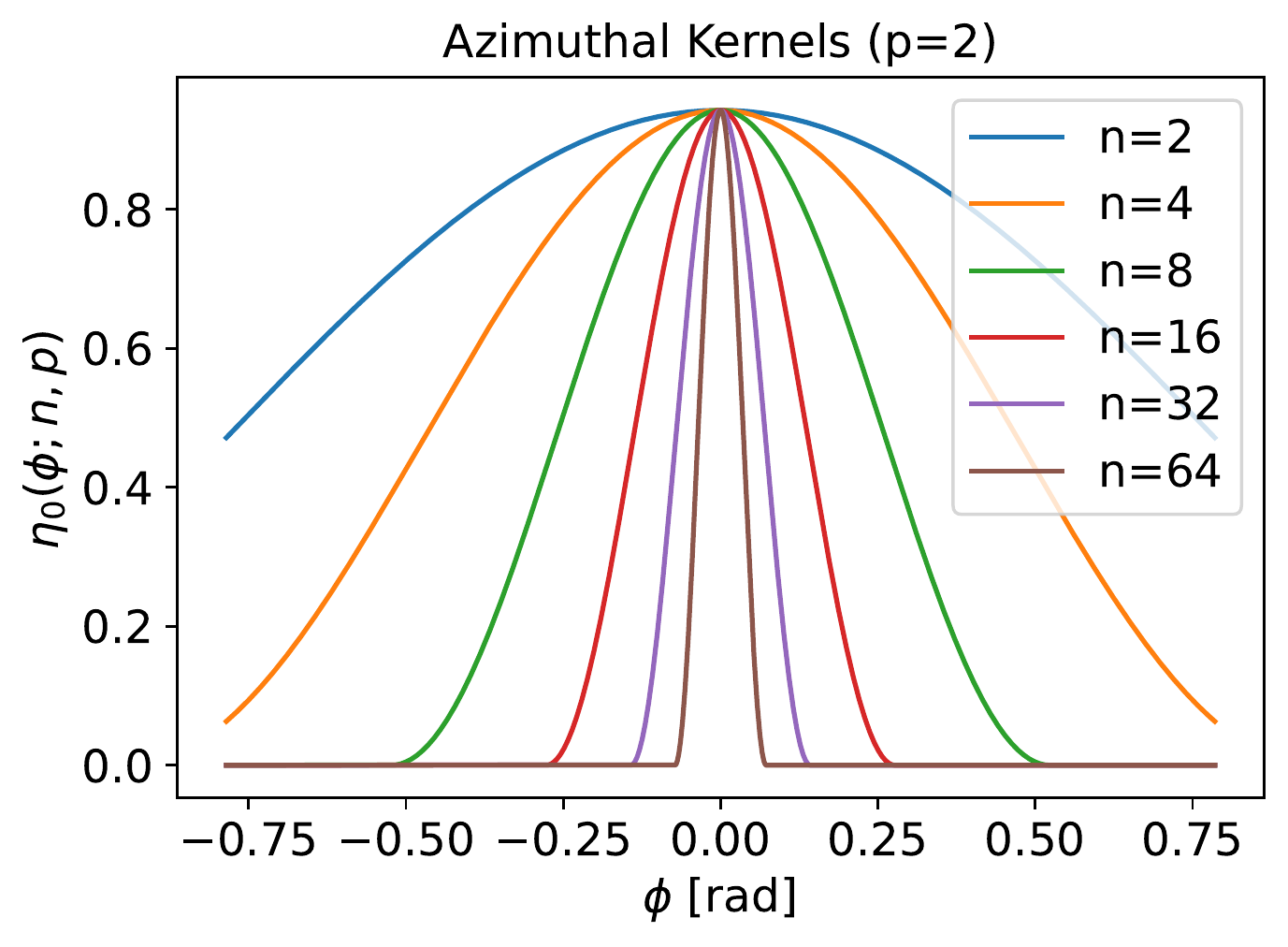} \\
    \vspace{20 pt}
    \includegraphics[width=\textwidth]{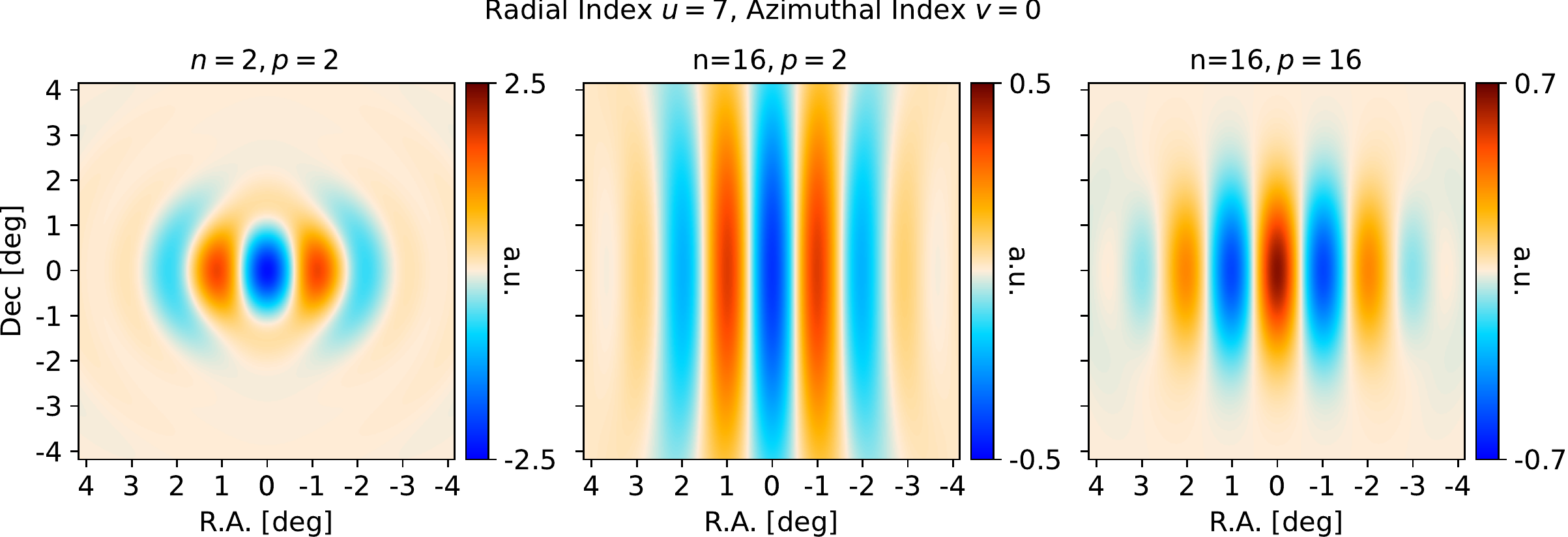}
    \caption{\textit{Top:} 1D profiles of various directional wavelet azimuthal kernels. The left plot demonstrates the decreasing Fourier space compactness with increasing shape parameter, $p$, values. The right plot demonstrates increasing Fourier space compactness with increasing number of kernels, $n+1$, values. \textit{Bottom}: Map space 2D kernels corresponding to $u=7, v=0$ for the kernel set used in this paper. The left map space kernel corresponds to the $n=2, p=2$ case, which has the most spatial locality and least directional locality. The middle map space kernel corresponds to the $n=16, p=2$ case, which has the least spatial locality and most directional locality. The right map space kernel corresponds to the $n=16, p=16$ case, which has intermediate spatial and directional locality.}
    \label{fig: fdw_kernels}
\end{figure}

We define the azimuthal kernels as follows:
\begin{align} \label{eq: apx_kernels}
    \eta_{u,v}(\phi) \rightarrow \eta_v(\phi; n_u, p_u) = \sum_{k=-\infty}^{\infty}
    \begin{cases}
        A_{k,p}\cos(\omega_{n,p}(\phi - \phi_{0,k,n,v}))^p, & \omega_{n,p}|\phi - \phi_{0,k,n,v}| < \pi/2 \\
        0, & \omega_{n,p}|\phi - \phi_{0,k,n,v}| \geq \pi/2
    \end{cases}
\end{align}
where:
\begin{equation}
    \begin{aligned}
        A_{k,p} &= i^p (-1)^{kp} \sqrt{\frac{2^{2p}}{(p+1)\binom{2p}{p}}} \\
        \omega_{n,p} &= \frac{n+1}{p+1} \\
        \phi_{0,k,n,v} &= (k + \frac{v}{n+1})\pi
    \end{aligned}
\end{equation}
with $n\geq 0$, $0\leq p\leq n$, $0\leq v\leq n$, and all indices are integers. One should interpret this as follows. For each radial index $u$, we choose two parameters for the set of associated azimuthal kernels: $n$ and $p$. $n$ sets the number ($n+1$) of azimuthal kernels, and $p$ is the kernel ``shape" parameter. $v$ is the kernel index, and it generates $n+1$ copies of the kernel equally spaced in azimuth (at each $\phi_{0,k,n,v}$). The kernel shape is the $p$-th power of a cosine centered on $\phi_{0,k,n,v}$, and truncated at either side of $\phi_{0,k,n,v}$ where the $\cos(...)^p$ function intersects the x-axis. The width of the cosine is set by $\omega_{n,p}$. The normalization ensures the kernels satisfy Equation \ref{eq: admissibility} (with $\tau_{j,\mathrm{b}} = \tau_{j,\mathrm{f}}*$) and respect reality conditions when applied to the Fourier-transformed map in \S\ref{sec: noise_models_fdw_algorithm}, step \ref{enum: fdw_algorithm, item: tiling}. Finally, the sum over $k$ ensures that this piecewise function is periodic in $\pi$. Example azimuthal kernels are shown in the top row of Figure \ref{fig: fdw_kernels}.

The guiding principle in the construction of $\eta_{u,v}$ is the tradeoff in locality between map space and Fourier space: a kernel compact in map space will necessarily be extended in Fourier space, and vice versa. Taking the radial kernels as a given (see Figure \ref{fig: fdw_kernel_1d}), all else being equal the large-scale wavelets will be therefore be extended in map space, while the small scale wavelets will be compact in map space. We counteract this by tuning the $n$ and $p$ parameters for each radial kernel. For a given number of azimuthal kernels $n+1$, increasing the shape parameter $p$ decreases the compactness of the kernel in Fourier space (and increases its compactness in map space). We note that $p=0$ corresponds to $n+1$ equally spaced, non-overlapping tophat functions, while $p=n$ corresponds to the steerable basis of \citet{Freeman1991}. On the other hand, for a given $p$, increasing $n$ increases the compactness of the kernels in Fourier space (and decreases its compactness in map space). This tradeoff is evident in the bottom row of Figure \ref{fig: fdw_kernels}. Therefore, we select low $n$ and high $p$ values for the large-scale wavelets, and the opposite for the small scale wavelets, subject to $0\leq p\leq n$. Meanwhile, computational interests favor fewer, more-compact Fourier space kernels at high $u$.

\begin{table}[h]
    \centering
    \movetableright=-24pt
    \begin{tabular}{c|*{16}{|c}}
        Rad. Kern. & 0 & 1 & 2 & 3 & 4 & 5 & 6 & 7 & 8 & 9 & 10 & 11 & 12 & 13 & 14 & 15 \\
        \hline
        \hline
        $\ell_{\mathrm{min}}$ & 0 & 7 & 11 & 17 & 27 & 43 & 69 & 110 & 176 & 282 & 451 & 721 & 1,153 & 1,845 & 2,952 & 4,723 \\
        
        $\ell_{\mathrm{max}}$ & 11 & 17 & 27 & 43 & 69 & 110 & 176 & 282 & 451 & 721 & 1,153 & 1,845 & 2,952 & 4,723 & 7,556 & $\infty$ \\
        
        $n$ & 0 & 6 & 6 & 6 & 6 & 12 & 12 & 12 & 12 & 24 & 24 & 36 & 36 & 36 & 36 & 0 \\
        
        $p$ & 0 & 6 & 4 & 2 & 2 & 12 & 8 & 4 & 2 & 12 & 8 & 2 & 2 & 2 & 2 & 0 \\
        
    \end{tabular}
    \caption{Parameters of the directional wavelet kernel set used in this paper. The radial kernels ($u$-indexed) and associated $\ell_{\mathrm{min}}$, $\ell_{\mathrm{max}}$ values are generated by supplying the parameters $\lambda=1.6$, $\ell_{\mathrm{min}}=10$, and $\ell_{\mathrm{max}}=5,300$ to the wavelet functions prescribed in  \citet{Wiaux2008, sd-wavelets}. We prescribe the $n$ and $p$ values to generate the azimuthal kernel set for each radial kernel.}
    \label{tab: fdw_kernels}
\end{table}

We found the parameters in Table \ref{tab: fdw_kernels}, which we used in \S\ref{sec: noise_models_fdw}, to be a good balance of all considerations. There are fewer radial kernels (16 vs. 26) compared to the isotropic wavelet model to make up for the increased total kernel count. The choice of $n$ and $p$ recalls the ``parabolic scaling" in \citet{curvelet-sd-wavelets}, such that the 2D kernels become increasingly directional (narrow) at larger $u$ (see Figure \ref{fig: fdw_kernel_2d}), except at the highest radial kernel. This allows the resolution of sufficient detail in 2D Fourier space (e.g., Figure \ref{fig: results_2d_ps_by_region}) without sacrificing large-scale detail in map space (e.g., Figure \ref{fig: results_ivar_by_scale_lowell}). We note that we accomplished this by increasing $n$ in large steps, while decreasing $p$ within each step. A similar narrowing of the 2D kernels with increasing radii could have been accomplished, for example, by more smoothly increasing $n$ while holding $p$ constant.

\clearpage
\section {Boolean Mask Construction} \label{apx: boolean_mask}
As discussed in \S\ref{sec: implementation_ssi}, we exclude some regions of the ACT footprint from our noise models and simulations. We do this through the application of a boolean mask, $\bm\mu_{\mathrm{bool}}$. We note that $\bm\mu_{\mathrm{bool}}$ can only have a value of 0 or 1 in a given pixel (corresponding to omitting the pixel, and permitting the pixel, respectively). Here, we motivate and define this boolean mask.

The most natural boolean mask is one that selects exactly the pixels that have non-zero hits; however, we construct a slightly more restrictive mask. In addition to unobserved pixels, we mask regions with low cross-linking that abut otherwise well cross-linked/deep regions. These regions are predominantly located on the left and right sides of the southern half of the ACT footprint (the solid black areas Figure \ref{fig: masks}). We found that masking these regions significantly improved the low-$\ell$ power spectra ratios with respect to the data (see \S\ref{sec: eval}). This can be understood as a fundamental limitation of our modeling: as with residual point sources, our models cannot resolve the step-function-like change from high cross-linking to approximately no cross-linking at the edges of these masked regions. In other words, if we did not mask these shallow, poorly cross-linked regions, their noise power would mix into adjacent deep, well cross-linked regions in the noise models. Importantly, these deep, well cross-linked regions are exactly those retained in the DR6 CMB power spectrum and CMB lensing pipelines --- the main consumers of our noise simulations. Therefore, we try to optimize noise model performance in these priority regions, while sacrificing as little data as possible.

We construct $\bm\mu_{\mathrm{bool}}$ separately in two disjoint regions. The first (northern) region includes all pixels north of $-6^\circ$ Dec between $180^\circ$ and $125^\circ$ R.A., north of $3^\circ$ Dec between $125^\circ$ and $-100^\circ$ R.A., and again north of $-6^\circ$ Dec between and $-100^\circ$ and $-180^\circ$ R.A.. The second (southern) region is the complement of the northern region. First, in both regions, we set $\bm\mu_{\mathrm{bool}}=1$ in pixels observed by \textit{all} eight splits in \textit{both} frequencies of a given modeled detector array, and 0 otherwise. Then, only in the southern region, we also set $\bm\mu_{\mathrm{bool}}=0$ in pixels whose cross-linking statistic is less than 0.001 in the point-source subtracted coadd map of \textit{any} DR6 array or frequency. A considerable fraction (0.65\%) of $\bm\mu_{\mathrm{est}}$ in the first region would be cut by the cross-linking threshold, hence why we only apply the cross-linking cut to the second region. In the second region, the fraction of $\bm\mu_{\mathrm{est}}$ cut by the cross-linking threshold is only 0.04\%. We add these pixels in the second region back to $\bm\mu_{\mathrm{bool}}$ such that $\bm\mu_{\mathrm{bool}}$ fully encompasses $\bm\mu_{\mathrm{est}}$. Together, this procedure is effective at removing the large areas of approximately zero cross-linking on the left and right sides of the southern region, while preserving the data used in downstream analyses. For PA5, for instance, this results in the blue outline in Figure \ref{fig: masks}. For the other arrays, the mask is visually indistinguishable.

The somewhat ad-hoc procedure of this section is only necessary given the properties of the ACT scanning strategy cross-linking. Future surveys can mitigate the need for a boolean mask by ensuring more uniform cross-linking.

\section{Model Smoothing} \label{apx: smoothing}
\begin{figure}
    \centering
    \includegraphics[width=0.477\textwidth]{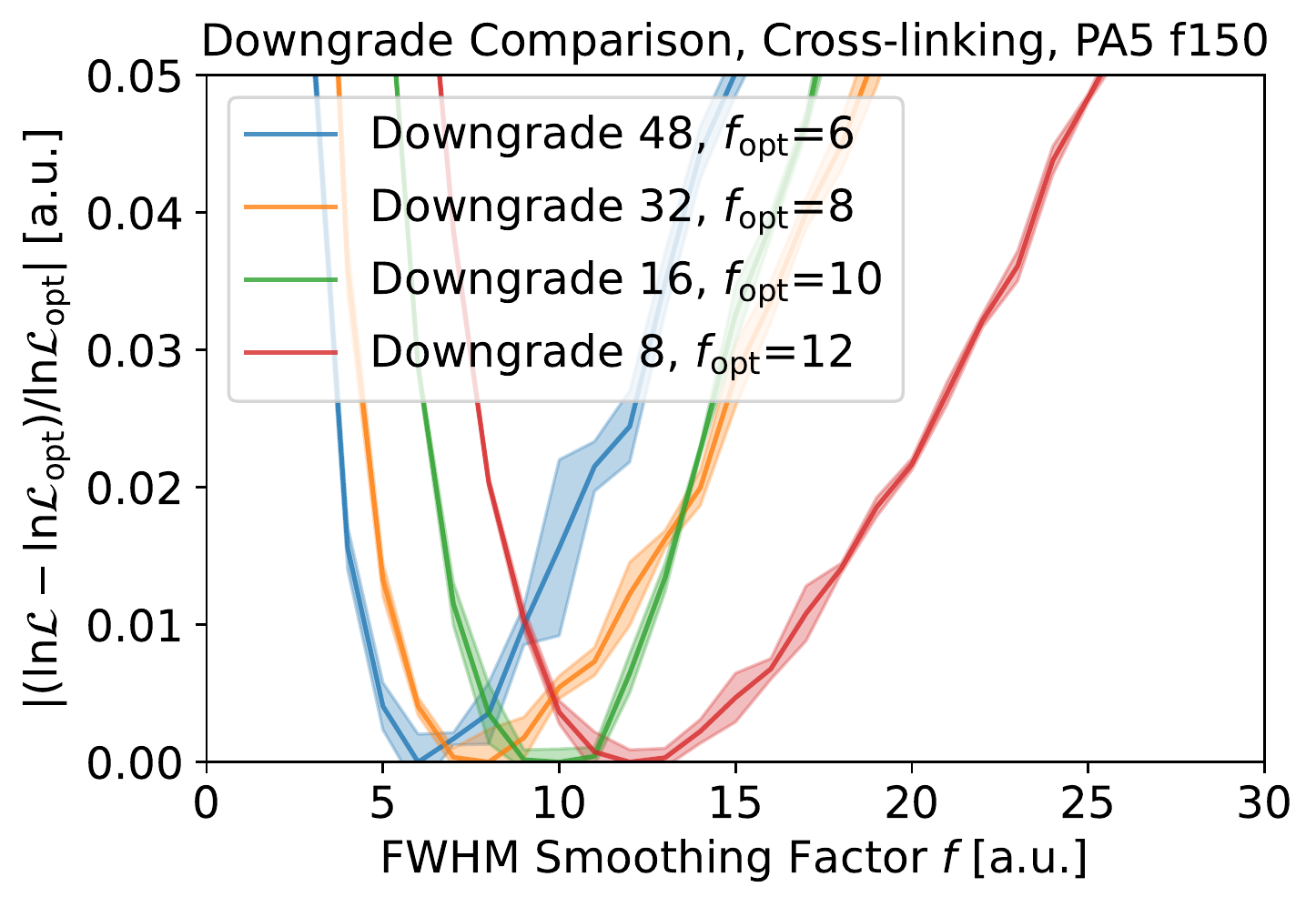}
    \includegraphics[width=0.497\textwidth]{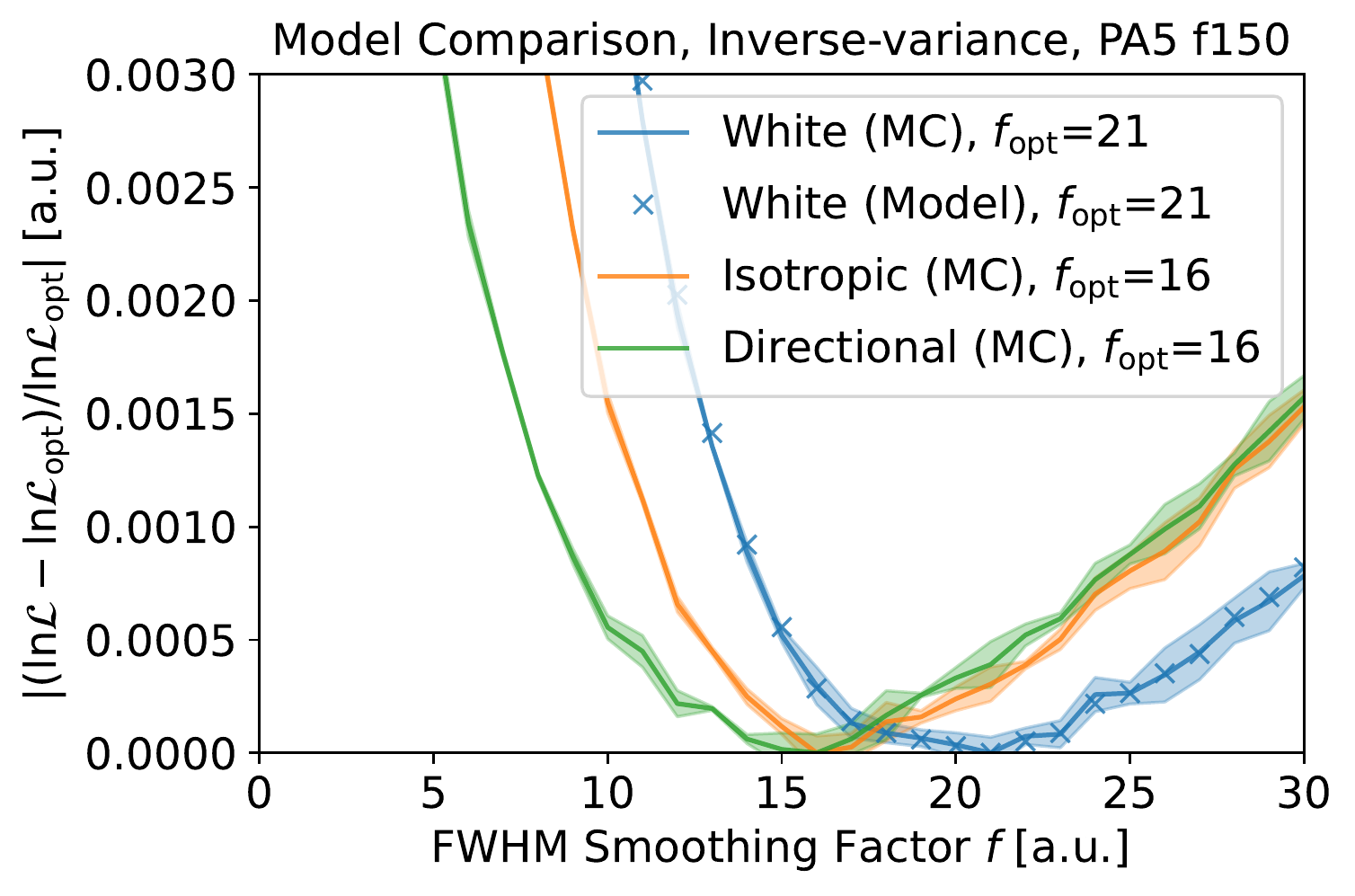}
    \caption{Model likelihoods ($-\mathrm{ln}\mathcal{L}$) as a function of smoothing factor $f$. Solid lines (shaded bands) denote the mean (1$\sigma$ range) over the $N_i=3$ noise realizations from the fiducial input covariance. Curves have been recentered and normalized relative to their minima to facilitate direct comparison. \textit{Left:} Likelihoods using the PA5 f150 cross-linking and inverse-variance combination map ($\bm\sigma_0^2=\mathbf f_p/\mathbf h$) as the input covariance. Different colors denote different choices of downgrade factors. \textit{Right:} Likelihoods for the set of noise modeling routines enumerated in the appendix, using the PA5 f150 inverse-variance map ($\bm\sigma_0^2=1/\mathbf h$) downgraded by a factor of four as the input covariance. The data labeled ``White (MC)" (``White (Model)") refer to the white-noise model's $\mathbf s_{i,\mathrm{MC}}^2$ map ($\mathbf s_{i,\mathrm{model}}^2$ map). The agreement between the curve and overplotted crosses demonstrates that 30 simulations are sufficient to reach convergence of Monte Carlo estimates. For the other models, we only have the option of using the Monte Carlo estimates. For both panels, the relatively small scatter over the $N_i=3$ noise realizations indicates the optima are robust.}
    \label{fig: smoothing_myth}
\end{figure}

This appendix expands on our discussion of how and why we smooth our wavelet noise models. As discussed in the text, we use a white-noise map space covariance, modulated by either the mapmaker cross-linking and/or inverse-variance maps, as the fiducial input to our optimization procedure. This simplification has a twofold advantage. First, we showed in \S\ref{sec: noise_properties} that the cross-linking and inverse-variance maps approximately capture the spatial noise structure at large and small scales, respectively. Since we do not know the true noise covariance of the DR6 noise, these products serve as realistic proxies. In particular, by downgrading these maps to specific, lower resolutions (and Nyquist bandlimits), we can emulate the noise covariance at specific wavelet scales. Second, by assuming the noise is white, we can evaluate the agreement between noise models and the input covariance using an analytic likelihood.

We proceed as follows:
\begin{enumerate}
    \item For a given array and frequency, select either the raw inverse-variance map alone, \mbf{h}, or the combination of \mbf{h} and the raw cross-linking map, \mbf{f_p}. Using Fourier space resampling, downgrade all maps by a given factor $d$, such that the Nyquist bandlimit of the maps is $\ell_{\mathrm{max}}=21,600/d$. In both cases, invert the downgraded inverse-variance map to obtain a ``variance" map, $1/\mathbf h$.
    \item In the first case, let the variance map define an an underlying, true, white-noise covariance map, $\bm\sigma_0^2=1/\mathbf h$. This map only has high variance when the inverse-variance is small. In the second case, let the underlying, true, white-noise covariance map be defined as $\bm\sigma_0^2=\mathbf f_p/\mathbf h$. Relative to the inverse-variance map, this map has high power when the cross-linking is also poor ($\mathbf f_p\sim 1$), following the discussion of Equation \ref{eq: var_nu_var_n_2}.
    \item Mask $\bm\sigma_0^2$ with the same boolean mask for the selected array as in \S\ref{sec: implementation_masks}. 
    \item Draw an uncorrelated, Gaussian realization from $\mathcal{N}(0, \bm\sigma_0^2)$, called \mbf{s_i}. Let this be analogous to the hypothetical $i$-th independent realization of the noise in the data.
    \item Feed \mbf{s_i} into one of several noise modeling routines, utilizing a fixed FWHM smoothing factor $f$ in all models:
    \begin{itemize}
        \item A white-noise model which is simply the pixel-wise square of \mbf{s_i}, $\mathbf s_{i,\mathrm{model}}^2$.
        \item The isotropic wavelet model of \S\ref{sec: noise_models_wav}, without the pseudospectrum-filtering step.
        \item The directional wavelet model of \S\ref{sec: noise_models_fdw}, without the pseudospectrum-filtering step.
    \end{itemize}
    \item Draw $N_{\mathrm{sim}}=30$ simulations, $\mathbf s_{i,\mathrm{sim}}$, and form the Monte Carlo estimated, white-noise covariance map $\mathbf s_{i,\mathrm{MC}}^2 = 1/N_{\mathrm{sim}}\sum\mathbf s_{i,\mathrm{sim}}^2$. For the two wavelet models, this is necessary because it is not possible to directly compare the model itself, which is in the wavelet basis, to the input covariance, $\bm\sigma_0^2$, which is in the map basis. For the white-noise model, as noted above, we retain both the Monte Carlo estimate, $\mathbf s_{i,\mathrm{MC}}^2$, as well the model itself, $\mathbf s_{i,\mathrm{model}}^2$, since the model is in the map basis. By comparing results using $\mathbf s_{i,\mathrm{MC}}^2$ to $\mathbf s_{i,\mathrm{model}}^2$ in the next step, this allows us to check that $N_{\mathrm{sim}}=30$ simulations is sufficient for the Monte Carlo estimates to converge (e.g., note the agreement between the blue crosses to the blue line in Figure \ref{fig: smoothing_myth}).
    \item For the white-noise model, the isotropic wavelet model, and the directional wavelet model, evaluate the likelihood of $\mathbf s_{i,\mathrm{MC}}^2$ given the true covariance $\bm\sigma_0^2$, $\mathcal{L}(\mathbf s_{i,\mathrm{MC}}^2 | \bm\sigma_0^2)$. For the white-noise model only, also evaluate $\mathcal{L}(\mathbf s_{i,\mathrm{model}}^2 | \bm\sigma_0^2)$. To avoid the poorly-observed edges of $\bm\sigma_0^2$, only perform the evaluation using pixels where the pseudospectrum mask from \S\ref{sec: implementation_masks} is nonzero. 
    \item Accumulate results over $N_i=3$ input noise realizations.
\end{enumerate}
The likelihood, $\mathcal{L}(\mathbf s_{i,\mathrm{MC}}^2 | \bm\sigma_0^2)$, follows the gamma distribution in each pixel $x$:
\begin{align}
    -\mathrm{ln}\mathcal{L}(\mathbf s_i^2 | \bm\sigma_0^2) = -\sum_{x}\mathcal{P}_{\mathrm{gamma}}(s_i^2 | \alpha=N_{\mathrm{sim}}/2, \theta=2\sigma_{0,x}^2/N_{\mathrm{sim}})
\end{align}
where $\mathcal{P}_{\mathrm{gamma}}$ is the gamma probability density with shape parameter $\alpha$ and scale parameter $\theta$. That is, for a given pixel with variance $\sigma_0^2$, we evaluate the likelihood that the average of $N_{\mathrm{sim}}=30$ independent samples of that variance equals the observed value. We minimize $-\mathrm{ln}\mathcal{L}$ as a function of the smoothing factor $f$, for various modeling routines and downgrade factors $d$. Example likelihoods are shown in the left panel of Figure \ref{fig: smoothing_myth}.

As a motivating example, we dispel the misconception that noise models with more wavelet kernels, and with thus with more parameters, require increased model smoothing. This view arises from the observation that, in the case of a single-frequency, unpolarized map, the directional model, with its 280 wavelet kernels, contains a factor $\sim$53 more parameters than the input data dimension. Compare that to the white-noise model we enumerated above, which has exactly the same number of parameters as the input data dimension. However, this argument overlooks that these 280 wavelet kernels are compact in Fourier space: their corresponding wavelet maps will therefore be smooth, with highly-correlated parameters. In fact, they will be correlated by exactly the right amount to guarantee that the number of degrees-of-freedom will be equal to the input data dimension. This is a consequence of Equation \ref{eq: admissibility}: with more kernels supporting a single element of the map space covariance matrix, Equation \ref{eq: admissibility} averages over equally more terms, and with proportionally smaller weights.

To confirm this expectation, we perform the above procedure for each of the listed noise modeling routines. We use inverse-variance maps downgraded by a factor of four as inputs, and examine the results in the right panel of Figure \ref{fig: smoothing_myth}. On the whole, we find the likelihood is generally insensitive to the noise modeling routine. In fact, the likelihood slightly prefers less smoothing as the number of wavelet kernels increases. This preference can also be understood as a consequence of the spatially-smooth wavelet kernels: they have a fundamentally limited ability to localize features in map space, and so enter the regime of ``over-smoothing" sooner than a perfectly-local white-noise model. We factor this limitation into our final choice of smoothing factors in \S\ref{sec: implementation_smoothing}.

\section{Benefits of Pseudospectrum Filter: Mode-decoupling and Stationarity} \label{apx: Nl_benefits}
\begin{figure}
    \centering
    \includegraphics[width=0.483\textwidth]{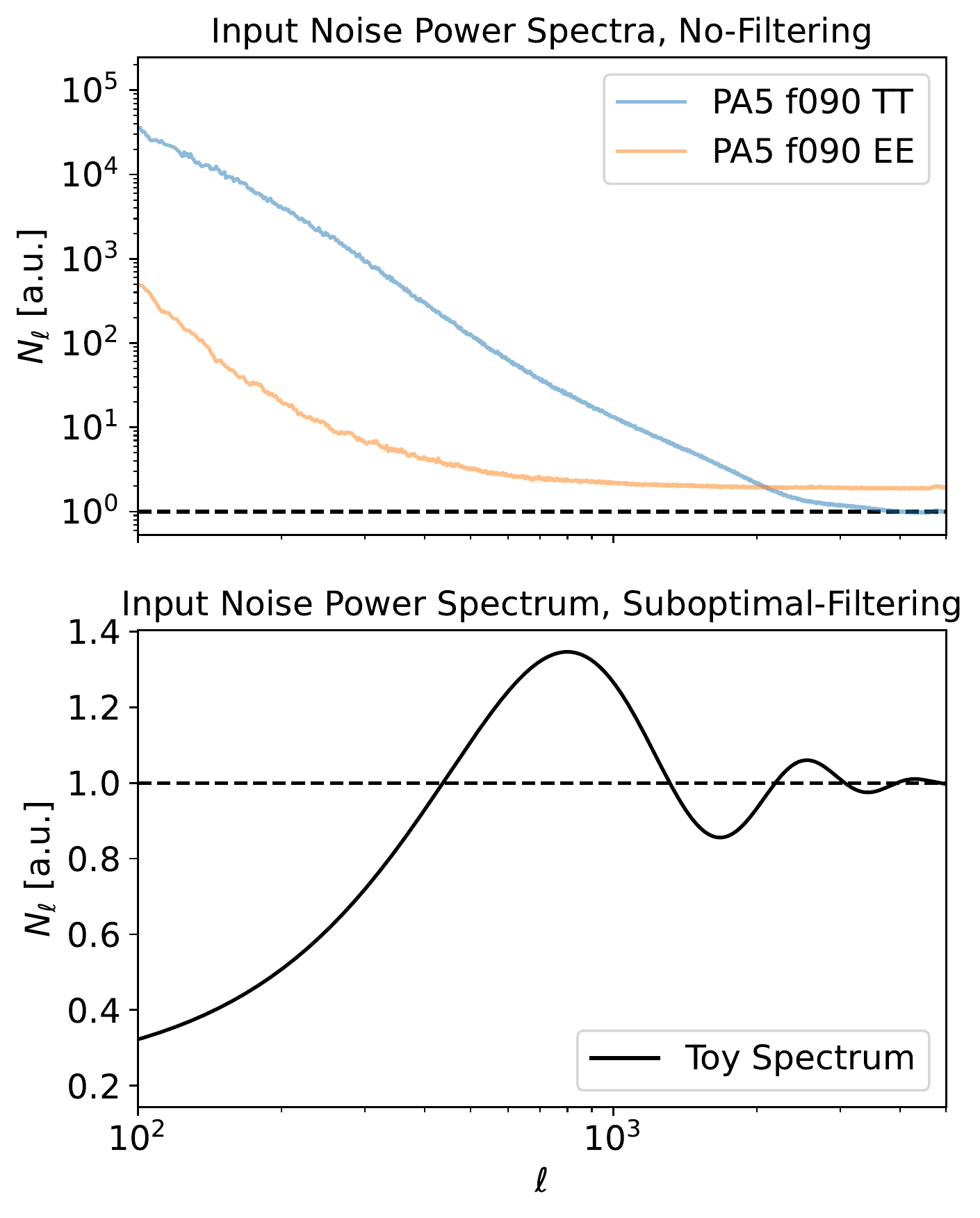}
    \includegraphics[width=0.49\textwidth]{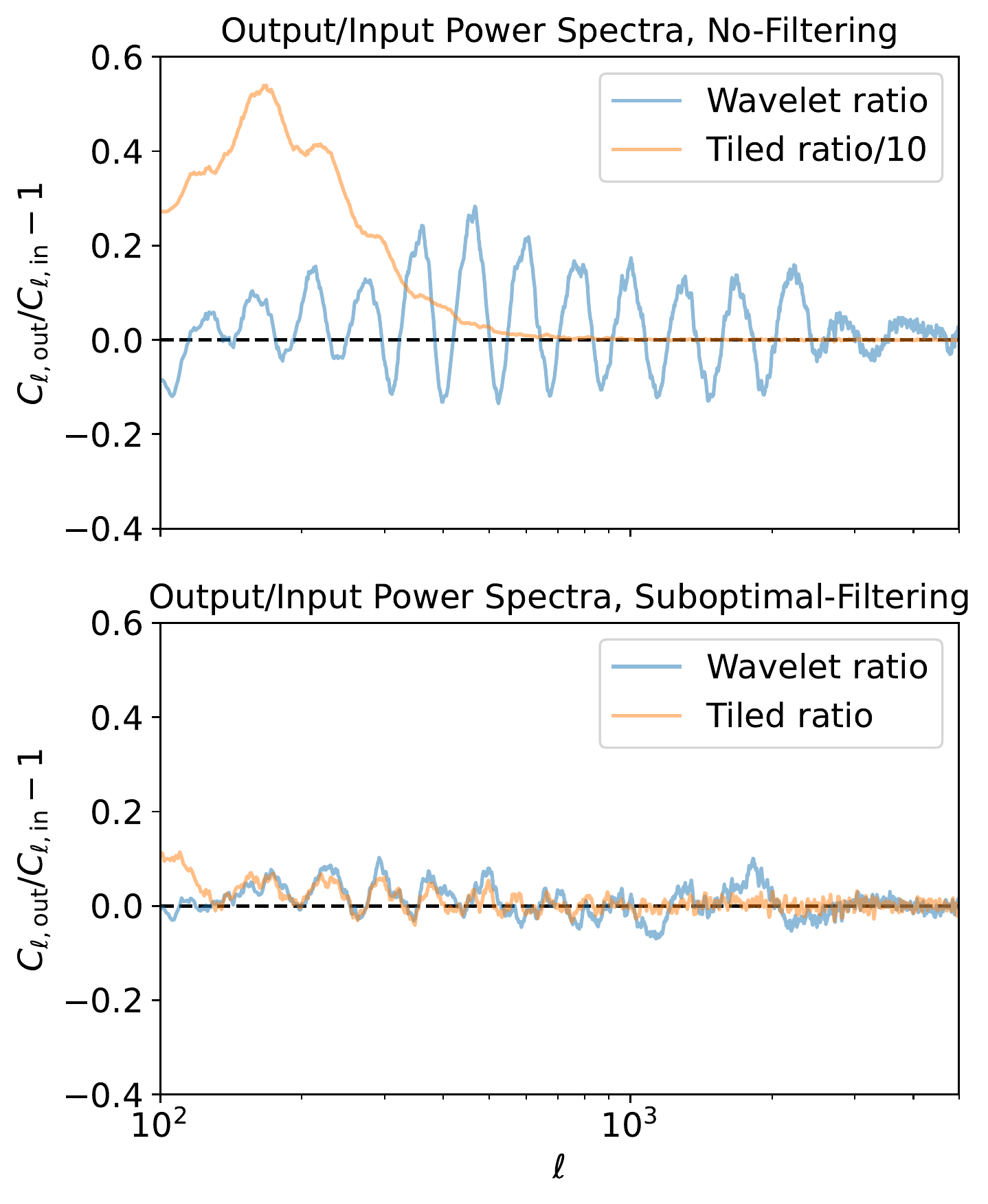}
    \caption{\textit{Top-left:} Noise pseudospectra from the actual data used as input power spectra for the no-filtering case. They are the mean over splits from the same curves as Figure \ref{fig: N_ell_r_ell}. The black dashed line indicates an arbitrary normalization to the spectra which has no effect on the results. \textit{Top-right}: Power spectra ratios between tiled and wavelet simulations, and their input noise realizations. The input realization from the tiled (wavelet) model is drawn from the EE (TT) spectrum. The tiled model deviation has been divided by 10 to fit on the same plot. Power spectra entering the ratios are smoothed with a tophat kernel of size $\Delta\ell=25$. \textit{Bottom-left:} Toy spectrum used as input for the suboptimal-filtering case. Now the black dashed line signifies that this spectrum represents the effect of an imperfect filter. \textit{Bottom-right:} Power spectra ratios for tiled and wavelet simulations, built off the same toy realization. The tiled model deviation is no longer divided by 10. Again, power spectra entering the ratios are smoothed with a tophat kernel of size $\Delta\ell=25$.} 
    \label{fig: Nl_benefits}
\end{figure}

The noise pseudospectrum filter is necessary for the construction of minimally biased noise models. We demonstrate that its exclusion leads to significant deviations from nominal in ratios of simulated versus input power spectra. We also demonstrate that our model estimation and simulation algorithms --- when including the filter --- suppress these deviations even if the filter is suboptimal. We examine the performance of a tiled and wavelet model in both cases.

We proceed as follows:
\begin{enumerate}
    \item We draw a full-sky, isotropic realization from an input fiducial noise power spectrum (bandlimited to $\ell_{\mathrm{max}}=5,400$). 
    \begin{itemize}
        \item For the no-filter, tiled (wavelet) model case, we use the PA5 f090 noise EE (TT) pseudospectrum.
        \item For the supoptimal-filter case, both models, we construct the input spectrum from scratch. The spectrum is much closer to flat than the actual noise spectra, but is not exactly flat (see Figure \ref{fig: Nl_benefits}). It is equivalent to the situation where we have drawn from the same unfiltered PA5 f090 noise spectra, and then applied an imperfect noise pseudospectrum filter, leaving some residual $\ell$-dependence to the filtered noise realization.
    \end{itemize} 
    \item We mask this realization with the same boolean footprint mask as the actual PA5 f090 data (see \S\ref{sec: implementation_masks}).
    \item From these realizations we generate a noise model as in \S\ref{sec: noise_models}, with some simplifications:
    \begin{itemize}
        \item For the tiled model, we exclude the inverse-variance map filter step since the input noise realization is isotropic. 
        \item For all cases, we exclude the pseudospectrum filter step.
        \item For all cases, we do not smooth the models at all. This is especially important for the tiled model whose sparse basis is in Fourier space: smoothing directly mixes power across scales and could also produce deviations in output/input power spectra ratios. For consistency, we do not smooth the wavelet model.
    \end{itemize}  
    \item From these models, we draw a single simulation as in \S\ref{sec: noise_models}, with the analogous simplifications.
    \item We compare the pseudospectra of the ouput noise simulation with the input noise realization, as measured in the apodized pseudospectrum mask of \S\ref{sec: implementation_masks}.
\end{enumerate}

The downsides of ignoring the pseudospectrum filter are self-evident in Figure \ref{fig: Nl_benefits}. As discussed in \S\ref{sec: noise_models_tile_algorithm}, multiplying each tile by an apodized mask in map space mixes modes in Fourier space. When the power spectrum of the signal in the tile is especially steep, as in the case of the PA5 f090 EE noise, a given mode receives excess power from nearby modes at slightly larger scales. In the simple demonstration here, we observe an excess in simulation power of $\sim$500\% near $\ell=200$ relative to the input noise power. Meanwhile, as discussed in \S\ref{sec: noise_models_wav_algorithm}, the wavelet model assumes stationary noise in harmonic space; this assumption is clearly violated by the strong $\ell$-dependence of the PA5 f090 TT noise power spectrum. Instead, the model projects out the stationary part of the noise --- that is, without an absolute $\ell$-dependence --- within each wavelet kernel. This produces a ``staircase-like" noise model in harmonic space, with a flat step on each wavelet kernel, rather than the smoothly decreasing input spectrum. Thus, we observe an oscillating ratio between the simulation and the input realization 

By including a (sub)optimal pseudospectrum filter we suppress these issues. As Figure \ref{fig: Nl_benefits} shows, even if the power spectrum of the input noise contains $\mathcal{O}$(1) variations over $\ell$, simulations drawn from the noise models are less biased over a large range of angular scales. Such a power spectrum could plausibly arise in \S\ref{sec: noise_models} after the pseudospectrum filtering step in regions of sky whose local noise power spectrum does not match the one used in the filter. We judge these improvements to be worth the cost of the filter's limited map space resolution discussed in \S\ref{sec: disc_conc}. Finally, we note that when applied to the real data, the pseudospectrum filter does not guarantee unbiased simulated power spectra, even if the filter perfectly matches the actual noise power spectrum. Rather, this appendix goes to show that such a filter is necessary, but not sufficient, for a performant noise model.

\end{document}